\documentclass[sigconf, nonacm, screen]{acmart}
\usepackage{float}
\usepackage{color}
\usepackage{multirow}
\usepackage{balance}
\usepackage{url}
\usepackage{graphicx}
\usepackage{booktabs}
\usepackage[shortlabels]{enumitem}
\setenumerate[1]{itemsep=0pt,partopsep=0pt,parsep=\parskip,topsep=1pt}
\setitemize[1]{itemsep=0pt,partopsep=0pt,parsep=\parskip,topsep=1pt}
\setdescription{itemsep=0pt,partopsep=0pt,parsep=\parskip,topsep=1pt}

\usepackage[justification=centering]{caption}
\usepackage{subcaption}
\usepackage{amsmath}
\usepackage{amsfonts}
\usepackage{textcomp}

\usepackage{algorithm}
\usepackage[noend]{algpseudocode}

\usepackage{amsthm}

\newtheorem{definition}{\bf Definition}
\newtheorem{example}{\bf Example}
\newtheorem{problem}{\bf Problem}

\newcommand\ExpCaption[1]{%
     \captionsetup{font=footnotesize}%
     \caption{#1}}
     
\renewcommand{\mathtt}{\texttt}

\begin{document}
\title{Towards Crowd-aware Indoor Path Planning (Extended Version)}

\author{Tiantian Liu$^{\dagger}$,  Huan Li$^{\dagger}$, Hua Lu$^{\ddagger}$, Muhammad Aamir Cheema$^{\S}$, Lidan Shou$^{\natural}$}
\affiliation{%
\institution{
{\fontsize{10}{10}\selectfont $^{\dagger}$}Department of Computer Science, Aalborg University, Denmark\\
{\fontsize{10}{10}\selectfont $^{\ddagger}$}Department of People and Technology, Roskilde University, Denmark\\
{\fontsize{10}{10}\selectfont $^{\S}$}Faculty of Information Technology, Monash University, Australia\\
{\fontsize{10}{10}\selectfont $^{\natural}$}College of Computer Science, Zhejiang University, China\\
{\fontsize{9}{9}\selectfont\ttfamily\upshape} \{liutt,lihuan\}@cs.aau.dk, luhua@ruc.dk, aamir.cheema@monash.edu, should@zju.edu.cn
}}

\begin{abstract}
Indoor venues accommodate many people who collectively form crowds. Such crowds in turn influence people's routing choices, e.g., people may prefer to avoid crowded rooms when walking from A to B. This paper studies two types of crowd-aware indoor path planning queries. The Indoor Crowd-Aware Fastest Path Query (\texttt{FPQ}) finds a path with the shortest travel time in the presence of crowds, whereas the Indoor Least Crowded Path Query (\texttt{LCPQ}) finds a path encountering the least objects en route. To process the queries, we design a unified framework with three major components. First, an indoor crowd model organizes indoor topology and captures object flows between rooms. Second, a time-evolving population estimator derives room populations for a future timestamp to support crowd-aware routing cost computations in query processing. Third, two exact and two approximate query processing algorithms process each type of query.
All algorithms are based on graph traversal over the indoor crowd model and use the same search framework with different strategies of updating the populations during the search process.
All proposals are evaluated experimentally on synthetic and real data. The experimental results demonstrate the efficiency and scalability of our framework and query processing algorithms.
\end{abstract}

\pagestyle{plain}
\maketitle

\renewcommand*{\shortauthors}{Tiantian Liu et al.}

\section{Introduction}
\label{sec:intro}

Indoor route planning queries are among the most fundamental queries underlying indoor location-based services (LBS)~\cite{lu2012foundation,shao2016vip,luo2016time,zhou2018indoor,li2018trips}. Such queries can facilitate people in need. For example, in an airport or a train station, passengers prefer to find the fastest path from their current position to the boarding gate.
In addition to the shortest or fastest paths, indoor routing supports many variations that meet practical needs.
For instance, customers in a shopping mall would like to find a path that can cover some given keywords like a coffee shop and shoes~\cite{feng2020indoor}.
Meanwhile, indoor venues accommodate many people who collectively form crowds that may in turn influence people's routing choices.
For example, crowds may influence one's moving speed, which will have an effect on the travel time of a path. In some places like an airport where passengers are sensitive to travel time, a topologically shortest path may still incur the too long time and result in missing flight if the path fails to consider the effect of crowds.
In other scenarios, people en route may prefer to encounter fewer people. For example, during the COVID-19 pandemic, people would like to find a path to avoid human contact as much as possible. As another example, autonomous objects (e.g., driverless cars in an airport) also prefer a path with fewer people en route to mitigate the interference and inconvenience caused by contact with people.

\begin{figure}[!ht]
    \centering
    \includegraphics[width=0.9\columnwidth]{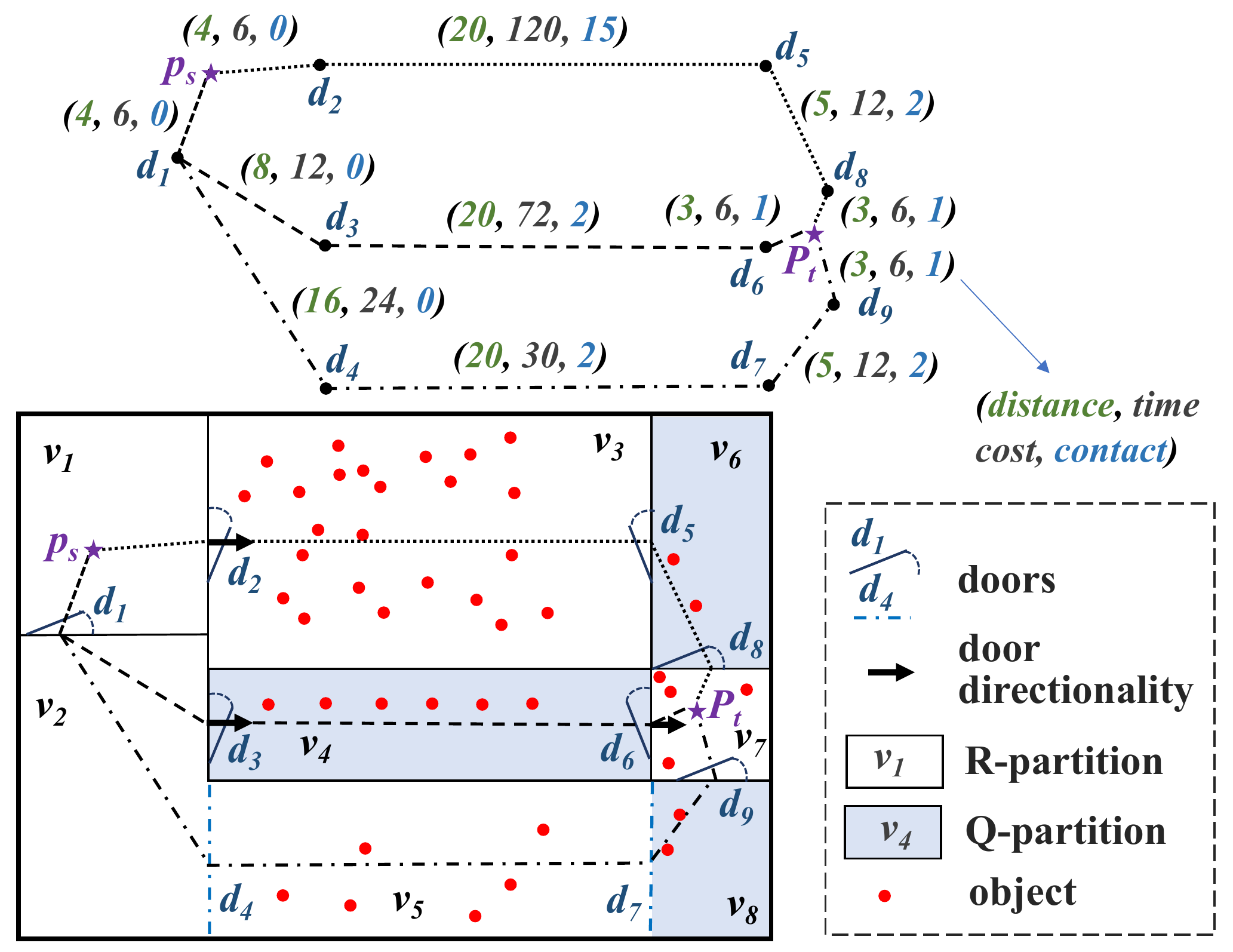}
    \caption{An Example of Floorplan at Query Time $t_q$}
    \label{fig:floorplan}
\end{figure}

In this paper, we formulate and study two crowd-aware indoor path planning queries. Referring to Figure~\ref{fig:floorplan}, given a source point $p_s$, a target point $p_t$, and a query time $t$, an Indoor Crowd-Aware Fastest Path Query (\texttt{FPQ}) returns a path with the shortest travel time in the presence of crowds, whereas an Indoor Least Crowded Path Query (\texttt{LCPQ}) returns a path that encounters the least objects en route. As an indoor path is essentially a series of indoor partitions (basic topological units like rooms), 
\texttt{FPQ}'s routing cost is partition-passing time, whereas an \texttt{LCPQ}'s is partition-passing contact.

We consider two types of indoor partitions.
A \emph{Queue Partition} (Q-partition) requests objects to enter and exit in a line, e.g., a security-check line in an airport or a ticketing entrance in a theater.
A \emph{Random Partition} (R-partition) refers to a more general case where there is no restriction on how to pass the partition but one's movement slows down when encountering a crowd.
Due to the different topological natures of the Q-partitions and R-partitions, partition-passing time for \texttt{FPQ} and partition-passing contact for \texttt{LCPQ} should be defined differently for the two partition types.

Existing techniques cannot handle the novel \texttt{FPQ} and \texttt{LCPQ}.
First, techniques for outdoor route planning~\cite{hall1986fastest, brodal2004time, wang2015efficient, ding2008finding, ardakani2012decremental, nannicini2012bidirectional, ardakani2015decremental, wei2020architecture} do not work for indoor spaces with distinct entities like doors, walls, and rooms, altogether forming a complex topology.
Second, the existing indoor route planning methods~\cite{lu2012foundation,shao2016vip,shao2018trip,feng2020indoor} do not consider the effect of crowds, lacking the modeling foundation for \texttt{FPQ} and \texttt{LCPQ}.
Third, some works study indoor flow and density~\cite{li2018finding,li2018search}, but they do not touch upon route planning.

To solve \texttt{FPQ} and \texttt{LCPQ} efficiently, we design a crowd-aware query processing framework. The framework is composed of three layers.
First, the indoor crowd model is the foundation layer of the framework. The model can handle three kinds of information, namely indoor topology, indoor geometry, and crowd-evolution.
A time-evolving population estimator derives the future flows and populations for indoor partitions. The estimated values are used as basic routing costs in \texttt{FPQ} and \texttt{LCPQ}.
The query processing layer consists of two parts. In part one, two functions, namely partition-passing time function and partition-passing contact function, calculate the routing cost for \texttt{FPQ} and \texttt{LCPQ}, respectively. Again, the differences in routing costs for different partition types are unified into a single computing process.
On top of that, part two provides two exact and two approximate path search algorithms that each can process both query types.
One of the exact searches uses a global estimator, whereas the other uses an estimator that only estimates a partition’s population by looking up its upstream partitions' flows.
The two approximate algorithms speed up query processing at the cost of query accuracy. One of them only derives populations for partial partitions, and the other derives populations only when necessary. All proposed techniques are experimentally evaluated on synthetic and real data. The experimental results demonstrate the efficiency and scalability of the proposed framework and query processing algorithms. The results also show the two approximate search algorithms achieve good routing accuracy.

This paper makes the following key contributions.
\begin{itemize}[leftmargin=*]
	\item We formulate Indoor Crowd-Aware Fastest Path Query (\texttt{FPQ}) and Indoor Least Crowded Path Query (\texttt{LCPQ}), and propose a unified processing framework for these queries (Section~\ref{sec:pre}).
	\item We design an indoor crowd model that organizes indoor topology and captures indoor partition flows and densities (Section~\ref{sec:model}). 
    \item We devise a time-evolving population estimator to derive future time-dependent flows and populations for partitions (Section~\ref{sec:cost_estimation}).
	\item We design two exact and two approximate query processing algorithms that each can process both query types (Section~\ref{sec:queries}).
	\item We conduct extensive experiments on synthetic and real datasets to evaluate our proposals (Section~\ref{sec:experiment}).
\end{itemize}
In addition, Section~\ref{sec:related} reviews the related work and Section~\ref{sec:conclusion} concludes the paper.

\section{Preliminaries}
\label{sec:pre}

Table~\ref{tab:notations} lists the notations frequently used in this paper.

\begin{table}[htbp]
\caption{Notations}\label{tab:notations}
\centering
\begin{tabular}{@{}l|l@{}}
\hline
\textbf{Symbol} & \textbf{Meaning}\\
\hline
$v$, $d$, $p$ & Partition, door, and indoor point\\
$o$, $O$ & Object, object set\\
$C$ & Indoor crowd\\
$t_{o \triangleright C}$, $t_{C \triangleright o}$ & The times $o$ joins and leaves $C$\\
$\mathit{RT}(d_i)$ & The sequence of $d_i$'s report timestamps \\
$\mathit{UTI}(v_k)$ & The set of $v_k$'s unit (update) time intervals \\
$\delta_{t_x, t_{x+1}}(v_k)$ & $v_k$'s density over $[t_x, t_{x+1}]$ \\
$\rho(v_k, t_c)$ & $v_k$'s lagging coefficient at time $t_c$ \\
$T(d_i, d_j, v_k, t_c)$ & The time to pass $v_k$ from $d_i$ to $d_j$ at $t_c$ \\
$\kappa(d_i, d_j, v_k, t_c)$ & The object contact to pass $v_k$ from $d_i$ to $d_j$ at $t_c$ \\
\hline
\end{tabular}
\end{table}

\subsection{Indoor Crowds}
\label{ssec:indoor_space}

An indoor space is divided by walls and doors into \emph{indoor partitions}.
A \textbf{Queue Partition} (Q-partition) requests objects to enter and leave \emph{sequentially}, while a \textbf{Random Partition} (R-partition) has no such a restriction and objects can enter and leave it \emph{randomly}.
Note that the type of partition is usually fixed and will change only when the space layout is redesigned.
The issue of topological change is out of the scope of this paper.

Within a partition, moving objects (e.g., persons) may form a crowd.
Corresponding to the partition types, we formally define an \emph{indoor crowd} as follows.

\begin{definition}[Indoor Crowd]
  An indoor crowd~$C_{t_s, t_e}(v_k)$~\footnote{When time is not of particular interest, we use $C_k$ to denote $v_k$'s associated crowd.} is a set of moving objects in a partition $v_k$ during a certain time interval $[t_s, t_e]$. $C.\tau$ denotes the type of crowd $C$.
  \begin{enumerate}[leftmargin=*, label=\arabic*)]
    \item In a \textbf{Queue Crowd} (Q-crowd), objects join and leave the crowd in the first-in-first-out (FIFO) manner. Formally, $\forall o_i, o_j \in C_k \wedge C_k.\tau = \mathsf{Q}, t_{o_i \triangleright C_k} \leq t_{o_j \triangleright C_k} \Rightarrow t_{C_k \triangleright o_i} \leq t_{C_k \triangleright o_j}$, where $t_{o_i \triangleright C_k}$ and $t_{C_k \triangleright o_i}$ is the time $o_i$ joins and leaves $C_k$, respectively.
    \item In a \textbf{Random Crowd} (R-crowd), objects join and leave the crowd without any ordering restrictions. Formally, $\exists o_i, o_j \in C_k \wedge C_k.\tau = \mathsf{R}, t_{o_i \triangleright C_k} < t_{o_j \triangleright C_k}, t_{C_k \triangleright o_i} \geq t_{C_k \triangleright o_j}$.
  \end{enumerate}
\end{definition}

A crowd changes as objects join and leave from time to time.
In other words, the object population and density of a partition are time-varying, rendering an object's routing cost passing the partition to change as well.
Therefore, for crowd-aware routing, it is of fundamental importance to know a crowd's dynamic population or density. This demands dynamic data from a localization system.

However, a localization system may not record the exact trajectory or join/leave time of each individual object due to computing/storage limitations and location privacy concerns.
Alternatively, a system may maintain the current number of objects in each partition (or a crowd) and records the number of objects joining and leaving during a time interval.
This can be easily achieved, e.g., by installing a counter at a door.
In our setting, each door counter reports objects' joining and leaving at a predefined frequency. This means that the object numbers in a crowd are updated at a number of discrete timestamps. Specifically, we use a time-ordered sequence $\mathit{RT}(d_i) = ( t_{i1}, \ldots, t_{in} )$ to denote the \textbf{report timestamps} of the counter at door $d_i$.
As a result, the \textbf{update timestamps} relevant to a partition $v_k$ is a time-ordered sequence $\mathit{UT}(v_k) = \bigcup_{d_j \in \mathit{P2D}(v_k)} \mathit{RT}(d_j)$ where $\mathit{P2D}(v_k)$ refers to all doors of partition $v_k$.
Each two consecutive timestamps in the sequence $\mathit{UT}(v_k)$ forms an \textbf{unit (update) time interval}. The set of all such intervals from $\mathit{UT}(v_k)$ is denoted by $\mathit{UTI}(v_k)$.

At the routing query time, it is necessary to know the flows in the future.
However, future exact object numbers from door counters are unavailable at that moment.
To this end, we employ door flow functions to model the crowd-evolution (detailed in Section~\ref{ssec:door_flow}).

We define a partition's \emph{time-parameterized density} as follows.
\begin{definition}[Time-Parameterized Density]
  Given a partition $v_k$ and its unit time interval $[t_x, t_{x+1}] \in \mathit{UTI}(v_k)$, its \emph{time-parameterized density} over $[t_{x}, t_{x+1}]$ is $\delta_{t_{x}, t_{x+1}}(v_k) = |C_k| / \textit{Area}(v_k)$, where $|C_k|$ is $v_k$'s population over $[t_{x}, t_{x+1}]$ and $\textit{Area}(v_k)$ is $v_k$'s area.
\end{definition}

The population and density in this paper are time-parameterized unless mentioned otherwise.
A partition $v_k$'s density at an arbitrary timestamp $t_c$ is estimated with respect to the unit time interval covering $t_c$. Specifically, we have
$\delta_{t_c}(v_k) = \delta_{t_{x}, t_{x+1}}(v_k)$ where $t_{x} \leq t_c < t_{x+1}, [t_{x}, t_{x+1}] \in \mathit{UTI}(v_k)$.

\subsection{Problem Formulation}
\label{ssec:problem}

In an indoor routing problem, a basic step is to move from one door to another through their in-between partition. To measure the cost to pass a partition, the intra-partition \textbf{door-to-door distance}~\cite{feng2020indoor} for two doors $d_i$ and $d_j$ is
\begin{equation}\label{equation:d2d}
\begin{aligned}
\mathit{d2d}(d_i, d_j) =
\begin{cases}
|d_i, d_j|_E,  & \mbox{if }\mathit{D2P}_\sqsupset(d_i) \cap \mathit{D2P}_\sqsubset(d_j) \not= \varnothing; \\
\infty, & \mbox{otherwise}.
\end{cases}
\end{aligned}
\end{equation}
where $\mathit{D2P}_\sqsupset(d_i)$ gives the set of partitions that one can enter through door $d_i$ and $\mathit{D2P}_\sqsubset(d_j)$ gives those that one can leave through door $d_j$.
Therefore, $\mathit{D2P}_\sqsupset(d_i) \cap \mathit{D2P}_\sqsubset(d_j) \not= \varnothing$ means $d_i$ and $d_j$ share a common partition that one can enter via $d_i$ and leave via $d_j$. In this case, the Euclidean distance is used between $d_i$ and $d_j$. Otherwise, the distance between them is set to infinite.

\begin{definition}[Indoor Path]
  An indoor path from the source $p_s$ to the target $p_t$ is $\phi = (p_s, d_x, \ldots, d_y, p_t)$, where $(d_x, \ldots, d_y)$ is a door sequence, $d_x$ is a leaveable door of $p_s$'s host partition, $d_y$ is an enterable door of $p_t$'s host partiton, and each two consecutive doors $d_n, d_{n+1}$ ($x \leq n < y$) on $\phi$ have $\mathit{D2P}_\sqsupset(d_n) \cap \mathit{D2P}_\sqsubset(d_{n+1}) \not= \varnothing$. Each two consecutive path nodes form a \emph{path segment}. The distance of $\phi$ is computed as $\mathit{dist}_\phi = |p_s, d_x|_E + \sum_{n=x}^{y-1}\mathit{d2d}(d_n ,d_{n+1}) + |d_y, p_t|_E$.
\end{definition}

When there is no crowd, the basic time cost of passing an in-between partition $v_k$ from $d_i$ to $d_j$ can be estimated based on the average object moving speed $\bar{s}$, i.e., $T^{(b)}(d_i, d_j) = \mathit{d2d}(d_i, d_j) / \bar{s}$.
To reflect a crowd's impact, we use the \textbf{lagging coefficient} $\rho(v_k, t_c)$ that takes into account the crowd's density and type as follows.
\begin{equation}\label{equation:lagging}
  \begin{aligned}
  \rho(v_k, t_c) =
  \begin{cases}
  1 + e^{\delta_{t_c}(v_k) / \mathsf{D}^\mathit{max}_k},  & \mbox{if~} C_k.\tau = \mathsf{Q};\\
  1 + e^{(\delta_{t_c}(v_k) / \mathsf{D}^\mathit{max}_k)^2}, & \mbox{otherwise}.
  \end{cases}
  \end{aligned}
\end{equation}
where $\delta_{t_c}(v_k)$ is $v_k$'s density at time $t_c$ and $\mathsf{D}^\mathit{max}_k$ corresponds to the maximum density\footnote{The maximum capacity (and therefore the maximum density) of a partition is usually known, such as the room capacity for fire safety.} of $v_k$.
For a Q-partition, the ratio $\delta_{t_c}(v_k) / \mathsf{D}^\mathit{max}_k$ is applied to reflect the crowding degree.
We modify the speed-density model~\cite{virkler1994pedestrian} to calculate the lagging coefficient in Equation~\ref{equation:lagging} which reflects real-world scenarios, e.g., in common sense, a crowd usually impacts people's moving speed and results in longer travel time. Equation~\ref{equation:lagging} guarantees that the coefficient is always greater than 1 and it increases monotonically as $v_k$'s density increases. For an R-partition, the square of the ratio is used because R-crowds incur less lagging effect.

Note that other forms of lagging coefficients can be defined and supported within our framework, e.g., lagging can be multiplied by the object number for a queue crowd. Since the lagging coefficient is \emph{not} our research focus, we simply apply Equation~\ref{equation:lagging} in this study.

Using the lagging coefficient, we can calculate our crowd-aware and time-dependent \textbf{partition-passing time} as follows.
\begin{equation}\label{equ:crossing_cost}
  T(d_i, d_j, v_k, t_c) = T^{(b)}(d_i, d_j) \cdot \rho(v_k, t_c)
\end{equation}
An object needs longer time to pass a more crowded partition.

As a special case, we replace $d_i$ with $p_s$ or replace $d_j$ with $p_t$ in Equation~\ref{equ:crossing_cost}, to estimate the cost of a path segment starting with $p_s$ or ending with $p_t$. Accordingly, $v_k$ is the host partition of $p_s$ or $p_t$.

With the partition-passing time, we can plan the fastest indoor path for users to avoid undesirable congestion caused by indoor crowds. An indoor path $\phi$'s \emph{overall travel time} $T_{\phi}$ is computed as the sum of the time of passing the partition between each path segment on $\phi$.
The fastest path query problem is defined as follows.
\begin{problem}[Indoor Crowd-Aware Fastest Path Query \texttt{FPQ}]
  Given a source $p_s$ and a target $p_t$, an \emph{indoor crowd-aware fastest path query} $\texttt{FPQ}(p_s, p_t, t)$ returns a path $\phi (p_s, d_i, \ldots, d_j, p_t)$ such that a) the overall travel time $T_\phi$ is minimized and b) $\phi$ is the shortest among all satisfying a). Formally, $\nexists \phi' \neq \phi$, $T_{\phi'} \leq T_{\phi} \wedge \mathit{dist}_{\phi'} < \mathit{dist}_{\phi}$.
\end{problem}

Note that the partition-passing time is determined by the time one arrives at that partition, while the arrival time, in turn, is dependent on the partition-passing time of the previous partition.
This calls for on-the-fly computation during the search to obtain the overall travel time $T_{\phi}$, which is to be detailed in Section~\ref{sec:queries}.

\begin{example}
  Figure~\ref{fig:floorplan} illustrates an indoor space at time $t_q$.
  The query time and crowd-evolution snapshot are considered. We indicate the distance, partition-passing time and object contact on each path segment in the top sketch.
  We suppose that there are some events in $v_7$, and $v_4$, $v_6$ and $v_8$ are Q-partitions for ID check before entering $v_7$.
  Given a query $\texttt{FPQ}(p_s, p_t, t_q)$, there are three candidate paths, namely $\phi_1(p_s, d_2, d_5, d_8, p_t)$, $\phi_2(p_s, d_1,$ $d_3, d_6,p_t)$, and $\phi_3 (p_s, d_1, d_4, d_7, d_9, p_t)$.
  Only considering the distance but not the impact from crowds, $\phi_1$ is the shortest with a length of 32 meters, while those of $\phi_2$ and $\phi_3$ are 35 meters and 48 meters, respectively.
  However, $\phi_1$ is not expected to be the fastest path when crowds are concerned. To be specific, $\phi_1$ goes through a highly crowded R-partition $v_3$, incurring a total travel time of 144 seconds. For $\phi_2$, the low-populated Q-partition $v_4$ with a long queue is involved, making the total time cost be 96 seconds. Among all, $\phi_3$ is expected to be the fastest with an overall cost of 78 seconds, though it is the longest distance passing 5 partitions.
\end{example}

Another practically interesting problem is to find the shortest path that contacts the least objects. E.g., it is useful to find a path that avoids human contact as much as possible in the COVID-19 case.
Given a path segment $(d_i, d_j)$ that goes through a partition $v_k$, we calculate the \textbf{partition-passing contact} as follows.
\begin{equation}\label{equ:crossing_contact}
  \begin{aligned}
    & \kappa(d_i, d_j, v_k, t_c) = \\ 
    & \begin{cases}
    (|d_i, d_j|_E \cdot \texttt{w}) \cdot \delta_{t_c}(v_k),  & \mbox{if }C_k.\tau = \mathsf{R}; \\
    (\texttt{w} / |d_i, d_j|_E) \cdot (\delta_{t_c}(v_k) \cdot \textit{Area}(v_k)), & \mbox{otherwise}.
    \end{cases}
  \end{aligned}
\end{equation}
Given a partition $v_k$, its enterable door $d_i$, and its leaveable door $d_j$, for any object reaching $d_i$ at time $t_c$, the partition-passing contact to pass $v_k$ and reach $d_j$ is defined in terms of the number of objects covered by the buffer of the path segment.
The contact to pass an R-partition is the partition density multiplied by the buffer area that is approximated as $|d_i, d_j|_E \cdot \texttt{w}$ where $\texttt{w}$ is the buffer width.
The contact to pass a Q-partition is the objects within the $\texttt{w}$ long queue line centered at the user's position, i.e., the proportion $\texttt{w} / |d_i, d_j|_E$ of the total objects in the queue.
This reflects common sense. For example, if we pass a random crowd, the close contacts are those who we meet in the buffer width. If we pass a queue crowd, we only have close distance with those in front of or behind us.

In our implementation, we set $\texttt{w}$ as the unit distance of 1m. For example, many countries suggest people keep a physical distance of 1m in the COVID-19 pandemic.
Similar to the computation of the overall travel time $T_\phi$, an indoor path $\phi$'s \emph{overall contact} $\kappa_\phi$ is computed as
the sum of the partition-passing contacts of path segments on $\phi$.
Likewise, Equation~\ref{equ:crossing_contact} applies to the path segment starting with $p_s$ and ending with $p_t$.
Accordingly, we formulate the least crowded path query as follows.

\begin{problem}[Indoor Least Crowded Path Query \texttt{LCPQ}]
  Given a source $p_s$ and a target $p_t$, an \emph{indoor least crowded path query} $\texttt{LCPQ}(p_s, p_t, t)$ returns a path $\phi (p_s, d_i, \ldots, d_j, p_t)$ such that a) the overall contact is the least, and b) $\phi$ is the shortest among all satisfying a). Formally, $\nexists \phi' \neq \phi$, $\kappa_{\phi'} \leq \kappa_{\phi} \wedge \mathit{dist}_{\phi'} < \mathit{dist}_{\phi}$.
\end{problem}

\begin{example}
  Consider a query $\texttt{LCPQ}(p_s, p_t, t_q)$ in Figure~\ref{fig:floorplan}, the candidates $\phi_1$, $\phi_2$ and $\phi_3$ involve 18, 3 and 5 contacts from the partitions which they pass, respectively.
  The query returns $\phi_2$ since it contacts the fewest objects.
\end{example}

\subsection{Solution Framework}

\begin{figure}[htbp]
    \centering
    \includegraphics[width=0.95\columnwidth]{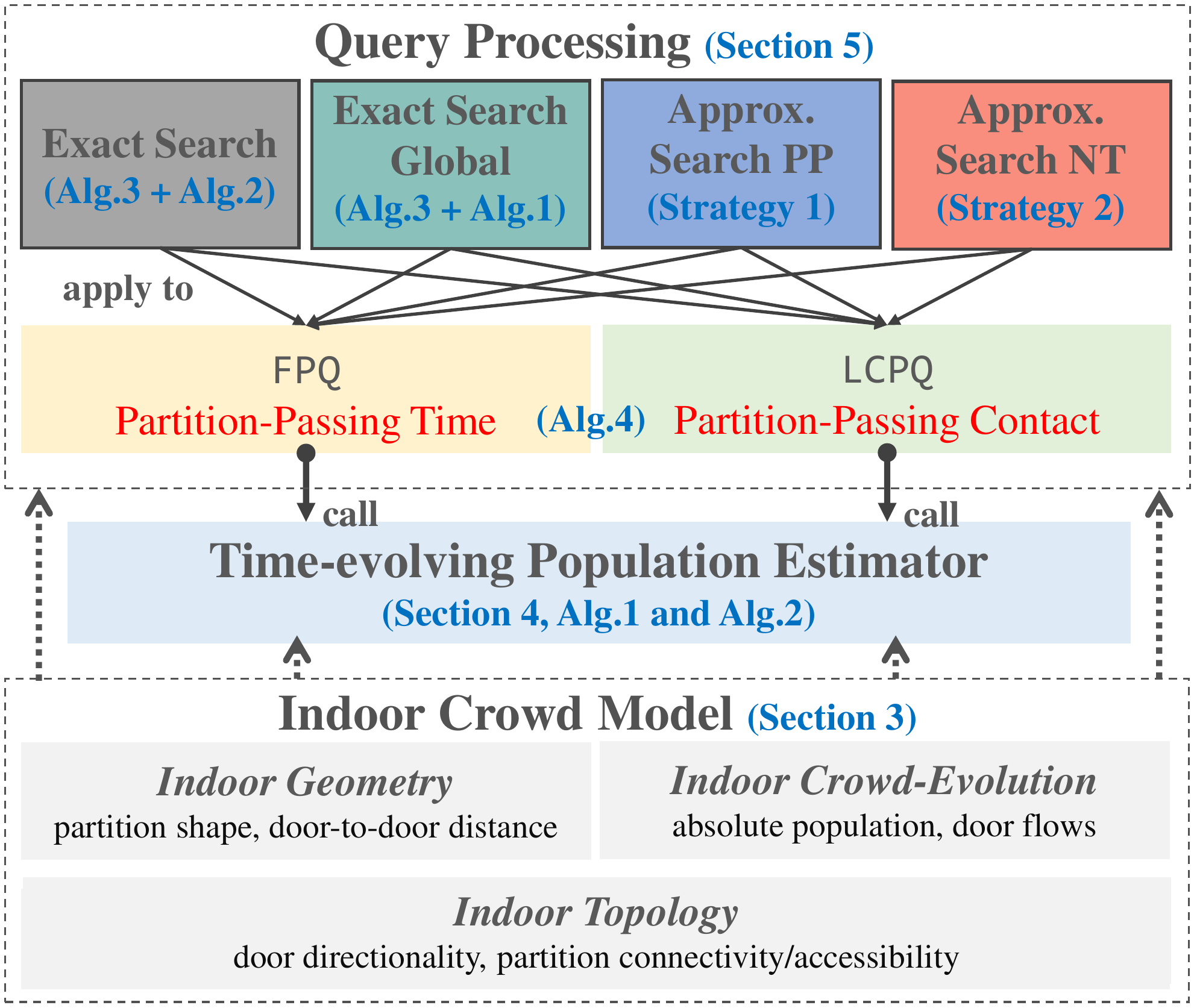}
    \caption{Crowd-Aware Path Planning Framework}
    \label{fig:architecture}
\end{figure}

We propose a crowd-aware query processing framework as illustrated in Figure~\ref{fig:architecture}.

In the bottom, an indoor crowd model (cf. Section~\ref{sec:model}) maintains the following aspects of an indoor space: \emph{Indoor Topology} that captures the directionality of doors and connectivity/accessibility of partitions, \emph{Indoor Geometry} that records the shapes of partitions and walking distances between two doors, and \emph{Indoor Crowd-Evolution} that models the objects joining and leaving the crowds.

Enabled by the indoor crowd model, a \emph{time-evolving population estimator} in the middle layer derives populations (and densities) of partitions at a future time and provides them to the query algorithms.
The population estimation process will be detailed in Section~\ref{sec:cost_estimation}.

In the top layer, crowd-aware search algorithms process \texttt{FPQ} and \texttt{LCPQ}.
Both algorithms are based on graph traversal over the indoor crowd model.
To expand to the next path node with the minimum cost, \texttt{FPQ}'s algorithm estimates the partition-passing time, while \texttt{LCPQ} search algorithm estimates the partition-passing contact.
Both costs are estimated based on the time-evolving populations derived in the middle layer.
For both queries, two exact and two approximate search algorithms are proposed.
Their main difference lies in the strategy of updating population(s) during the search.
All search algorithms will be presented in Section~\ref{sec:queries}.
Thanks to modular construction, our framework can be easily extended or reduced. For example, to support regular path planning, we only need the components Indoor Topology and an appropriate query processing algorithm that can be a variant of Algorithm ~\ref{alg:search}.

\section{Indoor Crowd Model}
\label{sec:model}

\subsection{Model Structure}
\label{ssec:structure}

As an extension of the accessibility graph~\cite{jensen2009graph}, the \emph{indoor crowd model} is a directed, labeled graph $G(V, E, L_V, L_E)$ where
\begin{enumerate}[leftmargin=*, label=\arabic*)]
	\item $V$ is the set of vertices, each for an indoor partition.
	\item $E$ is the set of directed edges such that each edge $e(v_i, v_j, d_k) \in E$ means one can reach $v_j$ from $v_i$ through a door $d_k$, i.e., $v_i \in \mathit{D2P}_\sqsubset(d_k)$ and $v_j \in \mathit{D2P}_\sqsupset(d_k)$. 
  \item $L_V$ is the set of vertex labels. Each label in $L_V$ is attached to a partition and captured as a five tuple $[v_i, \mathit{Area}(v_i), \mathsf{M}_\mathit{d2d}, \tau,$ $(\mathsf{P}^i_{t_l}, {t_l})]$.
  In particular, $v_i$ identifies the associated partition, $\mathit{Area}(v_i)$ is $v_i$'s area, $\mathsf{M}_\mathit{d2d}$ is a matrix that stores the intra-partition distance (See Equation~\ref{equation:d2d}) between each pair of doors of $v_i$.
  In addition, $\tau \in \{ \mathsf{R}, \mathsf{Q} \}$ indicates the type of $v_i$'s crowd and $(\mathsf{P}^i_{t_l}, {t_l})$ means that $v_i$'s absolute population at a latest timestamp $t_l$ is known as $\mathsf{P}^i_{t_l}$.
	In practice, the model can record the populations at historical timestamps, though only the latest population is relevant to a query.
  \item $L_E$ is the edge label set. For an edge $(v_i, v_j, d_k) \in E$, its label consists of a \textbf{door flow function} $f(v_i, v_j, d_k)$ that models the dynamic object flows from $v_i$ to $v_j$ via $d_k$ and a local array $\mathsf{F}[t]$ storing the actual object flows at each update timestamp $t$.
\end{enumerate}

\begin{figure}[htbp]
    \centering
    \includegraphics[width=0.95\columnwidth]{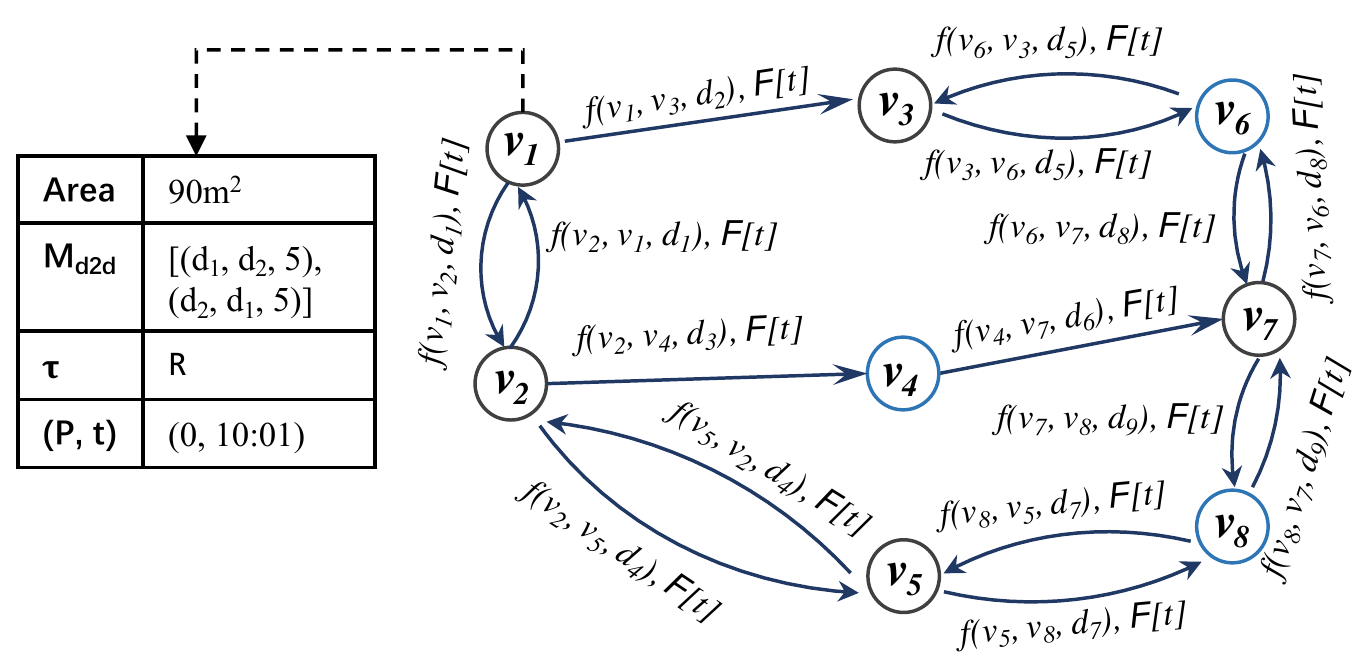}
    \caption{An Example of Indoor Crowd Model}
    \label{fig:model}
\end{figure}

Figure~\ref{fig:model} depicts the indoor crowd model corresponding to the space in Figure~\ref{fig:floorplan}.
Unlike a general time-dependent graph (GTG)~\cite{ding2008finding, yuan2019constrained}, our model represents doors as edges and partitions as vertices.
A GTG may model doors as vertices and partitions as edges and capture time-varying populations or distances as edge weights, but this way falls short in solving our problem.
First, GTG's vertices fail to capture the door directionality (e.g., unidirectional security check doors) directly.
Referring to Figure~\ref{fig:floorplan}, $d_2$ is unidirectional such that one can only go through $d_2$ from $v_1$ to $v_3$.
In a GTG, the edges cannot be directed because each edge connects two doors and one can always go from one door to any other door in the same room. 
E.g., one cannot go through $d_2$ from $v_3$ to $v_1$, but she can go to any door in $v_1$ from $d_2$ if she is in $v_1$.
The directionality information can be added in each node, e.g., that for node $d_2$ can be \{($v_1$, $v_3$)\}, and that for node $d_1$ can be \{($v_1$, $v_2$), ($v_2, v_1$)\}. However, it will result in considerably more space and search costs.
Second, a GTG will result in many door-to-door edges for the same partition, which will render the graph-based search inefficient (Further comparison is in Appendix~\ref{sec:appendix}.). The experimental comparison with GTG is reported in Section~\ref{sec:experiment}. 

The time-evolving function $f(v_i, v_j, d_k)$ models the number of objects flowing from $v_i$ to $v_j$ at each report time interval of $d_k$.
In practice, it can be implemented as a time-series prediction model driven by historical data such as ARIMA~\cite{contreras2003arima} and LSTM~\cite{liang2018geoman}, or it can be approximated by a queueing distribution function.
For the ease of presentation, in this paper, we employ a specific queueing distribution function to predict the door flows (Section~\ref{ssec:door_flow}).
Nevertheless, the door flow function can be replaced by other appropriate models or functions, which entails no change to any of the other parts in the overall computation framework (Figure~\ref{fig:architecture}).

\subsection{Door Flow Function}
\label{ssec:door_flow}

Following the classic Poisson distribution in queueing theory~\cite{consul1973generalization}, we design the following door flow function:
\begin{equation}
  f(v_i, v_j, d_k): t \mapsto \mathsf{P}_t, t \in \mathit{RT}(d_k), \mathsf{P}_t \sim \mathit{Poisson}(\lambda)
\end{equation}
where $t \in \mathit{RT}(d_k)$ is a report timestamp of $d_k$, $\mathsf{P}_t$ is the population that flows from $v_i$ to $v_j$ between $t$ and $d_k$'s next report timestamp, and $\lambda$ is the expected value of $\mathsf{P}_t$ under Poisson distribution.

The door flow function is parameterized by $\lambda$ and fitted based on a recent period of historical records in a format of $(t', \mathsf{P}_{t'})$.
In practice, for each door counter, the most recent timestamps' flows can be accessed from the local array $\mathsf{F}$ in the graph edge. An independent thread estimates $\lambda$ upon such most recent records.
Note that the focus of this paper is not to estimate $\lambda$ based on historical data. For its technical details, we refer readers to a previous work~\cite{boswell1966estimating}.
In our setting, at any query time, an up-to-date door flow function is ready to predict flows for future report timestamps.

\section{Time-evolving Populations}
\label{sec:cost_estimation}

\subsection{Rectifying Door Flows}
\label{ssec:population}

At a query time $t_q$, we can access a partition $v_k$'s latest population $\mathsf{P}^k_{t_l}$ at an earlier time $t_l \leq t_q$ from the indoor crowd model.
To enable the cost estimation for routing, we need to derive $v_k$'s time-evolving population and its future inflows/outflows based on $\mathsf{P}^k_{t_l}$.

Let $[t_0, t_1] \in \mathit{UTI}(v_k)$ be the unit time interval covering $t_l$. We have $\mathsf{P}^k_{t_0, t_1} = \mathsf{P}^k_{t_l}$, meaning that $v_k$'s population over $[t_0, t_1]$ is equal to $\mathsf{P}^k_{t_l}$. Subsequently, for a future unit time interval $[t_x, t_{x+1}] \in \mathit{UTI}(v_k)$, we compute its population as
\begin{equation}\label{equation:population}
  \mathsf{P}^k_{t_x, t_{x+1}} = \mathsf{P}^k_{t_{x-1}, t_x} - \mathit{out}(v_k, t_x) + \mathit{in}(v_k, t_x), \,\,\, x = 1, 2, \ldots \\
\end{equation}
where $\mathit{out}(v_k, t_x)$ and $\mathit{in}(v_k, t_x)$ are $v_k$'s estimated outflow and inflow at update timestamp $t_x$, respectively.

Suppose that all relevant door flow functions are ready at $t_q$. The inflow and outflow at a future update timestamp can be directly estimated based on the expected values $\lambda$. Formally,
\begin{equation*}
  \mathit{out}(v_k, t_x) = \sum_{d_i \in \mathit{P2D}_{\sqsubset}(v_k) \wedge t_x \in \mathit{RT}(d_i)} \sum_{v_p \in \mathit{D2P}_{\sqsupset}(d_i)}f(v_k, v_p, d_i).\lambda
\end{equation*}
\begin{equation*}
  \mathit{in}(v_k, t_x) = \sum_{d_j \in \mathit{P2D}_{\sqsupset}(v_k) \wedge t_x \in \mathit{RT}(d_j)} \sum_{v_q \in \mathit{D2P}_{\sqsubset}(d_j)}f(v_q, v_k, d_j).\lambda
\end{equation*}
where $d_i$ (resp. $d_j$) is a leaveable (resp. enterable) door updated at time $t_x$ and $v_p \in \mathit{D2P}_{\sqsupset}(d_i)$ (resp. $v_q \in \mathit{D2P}_{\sqsubset}(d_i)$) is its enterable (resp. leaveable) partition.

However, the estimated flows may be contrary to the real situation such that a partition's current population ($\mathsf{P}^k_{t_{x-1}, t_x}$ in Equation~\ref{equation:population}) cannot satisfy its outflow ($\mathit{out}(v_k, t_x)$ in Equation~\ref{equation:population}).
In this case, flows at doors should be rectified.

A basic idea is to rectify the expected outflow at each step such that it is not larger than the partition $v_k$'s current population.
Meanwhile, $v_k$'s inflow is naturally rectified as it is derived from the outflows of its adjacent partitions at the previous step.
In general, a dependency exists between partitions. It demands a suitable way to rectify the relevant outflows at the update timestamps.

\begin{figure*}[ht]
    \centering
    \includegraphics[width=\textwidth]{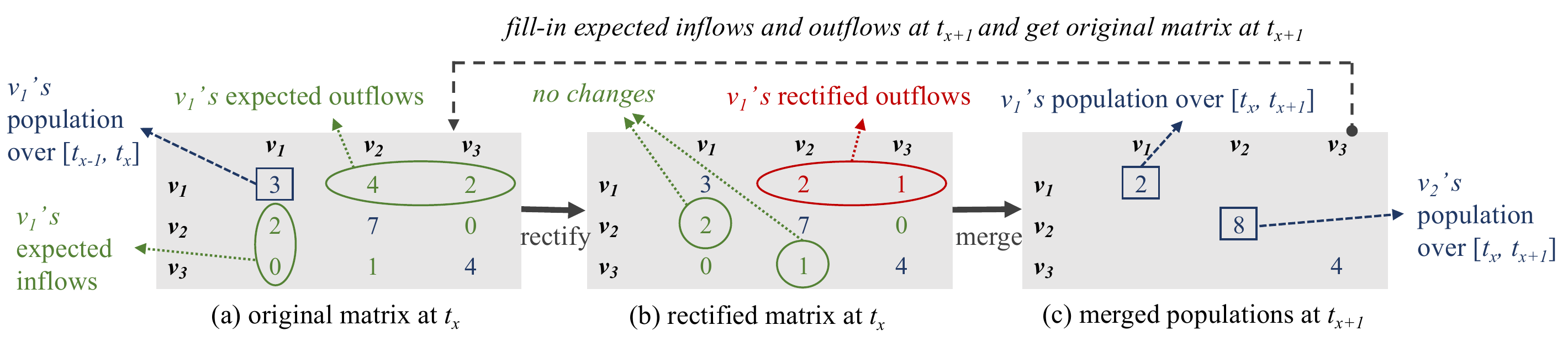}
    \caption{An Example of Rectifying Flows Globally}
    \label{fig:rectified_flows}
\end{figure*}

An example is depicted in Figure~\ref{fig:rectified_flows}, which rectifies the door flows globally.
To ease the presentation, at each particular update timestamp $t_x$ we put the absolute populations and door flows in a $|V| \times |V|$ matrix $\mathsf{M}$, where $|V|$ corresponds to the total number of partitions.
In particular, $\mathsf{M}[i, i]$ refers to partition $v_i$'s absolute population over unit time interval $[t_{x-1}, t_x]$, while $\mathsf{M}[i, j]$ ($i \neq j$) means the flow value from partition $v_i$ to $v_j$ over the next unit time interval $[t_x, t_{x+1}]$.
Referring to Figure~\ref{fig:rectified_flows}(a), partition $v_1$'s population over $[t_{x-1}, t_{x}]$ is 3 and that of $v_2$ is 7.
Besides, $v_1$'s expected outflows to $v_2$ and $v_3$ are 4 and 2, respectively; $v_2$'s inflow from $v_1$ and $v_3$ are 4 and 1, respectively.
Considering the space efficiency, in the implementation, we store the absolute populations on the graph nodes and the estimated flows on graph edges. That is, the space complexity at each update timestamp is $|V+E|$.

A rectification is then applied to each row of the original matrix as exemplified in Figure~\ref{fig:rectified_flows}(b).
Specifically, $v_1$'s current population (i.e., $3$) is less than the summation of its subsequent outflows (i.e., $4 + 2 = 6$).
In this case, we scale down the outflows at all doors to ensure that the actual number of objects outflowing is exactly equal to the current population.
That is, $\mathsf{M}'[1,2] = \mathsf{M}[1,2] \cdot (3/6) = 2$ and $\mathsf{M}'[1,3] = \mathsf{M}[1,3] \cdot (3/6) = 1$, where $\mathsf{M}$ and $\mathsf{M}'$ represent the original and rectified matrix, respectively.
Note that non-integer values may appear in the rectification. For the computation precision, we use non-integer values in the whole iterative derivation process.
Intuitively, the values in the matrix are a probability estimate, i.e., how likely an object will appear in or move to a certain partition.

After the rectification, each partition's population over the next interval $[t_x, t_{x+1}]$ is computed based on Equation~\ref{equation:population}.
In particular, partition $v_i$'s new population is obtained by deducting the overall outflows at the $i$-th row of $\mathsf{M}'$ and then adding the overall inflows at the $i$-th column of $\mathsf{M}'$.
Referring to Figure~\ref{fig:rectified_flows}(c), $v_1$'s new population is $3-3+2 =2$ while $v_2$'s is $7-2+3=8$.
After the merges on each partition, we fill in the matrix inflows and outflows at the timestamp $t_{x+1}$, and derive the populations iteratively.

\subsection{Implementation of Population Estimator}
\label{ssec:algorithm_population}

This section presents two versions of population estimators. The \emph{global estimator} estimates all partitions' populations globally (corresponding to the example in Figure~\ref{fig:rectified_flows}), whereas the \emph{local estimator}  only estimates a relevant partition's population by looking up its upstream partitions flows.

\begin{algorithm}
    \caption{\textsc{PopulationGlobal} (future timestamp $t^a$, indoor crowd model $G$)} \label{alg:population_global}
    \scriptsize
    \begin{algorithmic}[1]
    \State get the latest update timestamp $t^G_l$ from $G$
    \State $\mathit{UT}_G \gets \bigcup_{d_j \in G.D} \mathit{RT}(d_j)$
    \State $\mathsf{A} \gets \mathit{toArray}(\{t_c \mid t_c \in \mathit{UT}_G \wedge t_c \geq t^G_l \wedge t_c \leq t^a\})$
    \For {$t_c \in \mathsf{A}$}
      \For{$e(v_i, v_j, d_k) \in G.E$}
        \State \textbf{if} {$t_c \in \mathit{RT}(d_k)$} \textbf{then} $e.\mathsf{F}[t_c] \gets f(v_i, v_j, d_k).\lambda$
        \State \textbf{else} $e.\mathsf{F}[t_c] \gets 0$
      \EndFor
      \For{$v_i \in V$}
        \For {$d_k \in \mathit{P2D}_\sqsubset(v_i)$}
          \State $\mathit{out}_i \gets \mathit{out}_i + (v_i, v_j, d_k).\mathsf{F}[t_c]$
        \EndFor
        \State get $v_i$'s latest population record $(\mathsf{P}^i_{t}, t)$ from $G$
        \If{$\mathsf{P}^i_{t} - \mathit{out}_i < 0$}
          \For{$d_k \in \mathit{P2D}_\sqsubset(v_i)$}
            \State $(v_i, v_j, d_k).\mathsf{F}[t_c] \gets (v_i, v_j, d_k).\mathsf{F}[t_c] \cdot \mathsf{P}^i_{t}/\mathit{out}_i$
          \EndFor
        \EndIf
      \EndFor
      \For{$v_i \in V$}
        \For {$d_k \in \mathit{P2D}_\sqsubset(v_i)$}
          \State $\mathit{out}_i \gets \mathit{out}_i + (v_i, v_j, d_k).\mathsf{F}[t_c]$
        \EndFor
        \For {$d_k \in \mathit{P2D}_\sqsupset(v_i)$}
          \State $\mathit{in}_i \gets \mathit{in}_i + (v_j, v_i, d_k).\mathsf{F}[t_c]$
        \EndFor
        \State $\mathsf{P}^i_{t_c} \gets \mathsf{P}^i_{t} - \mathit{out}_i + \mathit{in}_i$; add $(\mathsf{P}^i_{t_c}, t_c)$ to $G.v_i$
      \EndFor
    \EndFor
    \end{algorithmic}
\end{algorithm}

The global estimator (Algorithm~\ref{alg:population_global}) takes the indoor crowd model $G$ as input and derives the populations from the latest update timestamp in $G$ to a future timestamp $t^a \geq t_q$.
In the beginning, the globally latest update timestamp $t^G_l$ over all partitions is obtained (line~1).
Then, the set $\mathit{UT}_G$ of all doors' report timestamps is obtained (line~2) and the period of interest is extracted out of $\mathit{UT}_G$ and organized into an array $\mathsf{A}$ (line~3).
The algorithm then progressively derives the populations for each timestamp in $\mathsf{A}$ (lines~4--20).
For each timestamp $t_c \in \mathsf{A}$, the algorithm iterates on each edge $e(v_i, v_j, d_k)$. If the corresponding door $d_k$ updates at $t_c$, $e$'s flow value at $t_c$, i.e., $e.\mathsf{F}[t_c]$, is assigned with the estimated flow of the corresponding flow function (line~6). Otherwise, the flow value $e.\mathsf{F}[t_c]$ is assigned with 0 (line~7).
Here $e.\mathsf{F}$ is a local array to maintain the rectified flow at each timestamp.
Subsequently, the algorithm goes through each partition $v_i$ and aggregates its expected outflow $\mathit{out}_i$ (lines~8--10).
If $\mathit{out}_i$ is greater than the population $\mathsf{P}^i_t$ at the latest update timestamp, the estimated flows at edges need to be rectified following the example in Figure~\ref{fig:rectified_flows} (lines~11--14).
Afterwards, the current timestamp $t_c$'s population for each partition is computed according to Equation~\ref{equation:population} and added to $G$ (lines~15--20).
Once the process is finished, all partitions' populations at each timestamp can be accessed from the edge nodes of $G$.

Algorithm~\ref{alg:population_global} needs to update the populations for all partitions. This incurs much unnecessary computation since a path planning search at a particular time only concerns a number of relevant partitions and their populations.
If these partitions' populations and flows can be precisely derived without a global update, the overall updating cost can be reduced substantially.

The local estimator is formalized in Algorithm~\ref{alg:population_local}. It derives a specific partition's population in a recursive manner.

\begin{algorithm}
    \caption{\textsc{PopulationLocal} (partition $v_i$, future timestamp $t^a$, indoor crowd model $G$)} \label{alg:population_local}
    \scriptsize
    \begin{algorithmic}[1]
    \State get $v_i$'s latest update timestamp $t_l$ from $G$
    \State $\mathit{UT}_G \gets \bigcup_{d_j \in G.D} \mathit{RT}(d_j)$
    \State $\mathsf{A} \gets \mathit{toArray}(\{t_c \mid t_c \in \mathit{UT}_G \wedge t_c \geq t_l \wedge t_c \leq t^a\})$, $t_{max} \gets \mathsf{A}.\textit{max}()$
    \While {$\mathsf{A}$ is not empty}
      \State $t_c \gets \mathsf{A}.\textit{max}()$; $\mathsf{A} \gets \mathsf{A} \setminus t_c$
      \If {$t_c$ = $t_l$}
        \State get $v_i$'s latest population record $(\mathsf{P}^i_{t_l}, t_l)$ from $G$
        \State $\mathsf{P}^i_{t_c} \gets \mathsf{P}^i_{t_l}$
      \Else ~$\mathsf{P}^i_{t_c} \gets \textsc{PopulationLocal}$ ($v_i$, $t_c$, $G$)
      \EndIf
      \For {$d_k \in \mathit{P2D}_\sqsubset(v_i)$}
        \If {$(v_i, v_j, d_k).\mathsf{F}[t_c]$ is null}
          \State $(v_i, v_j, d_k).\mathsf{F}[t_c] \gets f(v_i, v_j, d_k).\lambda$
        \EndIf
        \State $\mathit{out}_i \gets \mathit{out}_i + (v_i, v_j, d_k).\mathsf{F}[t_c]$
        \If{$\mathsf{P}^i_{t_c} - \mathit{out}_i < 0$}
          \For{$d_k \in \mathit{P2D}_\sqsubset(v_i)$}
            \State $(v_i, v_j, d_k).\mathsf{F}[t_c] \gets (v_i, v_j, d_k).\mathsf{F}[t_c] \cdot \mathsf{P}^i_{t_c}/\mathit{out}_i$
          \EndFor
          \State $\mathit{out}_i \gets \mathsf{P}^i_{t_c}$
        \EndIf
      \EndFor
      \For {$d_k \in \mathit{P2D}_\sqsupset(v_i)$}
        \If {$(v_j, v_i, d_k).\mathsf{F}[t_c]$ is null}
          \State $\textsc{PopulationLocal}$ ($v_j$, $t_c$, $G$)
        \EndIf
        \State $\mathit{in}_i \gets \mathit{in}_i + (v_j, v_i, d_k).\mathsf{F}[t_c]$
      \EndFor
      \State $\mathsf{P}^i_{t_c} \gets \mathsf{P}^i_{t_l} - \mathit{out}_i + \mathit{in}_i$
    \EndWhile
    \State $t \gets t_{max}$
    \State \Return $\mathsf{P}^i_{t}$
    \end{algorithmic}
\end{algorithm}

Its preparation (lines~1--3) is almost the same as the counterpart in Algorithm~\ref{alg:population_global}, except that the latest update timestamp $t_l$ in line~1 is with respect to the input partition $v_i$.

Next, the algorithm derives $v_i$'s population in reverse temporal order (lines~4--24).
Specifically, the newest update timestamp $t_c$ in $\mathsf{A}$ is archived and removed from $\mathsf{A}$ (line~5).
If $t_c$ just equals $t_l$, the population is directly obtained from $G$ (lines~6--8).
Otherwise, $t_c$ is an earlier timestamp to $t_l$, and the algorithm recursively calls Algorithm~\ref{alg:population_local} to obtain $v_i$'s population $\mathsf{P}^i_{t_c}$ at $t_c$ (line~9).
Once $\mathsf{P}^i_{t_c}$ is derived, the expected flow from each upstream door flow function (see lines~10-12) is compared to $v_i$'s population.
Note that the intermediate results are maintained in each edge's local array $\mathsf{F}[t]$ to avoid repeated computations (line~11).
If an expected outflow is larger than $\mathsf{P}^i_{t_c}$, a rectification is performed (lines~14--16). In this case, the rectified outflow is assigned with $\mathsf{P}^i_{t_c}$ (line~17).
Then, the inflows from all enterable doors are also derived (lines~18-22).
For each enterable door $d_k$, if its inflow has not been derived, Algorithm~\ref{alg:population_local} is recursively called to get the adjacent partition $v_j$'s population at time $t_c$ (line~18-20).
Note that the inflow from $v_j$ to $v_i$ will be rectified in this recursion.
After that, $v_i$'s overall inflow is obtained (line~21) and its population is computed (line~22).
The last two lines of Algorithm~\ref{alg:population_local} return the population nearest to the input time $t^a$.

Once the partition's population at a particular arrival time is derived, the corresponding partition-passing cost (time or contact) can be computed easily according to Equation~\ref{equ:crossing_cost} or~\ref{equ:crossing_contact}.
Both global and local population estimators can be utilized by the exact search presented in Section~\ref{ssec:exact_search} (see line~17 in Algorithm~\ref{alg:search}).

\section{Query Processing Algorithms}
\label{sec:queries}

\subsection{Exact Algorithms for \texttt{FPQ} and \texttt{LCPQ}}
\label{ssec:exact_search}

On top of the indoor crowd model (Section~\ref{sec:model}), we propose an indoor path search process in Algorithm~\ref{alg:search}. Following the spirit of Dijkstra's algorithm, our algorithm can handle both \texttt{FPQ} and \texttt{LCPQ}, as long as a cost measure corresponding to a specific query type is set as the routing cost of the graph traversal.
Algorithm~\ref{alg:search} returns an indoor path $\phi$ from the source $p_s$ to the target $p_t$ that satisfies the query type $\mathtt{QT}$ for a particular query time $t_q$.

\begin{algorithm}
    \caption{\textsc{Search} (source $p_s$, target $p_t$, query time $t_q$, indoor crowd model $G$, query type $\mathtt{QT}$)} \label{alg:search}
    \scriptsize
    \begin{algorithmic}[1]
    \State initialize a priority queue $\mathsf{Q}$
    \For {each door $d_i \in G.D$}
        $\mathit{prev}[d_i] \gets$ \textit{null}
    \EndFor
    \State $\mathit{prev}[p_s] \gets \textit{null}$; $\mathit{prev}[p_t] \gets \textit{null}$

    \State $\mathit{UT}_G \gets \bigcup_{d_j \in G.D} \mathit{RT}(d_j)$
    \State $t_l \gets \max\{t \in \mathit{UT}_G \mid t \leq t_q\}$; $t^a \gets \varnothing$ \Comment{latest update time $t_l$ and arrival time $t^a$}

    \State \textbf{if} $\mathtt{QT} = \texttt{LCPQ}$ \textbf{then} $\mathit{cost}_0 \gets (0, 0, 0)$ \textbf{else} $\mathit{cost}_0 \gets (0, 0)$ \Comment{(distance, time, contact) for \texttt{LCPQ} and (distance, time) for \texttt{FPQ}}
    \State $S_0 \gets (p_s, \mathit{cost}_0)$ \Comment{S(node, cost)}
    \State $\mathsf{A}_S[p_s] \gets S_0$; $\mathsf{Q}.\textit{push}(S_0)$
    \While {$\mathsf{Q}$ is not empty}
        \State $S_i \gets \mathsf{Q}.\textit{pop}()$; $d_i \gets S_i.\text{node}$
        \If {$d_i = p_t$}
            \textbf{return} \textsc{GetPath}($p_t$, $\mathit{prev}$, $p_s$)
        \EndIf
        \If {$d_i = p_s$}
            $v \gets \textit{host}(p_s)$
        \EndIf
        \State \textbf{else} $v \gets \mathit{D2P}_\sqsupset(d_i) \setminus$ $d_i$'s previous partition
        \State mark $d_i$ as visited
        \State $t^a \gets \max\{t \in \mathit{UT}_G \mid t \leq t_q + S_i.\text{cost}.\text{time}\}$
        \If {$t^a > t_l$} \Comment{further derive populations}
                \State $\textsc{Population}(t^a, G)$
                \State $t_l \gets t^a$
        \EndIf
        \If {$d_i \in \mathit{P2D}_\sqsupset(\textit{host}(p_t))$}
            \State $\textsc{Expand}(d_i, p_t, G, v, t^a, S_i, \mathtt{QT})$ \Comment{towards target $p_t$}
        \EndIf
        \For {each unvisited door $d_j \in \mathit{P2D}_\sqsubset(v)$}
            \State $\textsc{Expand}(d_i, d_j, G, v, t^a, S_i, \mathtt{QT})$
        \EndFor
    \EndWhile
    \end{algorithmic}
\end{algorithm}

The algorithm starts with initializing a priority queue $\mathsf{Q}$ (line~1) whose priority is the minimum travel time for \texttt{FPQ} and the minimum contact for \texttt{LCPQ}.
Also, an array $\mathit{prev}$ is initialized to record each path node's previous node in the search (lines~2--3).
Then, the full set $\mathit{UT}_G$ of the report timestamps over all doors are obtained (line~4).
With respect to the query time $t_q$, the latest update timestamp in $\mathit{UT}_G$ is found and assigned to $t_l$ (line~5).
Variable $t_l$ is the latest population derivation time in the search.
Next, the cost of the current search is initialized (line~6).
The cost for \texttt{LCPQ} consists of overall travel distance, overall travel time, and overall contact value. The cost for \texttt{FPQ} only contains the first two.
The source and initial cost are put into a stamp $S_0$, and $S_0$ is pushed into $\mathsf{Q}$ and maintained in an array $\mathsf{A}_S$ as well (lines~7--8).

After the preparation, the algorithm explores the next path node towards $p_t$ in an order controlled by $\mathsf{Q}$ (lines~9--22).
Specifically, the stamp $S_i$ with the lowest cost is popped from $\mathsf{Q}$, and the corresponding path node $d_i$ is obtained (line~10).
If $d_i$ is $p_t$, i.e., the searched is complete, the algorithm calls a function \textsc{GetPath}($p_t$, $\mathit{prev}$, $p_s$) to return the reverse path from $p_t$ to $p_s$ (line~11). Otherwise, the algorithm explores the next path node as follows.

First, if the current node is the source $p_s$, the current partition $v$ is obtained as the host of $p_s$ (line~12). Otherwise, $v$ is assigned as $d_i$'s enterable partition\footnote{To ease presentation, here we only show the case that a door connects two partitions. A complex space can be handled by maintaining a collection of enterable partitions.} (line~13).
Then, $d_i$ is marked as visited (line~14).
Next, the estimated cost to pass $v$ is obtained as $S_i.\text{cost}.\text{time}$ and it is added to the query time $t_q$ to get the arrival time $t^a$ to the next path node (line~15). An alignment to the update timestamps in $\mathit{UT}_G$ is needed for $t^a$.
The algorithm then determines if the population needs to be derived to meet the next arrival time $t^a$ (lines~16--18). If the latest derivation time $t_l$ is earlier than $t^a$, a population estimator is invoked to get $v$'s derived populations up to $t^a$ (line~17).
Here, either the global (Algorithm~\ref{alg:population_global}) or local estimator (Algorithm~\ref{alg:population_local}) can be used.
The performance difference of these two ways will be experimentally studied in Section~\ref{sec:experiment}.

Afterward, it expands to the next node from the current node $d_i$.
If $d_i$ is an enterable door of $p_t$'s host partition, the expansion goes towards $p_t$ (lines~19-20). Regardless of whether the current path reaches $p_t$'s host partition, the expansion should also reach each unvisited leaveable door of the current partition $v$ (lines~21--22). This ensures that the planned path can leave and re-enter $p_t$'s host partition when the host is currently too crowded.

The function $\textsc{Expand}$ is formalized in Algorithm~\ref{alg:expand}, which expands from the current node $p_1$ to the next possible node $p_2$ through partition $v$.
First, it estimates the cost to reach $p_2$ from $p_1$ through an inline function $\textsc{Cost}$ (see lines~7--16).
In particular, the distance between $p_1$ and $p_2$ is obtained as the door-to-door distance if both are doors, or Euclidean distance if either is an indoor point (lines~8--10).
Then, the population of the partition $v_k$ to pass is obtained from $G$ (line~11), and the partition-passing time and contact are computed based on Equations~\ref{equ:crossing_cost} and~\ref{equ:crossing_contact}, respectively (lines~12 and 14).
The corresponding cost is then returned according to the query type $\mathtt{QT}$ (lines~13 and 15--16).

Back to $\textsc{Expand}$ in Algorithm~\ref{alg:expand}, once the cost to pass $v$ is obtained, it is added to the current stamp $S_i$'s cost to get the overall cost in the current expansion, i.e., $\mathit{cost}_c$ (line~1). The tuple-form cost is summed in an element-wise way.
Next, the estimated cost to reach $p_2$ so far is obtained from the array $\mathsf{A}_S$ (line~2).
The algorithm compares the current estimated cost $\mathit{cost}_c$ to the previously recorded cost $\mathit{cost}_{pre}$.
If $\mathit{cost}_{pre}$ does not exist or $\mathit{cost}_c$ is lower, a valid expand is performed (lines~3--6).
Specifically, a new stamp $S'$ is formed with the next path node $p_2$ and the new cost $\mathit{cost}_c$.
It is then pushed to $\mathsf{Q}$. If an old stamp exists with the same node, the old stamp is updated by $S'$ (line~5).
Then, $S'$ is inserted into $\mathsf{A}_S$ and $p_2$'s previous path node is recorded as $p_1$.

\begin{algorithm}
    \caption{\textsc{Expand} (start node $p_1$, end node $p_2$, indoor crowd model $G$, partition $v$, arrival time $t^a$, stamp $S_i$, query type $\mathtt{QT}$)} \label{alg:expand}
    \scriptsize
    \begin{algorithmic}[1]
      \State $\mathit{cost}_c \gets S_i.\text{cost} + \textsc{Cost}(p_1, p_2, v, t^a, G)$ \Comment{element-wise}
      \State $\mathit{cost}_\mathit{pre} \gets \mathsf{A}_S[p_2].\text{cost}$

      \If{$\mathit{cost}_\mathit{pre}$ is $\textit{null}$ or $\mathit{cost}_c < \mathit{cost}_\mathit{pre}$}
        \State $S' \gets (p_2, \mathit{cost}_c)$
        \State $\mathsf{Q}.\textit{push}(S')$ \Comment{update if exists}
        \State $\mathsf{A}_S[p_2] \gets S'$; $\mathit{prev}[p_2] \gets p_1$
      \EndIf

      \Function{Cost} {$p_1$, $p_2$, $v_k$, $t^a$, $G$}
        \State $\mathit{dist} \gets \varnothing$
        \State \textbf{if} {$p_1, p_2$ are both doors} \textbf{then} $\mathit{dist} \gets v_k.\mathsf{M}_\mathit{d2d}(p_1, p_2)$
        \State \textbf{else} $\mathit{dist} \gets |p_1, p_2|_E$
        \State get $\mathsf{P}^k_{t^a}$ from $G$
        \State $\mathit{time} \gets T(p_1, p_2, v_k, t^a)$ \Comment{Equation~\ref{equ:crossing_cost}}
        \If{$\mathtt{QT} = \texttt{LCPQ}$}
          \State $\mathit{contact} \gets \kappa(p_1, p_2, v_k, t^a)$ \Comment{Equation~\ref{equ:crossing_contact}}
          \State \Return $(\mathit{dist}, \mathit{time}, \mathit{contact})$
        \EndIf
        \State \textbf{else} \Return $(\mathit{dist}, \mathit{time})$
      \EndFunction
    \end{algorithmic}
\end{algorithm}

In the exact search, the time-evolving populations are rectified and computed rigidly timestamp by timestamp. This may result in a bottleneck in the graph traversal.
We intend to reduce the workload for population derivation by approximation.
On the one hand, the severity of population derivation can be skipped for those less important partitions, e.g., those far away from the current partition to pass.
On the other hand, some of the update timestamps in the iterative derivation can be skipped if the population changes within that iteration period is relatively stable.
We proceed to introduce two approximate search algorithms for \texttt{FPQ} and \texttt{LCPQ}.

\subsection{Approximate Algorithms for \texttt{FPQ} and \texttt{LCPQ}}
\label{ssec:approx_search}

We design two strategies to derive approximate populations.

\noindent\textbf{Strategy 1: Population Derivation for Partial Partitions (\texttt{PP})}.
Recall that the population derivation in Algorithm~\ref{alg:population_local} (see line~20) recursively obtains its adjacent partition's population to ensure the overall derivation is fully precise.
This recursion terminates when the outflows of all relevant partitions at all relevant update timestamps have been rectified.
In fact, the door flows from a long distance or at a very old timestamp only have a slight impact on a partition's current population.
Therefore, one option is to rectify only the outflows of the current relevant partition without strictly processing the outflows of its upstream partitions (i.e., the inflows to the current relevant partition).
To this end, only a minor change is made to Algorithm~\ref{alg:population_local}: line~20 is modified to $(v_j, v_i, d_k).\mathsf{F}[t_c] \gets f(v_j, v_i, d_k).\lambda$.
That is, the outflow of an adjacent partition $v_j$ is directly obtained from the corresponding door flow function.

\noindent\textbf{Strategy 2: Population Derivation at Necessary Timestamps (\texttt{NT})}.
To further speed up the population derivation for individual partitions, we consider reducing the workload by only calling Algorithm~\ref{alg:population_local} at some necessary timestamps.
The general idea is that when we observe that the historical flows of a partition are relatively stable, we skip the iterative population computations and directly estimate its population at the arrival time $t^a$. Note that here Strategy 2 is used in combination with Strategy 1 to achieve the maximum effect of acceleration.

In particular, when the search visits a partition $v_k$, the mean $\mu$ and standard deviation $\sigma$ of its flow difference (i.e., inflow deducts outflow) in the historical timestamps are computed as follows.
\begin{equation*}\label{equation:ave}
  \mu = \Big(\sum\nolimits_{t_x \in \mathsf{UT}_\mathit{past}} \big(\mathit{in}(v_k, t_x) - \mathit{out}(v_k, t_x)\big) \Big) /|\mathsf{UT}_\mathit{past}| \\
\end{equation*}
\vspace*{-5pt}
\begin{equation*} \label{equation:sqrt}
    \sigma = \Big( \big(\sum\nolimits_{t_x \in \mathsf{UT}_\mathit{past}}(\mathit{in}(v_k, t_x) - \mathit{out}(v_k, t_x) - \mu)^2\big)/|\mathsf{UT}_\mathit{past}| \Big)^{1/2}
\end{equation*}
where $\mathsf{UT}_\mathit{past}$ is a set of the historical update timestamps of $v_k$.
The update timestamps in $\mathsf{UT}_\mathit{past}$ will be obtained from the local array that we maintain for fitting door flow function in Section~\ref{ssec:door_flow}.

If $\sigma$ is smaller than a pre-defined threshold value $\eta$, it indicates that $v_k$'s historical flows change only slightly. Thus, we directly estimate $v_k$'s population according to its historical trend as follows.
\begin{equation} \label{equation:arrPop}
    \mathsf{P}^k_{t^a} = \mathsf{P}^k_{t_l} + \mu \cdot |\{t \in \mathit{UT}(v_k) \mid t_l < t \wedge t \leq t^a \}|
\end{equation}
where $t_l = \max\{t_x \in \mathit{UT}_G \mid t_x \leq t_q\} $ is the latest population update time as in line~5 of Algorithm~\ref{alg:search}, $\mu$ is the mean of historical flow differences, and $|\{t \in \mathit{UT}(v_k) \mid t_l < t \wedge t \leq t^a \}|$ is the number of skipped update timestamps from $t_l$ to $t^a$.
In our experiment, $\eta=3$ achieves the best performance approaching exact search results.

Otherwise, the search has to call Algorithm~\ref{alg:population_local} (applied with Strategy 1) to derive population for $v_k$.

\subsection{Complexity Analysis}

The main difference of the four algorithms' complexity is related to population derivation. Therefore, we focus on comparing the four ways of population derivation.
Assume that we estimate a partition's population at a future timestamp, and the derivation involves $k$ unit time intervals.

The time complexity of the global population estimator (Algorithm~\ref{alg:population_global}) is $k|V| \cdot \mathtt{u}$, where $\mathtt{u}$ is the unit computational cost for a partition at an update timestamp.
For the local estimator (Algorithm~\ref{alg:population_local}) which only considers the current partition and its upstream partitions, its time complexity is $\big(k|V|$ $-$ $((k-1)n^k + (k-2)n^{k-1} + \dots + n^2)\big) \cdot \mathtt{u}$, where $n$ is the average number of enterable door per partition.

For two approximate strategies, \texttt{PP} rectifies only the outflows of the current certain partition without strictly processing the outflows of its upstream partitions, so the time complexity of \texttt{PP}'s population derivation per partition is only $k\mathtt{u}$.
\texttt{NT} omits population estimations for some partitions with the relatively stable flow. Thus for a partition in consideration, the time complexity depends on its flow stability. That is, if it is stable, we do not estimate the population; otherwise, the complexity is also $k\mathtt{u}$.

\section{Experiments}
\label{sec:experiment}

For either \texttt{FPQ} or \texttt{LCPQ}, we implement four search algorithms. Specifically, \texttt{*PQ} is Algorithm~\ref{alg:search} calling Algorithm~\ref{alg:population_local}, \texttt{*PQ-G} is Algorithm~\ref{alg:search} calling Algorithm~\ref{alg:population_global}, \texttt{*PQ-PP} is the approximate search using Strategy \texttt{PP}, and \texttt{*PQ-NT} is the approximate search using Strategy \texttt{NT}.
All algorithms are implemented in Java and run on a PC with a 2.30GHz Intel i5 CPU and 16 GB memory.

\subsection{Results on Synthetic Data}

\subsubsection{Settings}

\textbf{Indoor Space.}
Using a real-world floorplan\footnote{\url{https://longaspire.github.io/s/fp.html}}, we generate a multi-floor indoor space where each floor takes 1368m $\times$ 1368m. The irregular hallways are decomposed into smaller but regular partitions following the decomposition algorithm in~\cite{xie2014distance}.
As a result, we obtain 141 partitions and 216 doors on each floor. We duplicate the floorplan 3, \textbf{5}, 7, or 9 times to simulate different indoor spaces. All parameter settings are listed in Table~\ref{tab:syn_parameters} with default values in bold. The four staircases of each two adjacent floors are connected by stairways, each being 20m long. On each floor, we randomly pick 14 out of all those partitions having two doors as the Q-partitions while regard all others as R-partitions.

\begin{table}[htbp]
\caption{Parameter Settings}\label{tab:syn_parameters}
\scriptsize
\centering
\begin{tabular}{|c|c|}
\hline
Parameters & Settings \\
\hline\hline
$\mathit{floor}$ & 3, \textbf{5}, 7, 9 \\
\hline
$|o|$ & 300, \textbf{600}, 900, 1200, 1500\\
\hline
$\mathit{TI}$ (s) & 5, \textbf{10}, 15, 20\\
\hline
$\mathit{s2t}$ (m) & 900, 1100, \textbf{1300}, 1500, 1700 \\
\hline
\end{tabular}
\end{table}

\textbf{Populations and Flows.}
We generate each partition's population at an initial time randomly from 0 to $|o|$ (see Table~\ref{tab:syn_parameters}).
We set the max capacity of a partition $v$ as $Area(v) \cdot \beta$ ($\beta$ is 1 per m$^2$ in this paper). Note that the initial population will not exceed the max capacity.
The parameter $\lambda$ of each door flow function is varied from 0 to 3~\footnote{The value is set according to our analysis of real data. The door flow of a hallway/staircase is relatively more than that of a room.}.
We use a variable $\mathit{TI}$ (5, \textbf{10}, 15, or 20 seconds) to control the length of the unit update time interval of partitions.
To this end, all doors' initial report timestamps are aligned and they only report the flows in every $n\cdot\mathit{TI}$ seconds ($n = 1, 2, \ldots, 5$ is randomly decided for each door counter).

\textbf{Query Instances.}
We use a parameter $\mathit{s2t}$ to control the shortest distance from the source $p_s$ and the target $p_t$: First, we randomly select a point $p_s$ from the indoor space.
Second, we find a door $d$ whose indoor distance to $p_s$ approximates $\mathit{s2t}$.
Then, we expand from $d$ to find a random point $p_t$ whose indoor distance to $p_s$ approaches $\mathit{s2t}$.
For each $\mathit{s2t}$ value, we generate 100 different pairs to form query instances.

\textbf{Baseline Methods.}
We use a general time-dependent graph (GTG) to form a baseline. Each vertex in GTG represents a door and the weight of each edge is the cost between two doors, i.e., the time cost for \texttt{FPQ} or the contact for \texttt{LCPQ}. To be fair, we employ a Dijkstra-based algorithm (\texttt{*PQ-GTG}) without precomputation and combine it with our exact population estimator to process queries.
Since GTG fails to represent the door directionality directly, we assume all doors are bidirectional in comparative experiments.
Another baseline is the adaptive method based on indoor crowd model (\texttt{*PQ-A}) that keeps updating and recomputing the optimal route at every node until the user gets to the target.

\textbf{Performance Metrics.}
To compare the efficiency of different search algorithms, we run each query instance ten times and measure their average running time and memory cost.
We also look into the accuracy of the four searches.
In particular, the query \emph{hit rate} is the fraction of query instances whose search result equals its gold standard result among all 100 instances.
The gold result is returned by searching over the detailed simulated trajectories.
Moreover, we measure the \emph{relative error} of the estimated routing cost against the gold result. The estimated cost refers to overall travel time $T_\phi$ for \texttt{FPQ} and overall contact $\kappa_\phi$ for \texttt{LCPQ}.
Taking \texttt{FPQ} as an example, the relative error is $\gamma = {|T^{(E)}_\phi - T^{(G)}_\phi|}/{T^{(G)}_\phi}$ where $T^{(E)}_\phi$ and $T^{(G)}_\phi$ is the overall travel time corresponding to the exact search and gold result, respectively.

\subsubsection{Search Performance of \texttt{FPQ}}

\textbf{Comparison in default setting.}
The measures of different \texttt{FPQ} algorithms are reported in Table~\ref{tab:comparison_SYN}.
\texttt{FPQ-NT} performs the best in terms of the running time and memory because it skips the iterative population computations and directly estimates its population at the arrival time in each node.
\texttt{FPQ} and \texttt{FPQ-G} perform similarly as two exact searches, implying that the two exact estimators achieve similar efficiency in the default setting.
Besides, they are the best in terms of hit rate and relative error.
The baseline \texttt{FPQ-GTG} uses the exact estimator that we propose, so its accuracy is the same as \texttt{FPQ} and \texttt{FPQ-G}. However, \texttt{FPQ-GTG} incurs the highest time and memory costs due to the large size of GTG (cf. Section~\ref{sec:model}).
\texttt{FPQ-PP} works as accurately as the exact algorithms, which reflects the effectiveness of Strategy \texttt{PP}. Also, \texttt{FPQ-PP} saves some time and memory.
\texttt{FPQ-NT} and \texttt{FPQ-A} perform worse in terms of hit rate and relative error.
\texttt{FPQ-NT} skips some intermediate update timestamps, making its population derivation less accurate.
\texttt{FPQ-A} expands to next nodes by reevaluation, making its result only optimal locally rather globally.
Note that the running time (and memory) of \texttt{FPQ-A} is the sum of that at all nodes in a path. Indeed, \texttt{FPQ-A} keeps updating until a user gets to the target, while other methods return the path before departure.
We omit \texttt{FPQ-GTG} and \texttt{FPQ-A} in the subsequent experiments as the comparison results show a similar trend to that here.

\begin{table*}[]
\scriptsize
\caption{Comparison of Algorithms for \texttt{FPQ} and \texttt{LCPQ} on Synthetic Data (best result in bold)}\label{tab:comparison_SYN}
\centering
\resizebox{17.5cm}{0.75cm}{
\begin{tabular}{|l||l|l|l|l|l|l||l|l|l|l|l|l|}
\hline
               &\texttt{FPQ} &\texttt{FPQ-G} &\texttt{FPQ-PP} &\texttt{FPQ-NT} &\texttt{FPQ-GTG} &\texttt{FPQ-A}&\texttt{LCPQ}&\texttt{LCPQ-G}&\texttt{LCPQ-PP}&\texttt{LCPQ-NT}&\texttt{LCPQ-GTG}&\texttt{LCPQ-A} \\ \hline
Running Time (ms) & 584      & 585      & 208      & \textbf{25} & 2857     & 189      & 446      & 461      & 131      & \textbf{20} & 2532     & 163\\ \hline
Memory (KB)       & 115      & 112      & 111      & \textbf{12} & 278      & 14     & 182      & 192      & 144      & \textbf{7}  & 257      & 8  \\ \hline
Hit Rate (\%)     & \textbf{98}&\textbf{98}&\textbf{98}& 95          &\textbf{98}& 94 & 83 & 83 & 83 & 60       & 83 & \textbf{87}       \\ \hline
Relative Error    &\textbf{4.37E-08}&\textbf{4.37E-08}&\textbf{4.37E-08}& 8.09E-08&\textbf{4.37E-08}& 0.1233 & \textbf{0.0128} & \textbf{0.0128} & 0.0129 & 0.1113 & \textbf{0.0128} & 0.1256\\ \hline
\end{tabular}}
\end{table*}

\begin{figure*}[!ht]
\centering
\begin{minipage}[t]{0.245\textwidth}
\centering
\includegraphics[width=\textwidth, height = 3cm]{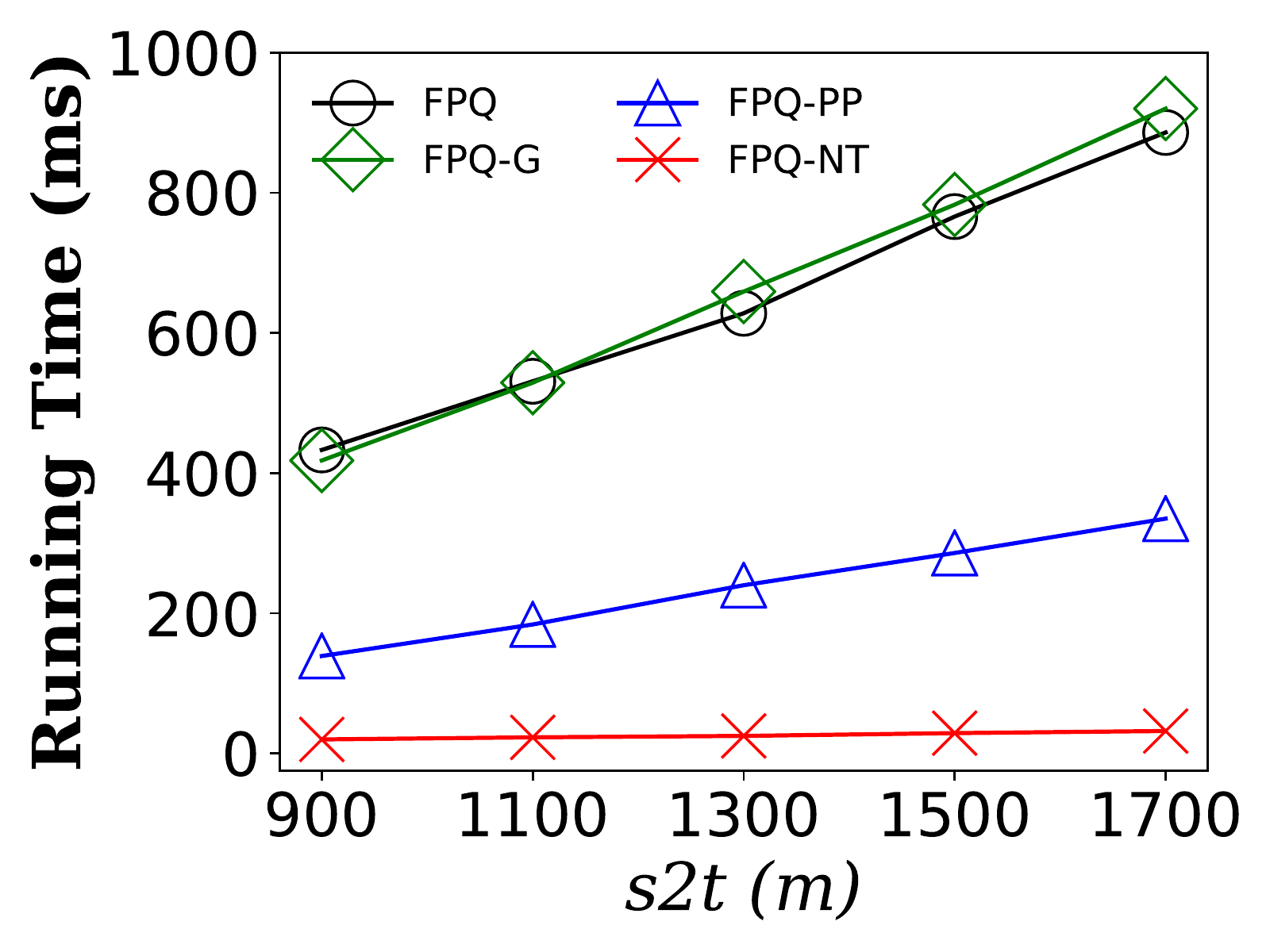}
\ExpCaption{\texttt{FPQ} Time vs. $\mathit{s2t}$}\label{fig:FPQ_s2t_time}
\end{minipage}
\begin{minipage}[t]{0.245\textwidth}
\centering
\includegraphics[width=\textwidth, height = 3cm]{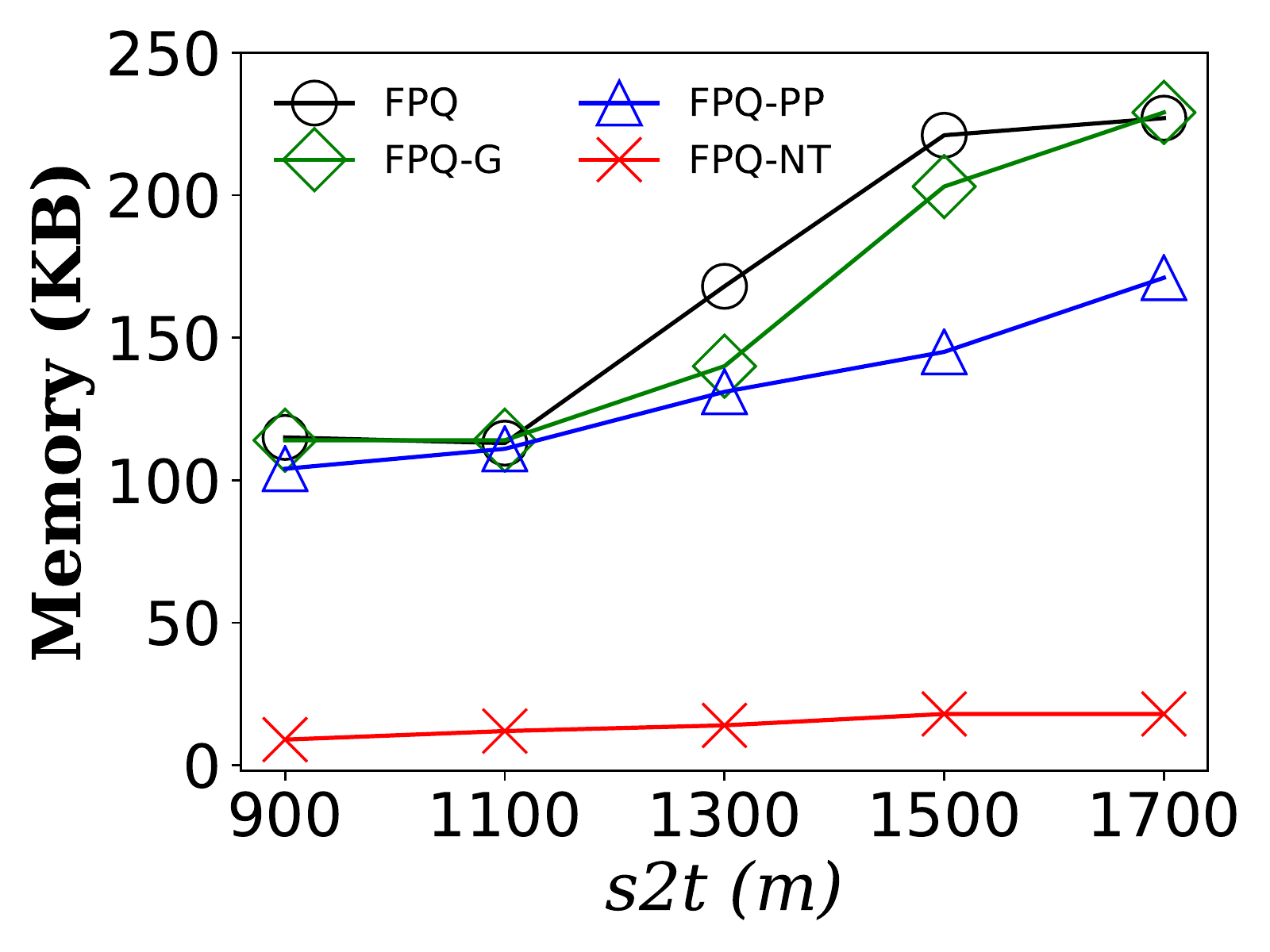}
\ExpCaption{\texttt{FPQ} Memory vs. $\mathit{s2t}$}\label{fig:FPQ_s2t_mem}
\end{minipage}
\begin{minipage}[t]{0.245\textwidth}
\centering
\includegraphics[width=\textwidth, height = 3cm]{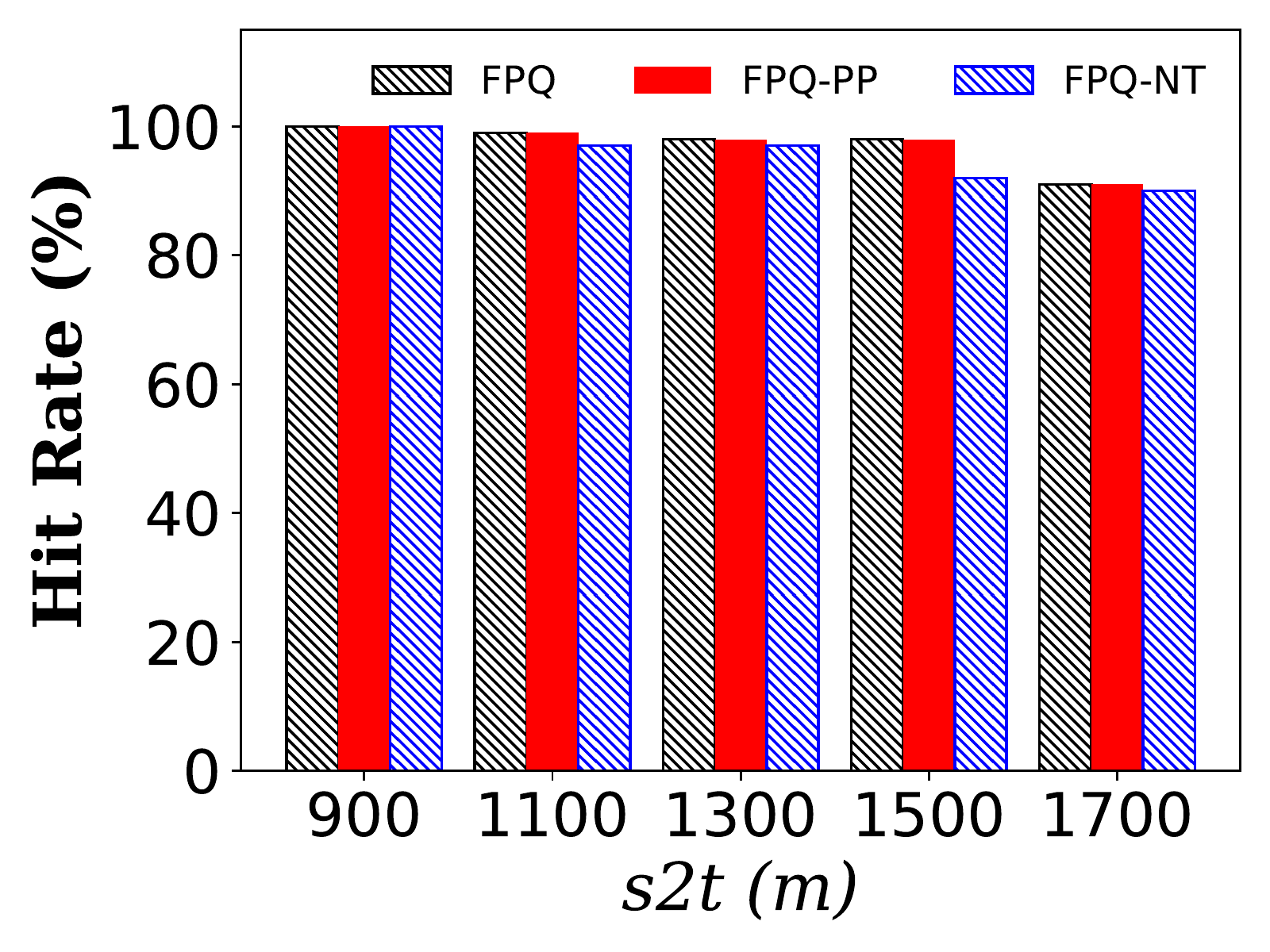}
\ExpCaption{\texttt{FPQ} Hit Rate vs. $\mathit{s2t}$}\label{fig:FPQ_s2t_hit}
\end{minipage}
\begin{minipage}[t]{0.245\textwidth}
\centering
\includegraphics[width=\textwidth, height = 3cm]{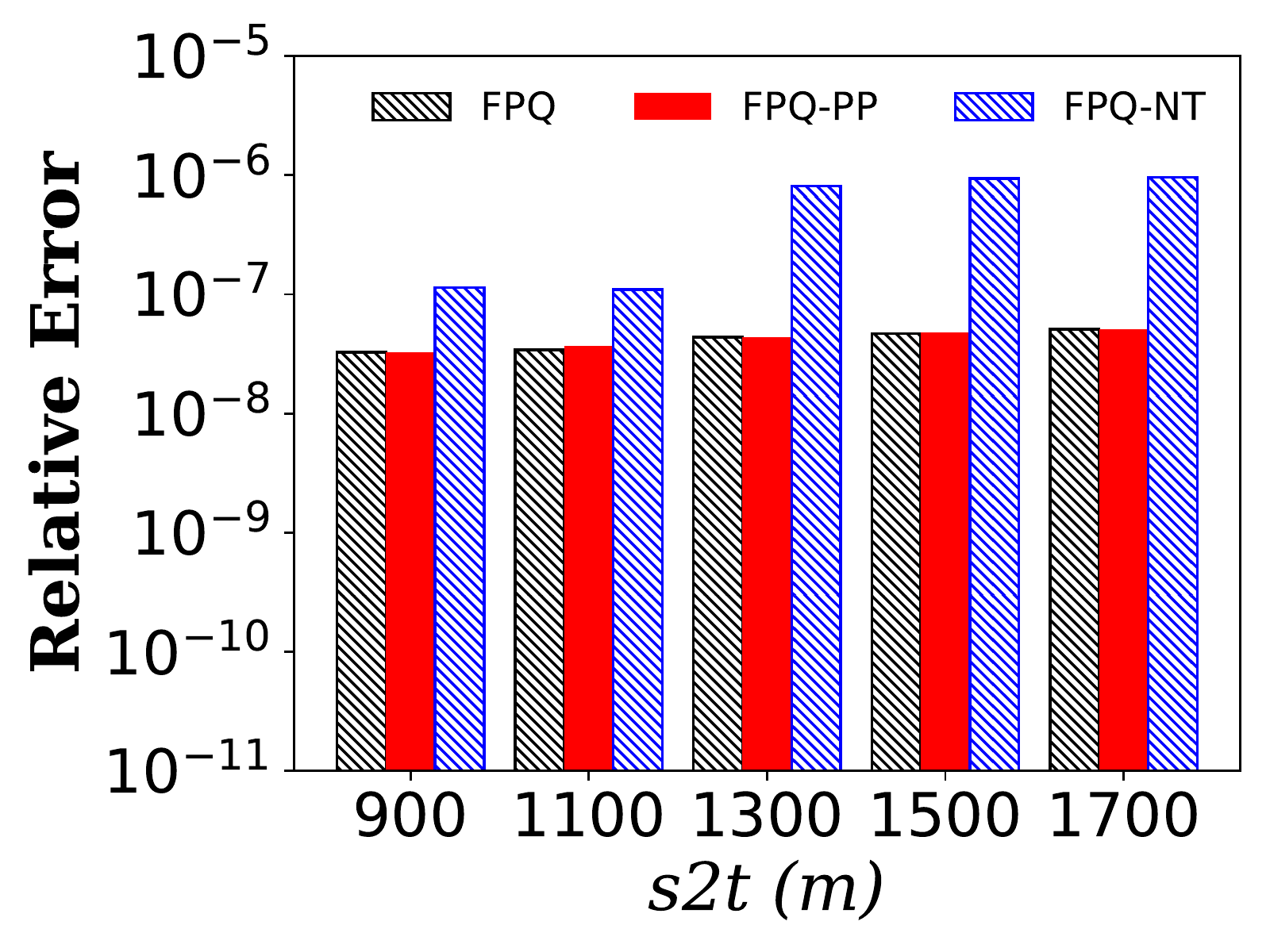}
\ExpCaption{\texttt{FPQ}'s $\gamma$ vs. $\mathit{s2t}$}\label{fig:FPQ_s2t_error}
\end{minipage}
\end{figure*}

\begin{figure*}[!ht]
\centering
\begin{minipage}[t]{0.245\textwidth}
\centering
\includegraphics[width=\textwidth, height = 3cm]{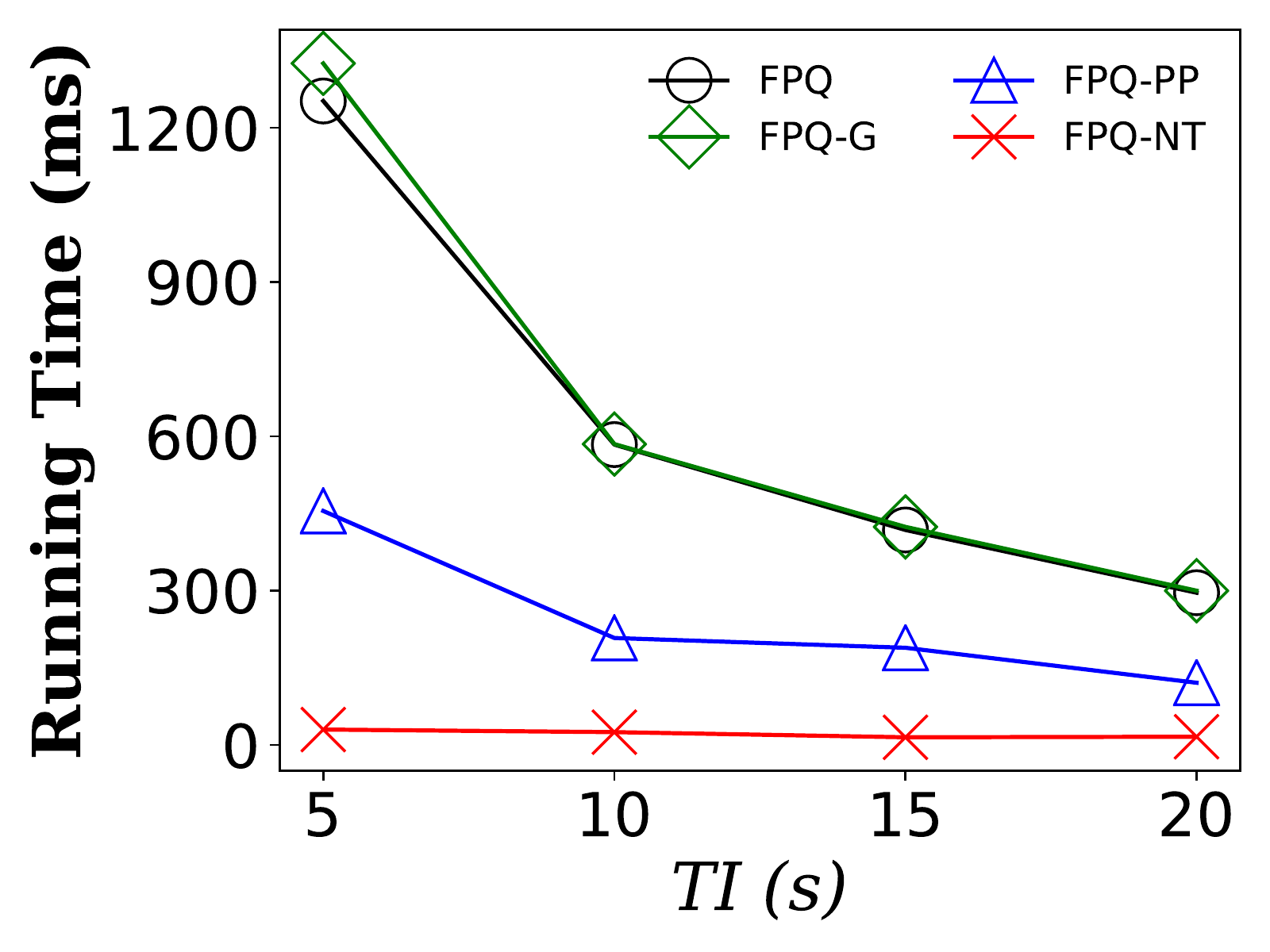}
\ExpCaption{\texttt{FPQ} Time vs. $\mathit{TI}$}\label{fig:FPQ_TI_time}
\end{minipage}
\begin{minipage}[t]{0.245\textwidth}
\centering
\includegraphics[width=\textwidth, height = 3cm]{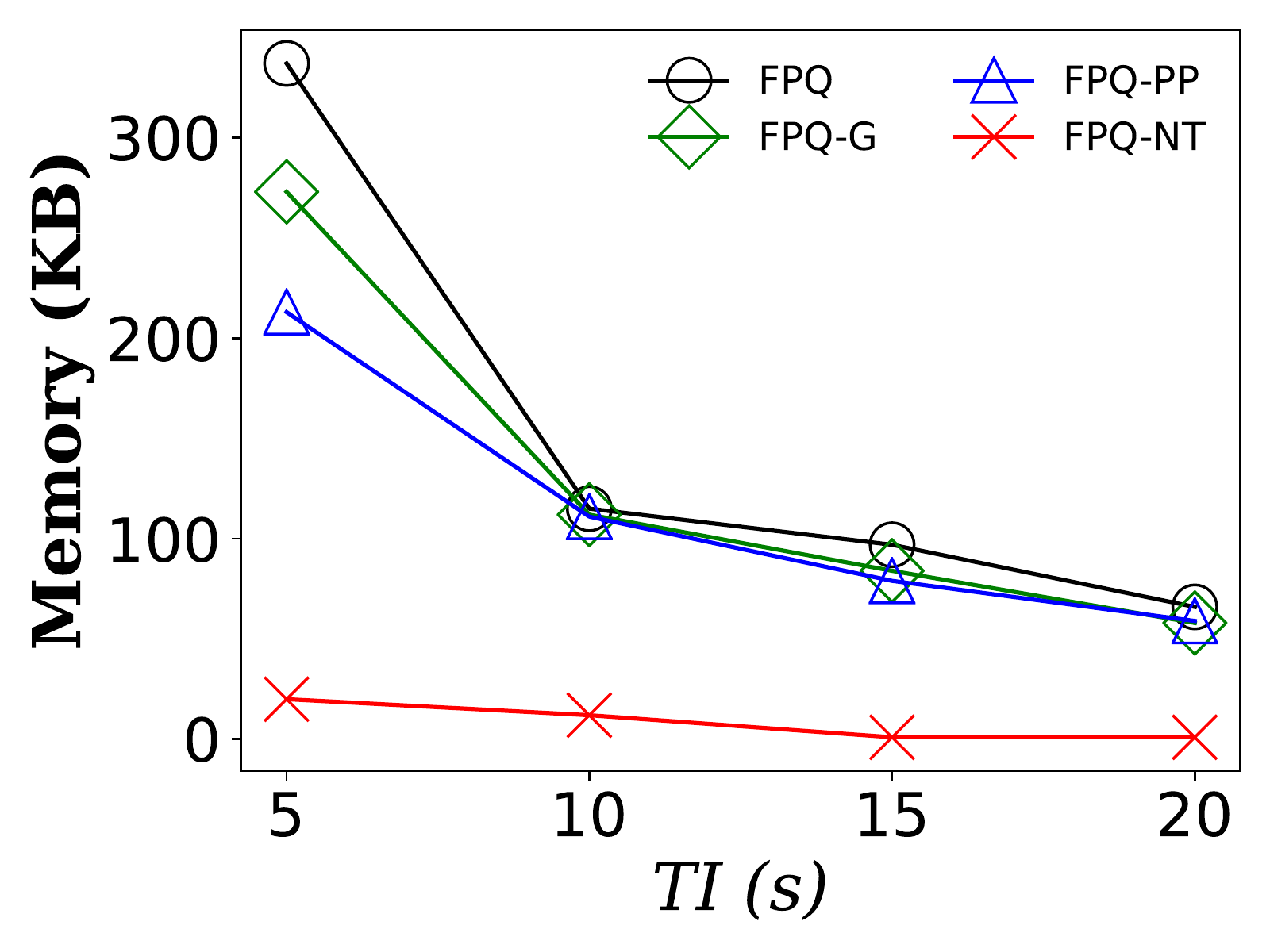}
\ExpCaption{\texttt{FPQ} Memory vs. $\mathit{TI}$}\label{fig:FPQ_TI_mem}
\end{minipage}
\begin{minipage}[t]{0.245\textwidth}
\centering
\includegraphics[width=\textwidth, height = 3cm]{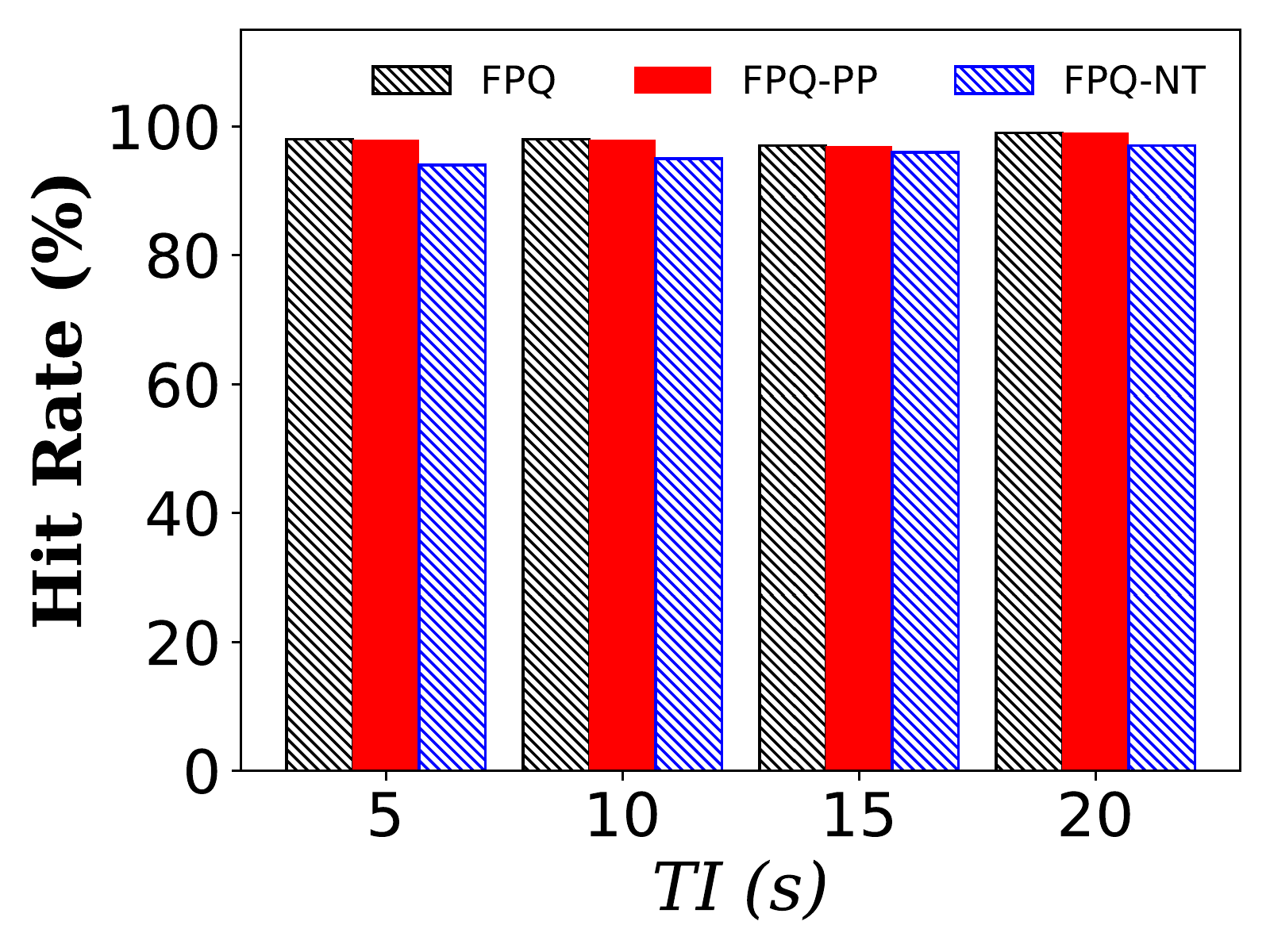}
\ExpCaption{\texttt{FPQ} Hit Rate vs. $\mathit{TI}$}\label{fig:FPQ_TI_hit}
\end{minipage}
\begin{minipage}[t]{0.245\textwidth}
\centering
\includegraphics[width=\textwidth, height = 3cm]{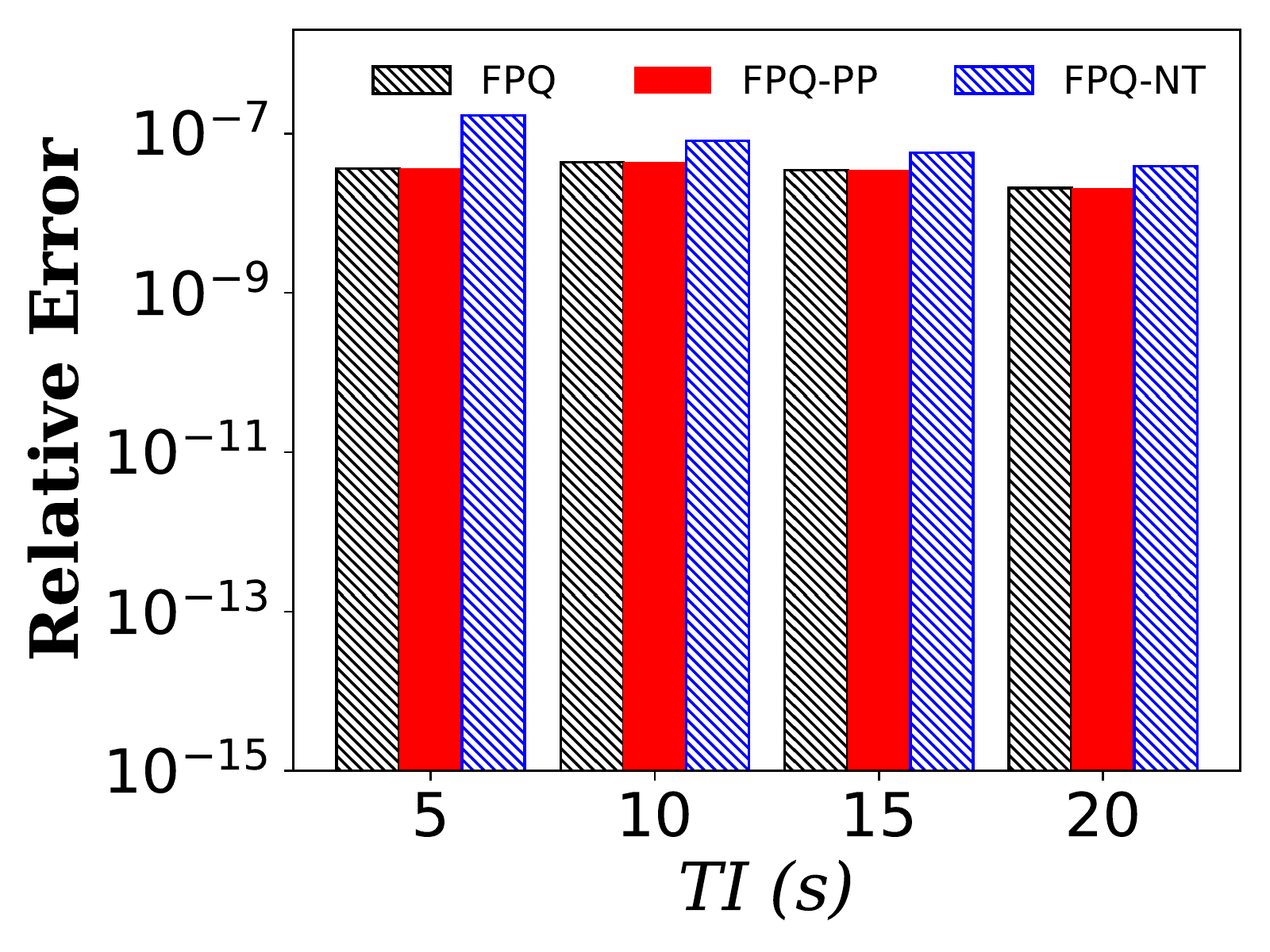}
\ExpCaption{\texttt{FPQ}'s $\gamma$ vs. $\mathit{TI}$}\label{fig:FPQ_TI_error}
\end{minipage}
\end{figure*}

\begin{figure*}[!ht]
\centering
\begin{minipage}[t]{0.245\textwidth}
\centering
\includegraphics[width=\textwidth, height = 3cm]{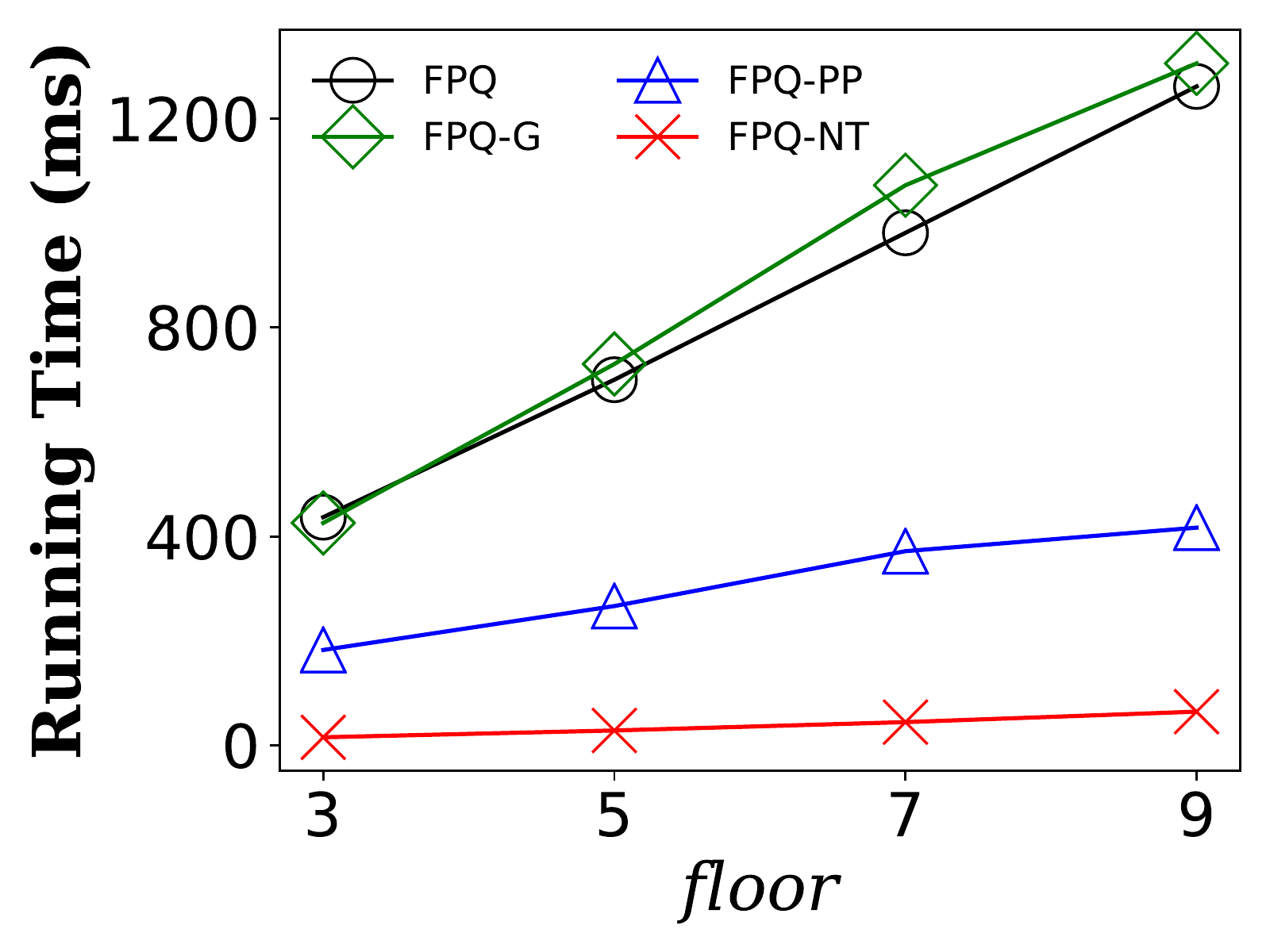}
\ExpCaption{\texttt{FPQ} Time vs. $\mathit{floor}$}\label{fig:FPQ_floor_time}
\end{minipage}
\begin{minipage}[t]{0.245\textwidth}
\centering
\includegraphics[width=\textwidth, height = 3cm]{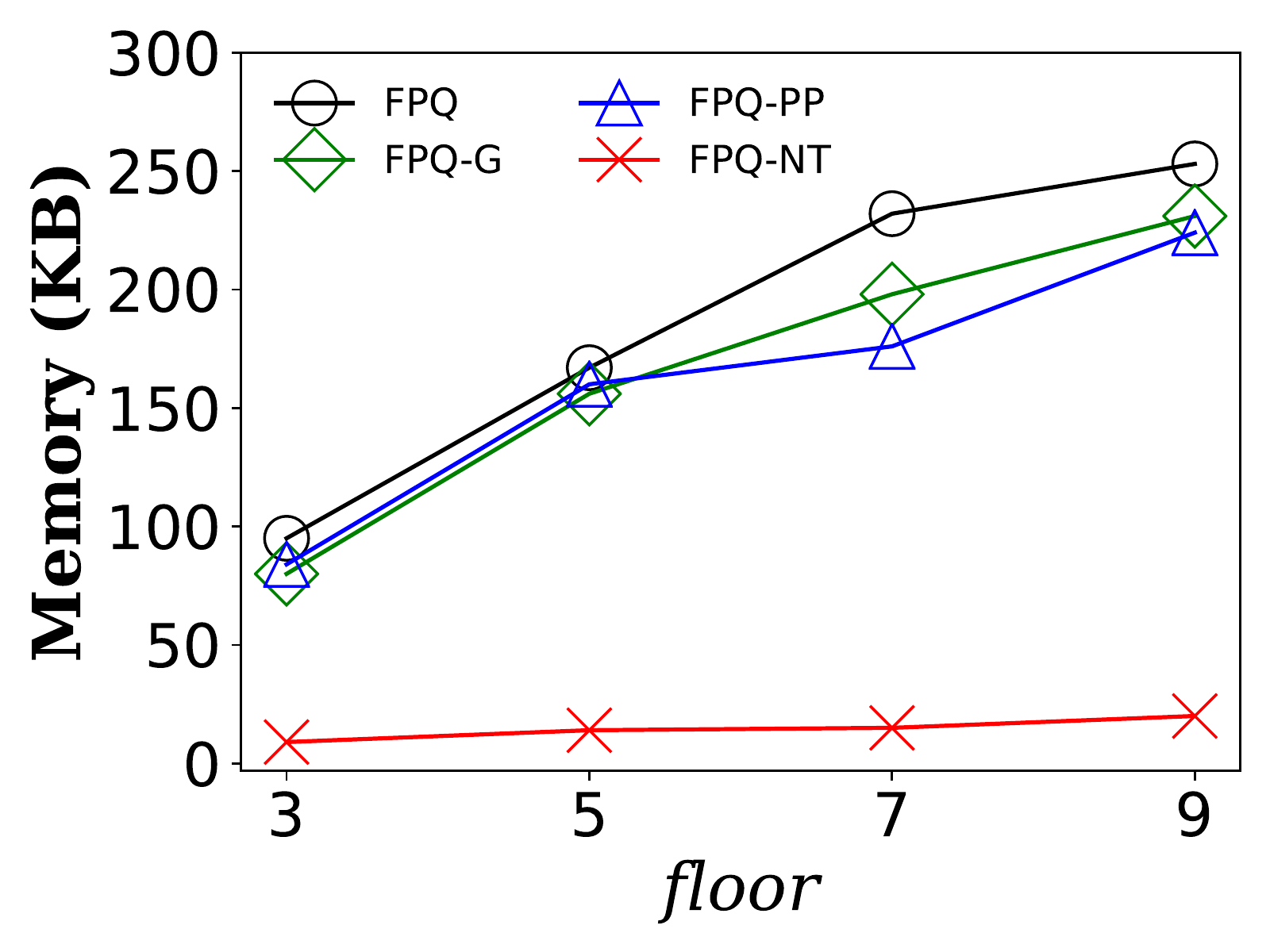}
\ExpCaption{\texttt{FPQ} Memory vs. $\mathit{floor}$}\label{fig:FPQ_floor_mem}
\end{minipage}
\begin{minipage}[t]{0.245\textwidth}
\centering
\includegraphics[width=\textwidth, height = 3cm]{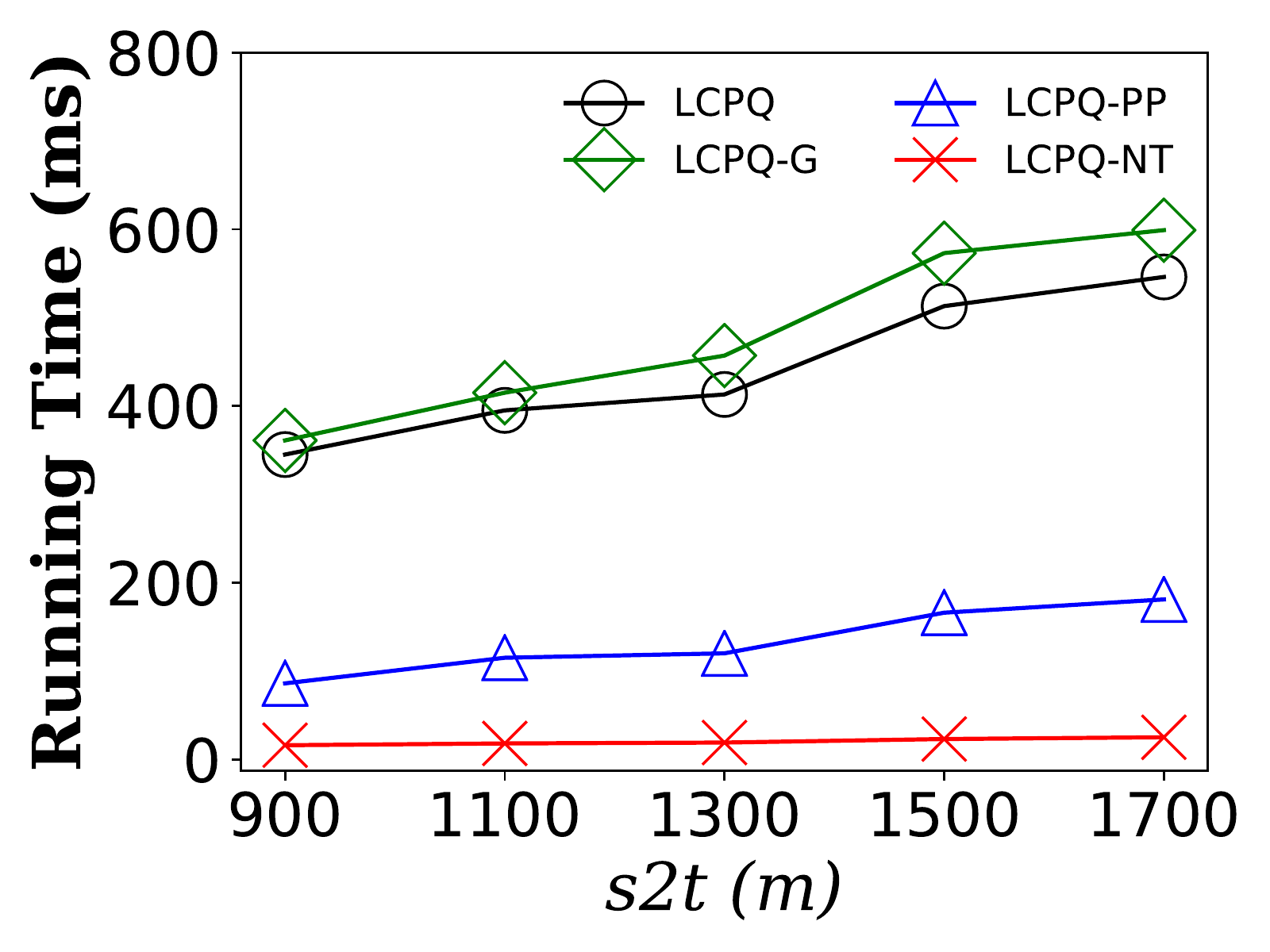}
\ExpCaption{\texttt{LCPQ} Time vs. $\mathit{s2t}$}\label{fig:LCPQ_s2t_time}
\end{minipage}
\begin{minipage}[t]{0.245\textwidth}
\centering
\includegraphics[width=\textwidth, height = 3cm]{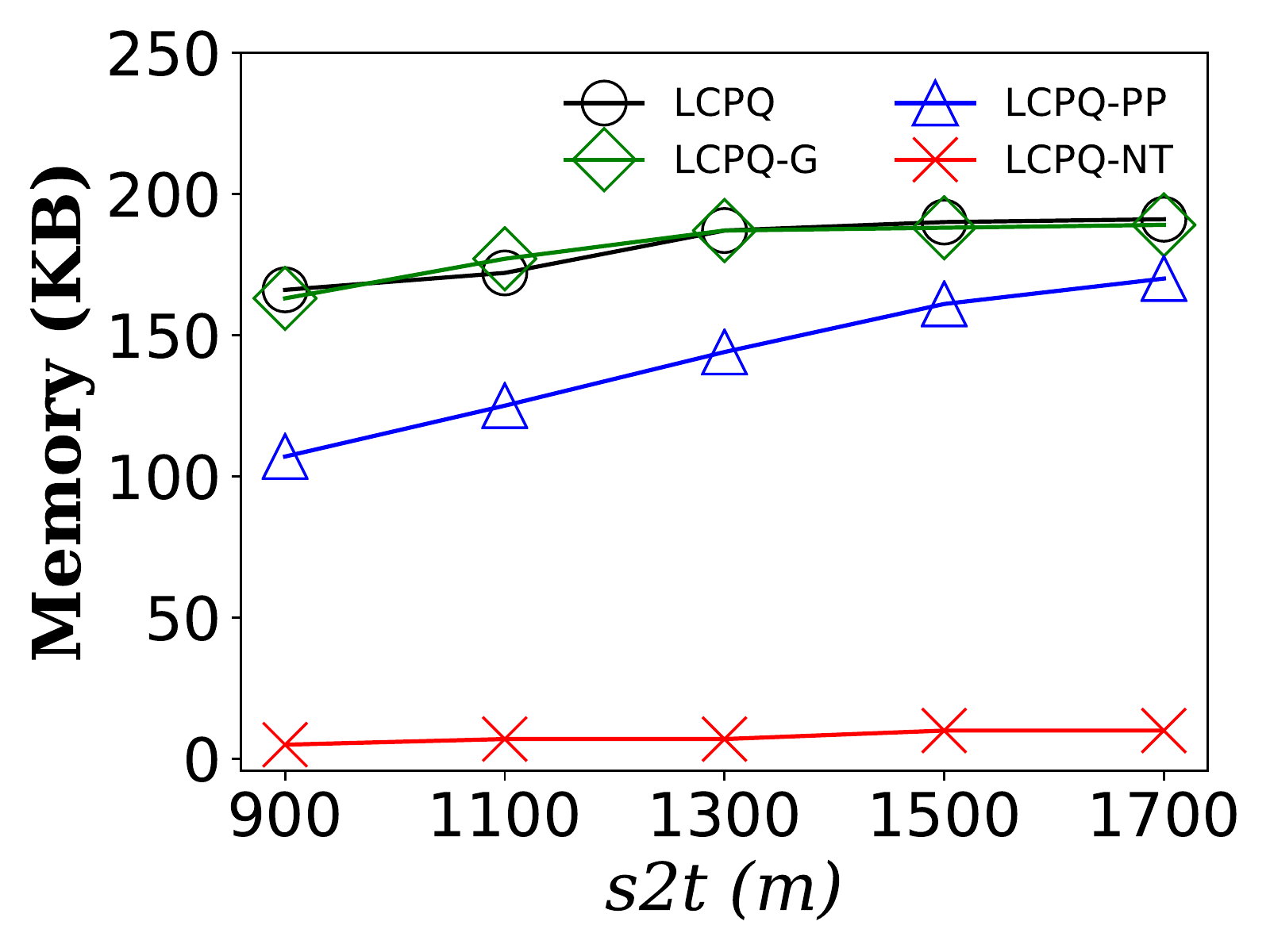}
\ExpCaption{\texttt{LCPQ} Memory vs. $\mathit{s2t}$}\label{fig:LCPQ_s2t_mem}
\end{minipage}
\end{figure*}

\begin{figure*}[!ht]
\centering
\begin{minipage}[t]{0.245\textwidth}
\centering
\includegraphics[width=\textwidth, height = 3cm]{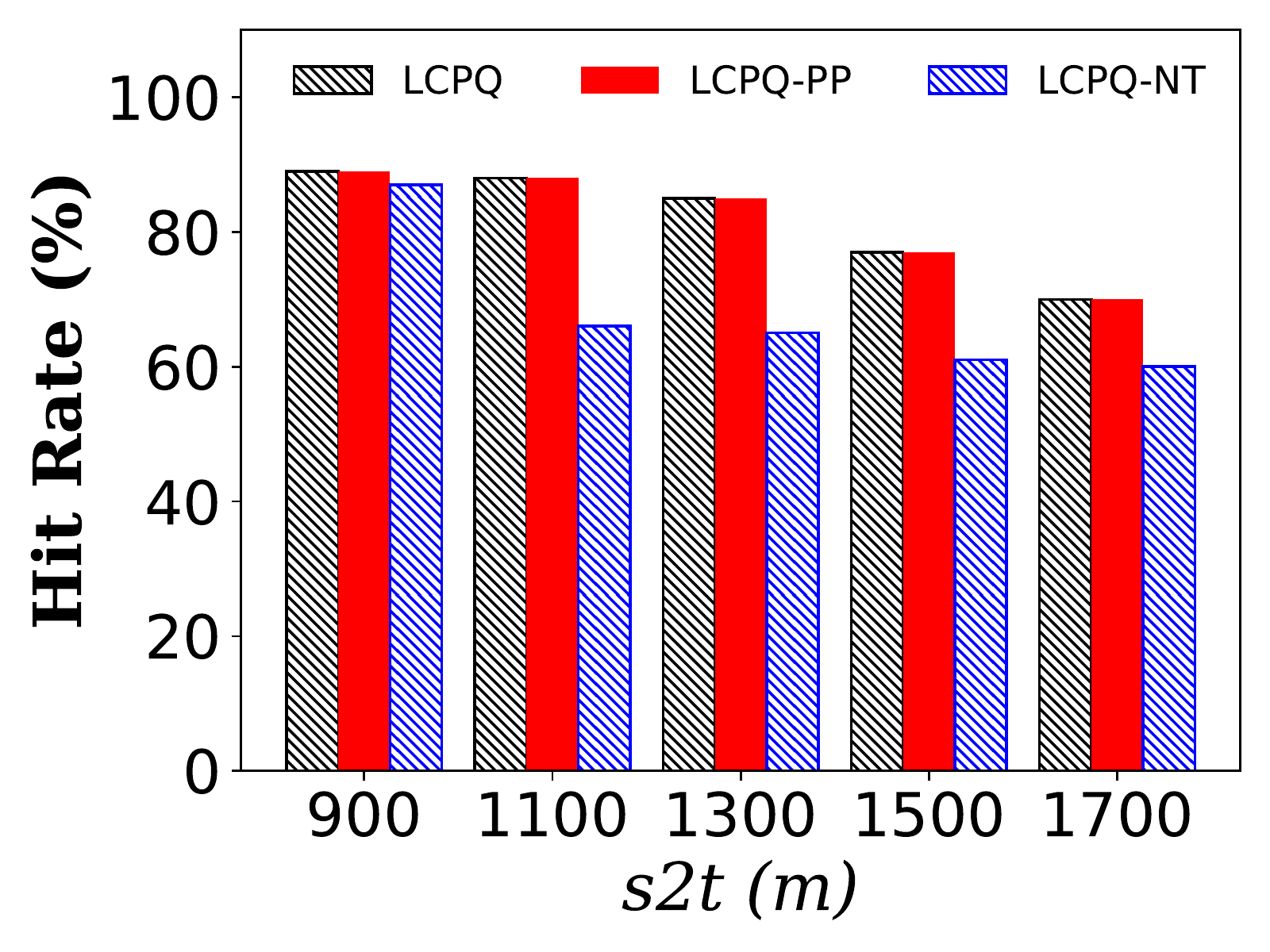}
\ExpCaption{\texttt{LCPQ} Hit Rate vs. $\mathit{s2t}$}\label{fig:LCPQ_s2t_hit}
\end{minipage}
\begin{minipage}[t]{0.245\textwidth}
\centering
\includegraphics[width=\textwidth, height = 3cm]{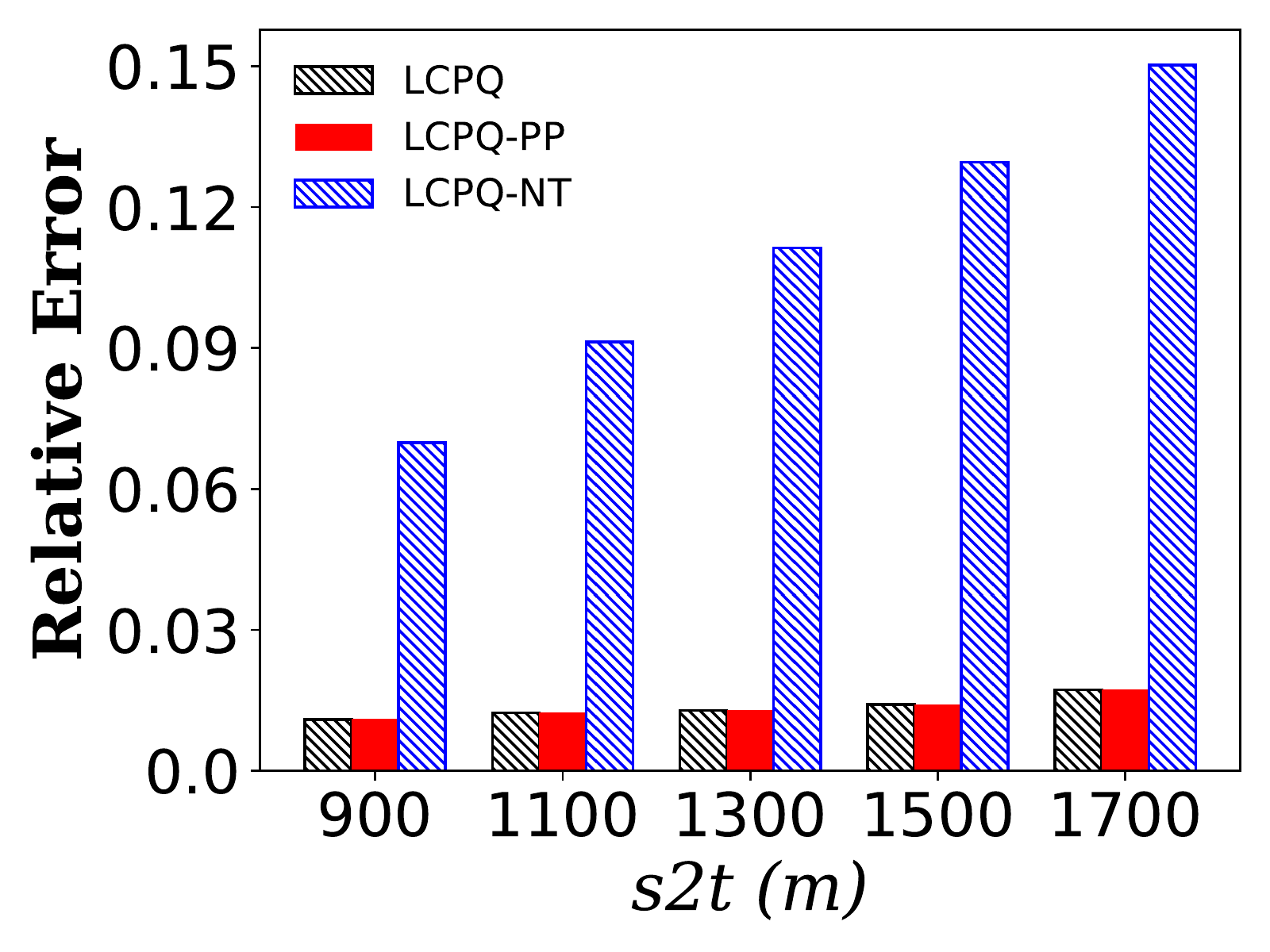}
\ExpCaption{\texttt{LCPQ}'s $\gamma$ vs. $\mathit{s2t}$}\label{fig:LCPQ_s2t_error}
\end{minipage}
\begin{minipage}[t]{0.245\textwidth}
\centering
\includegraphics[width=\textwidth, height = 3cm]{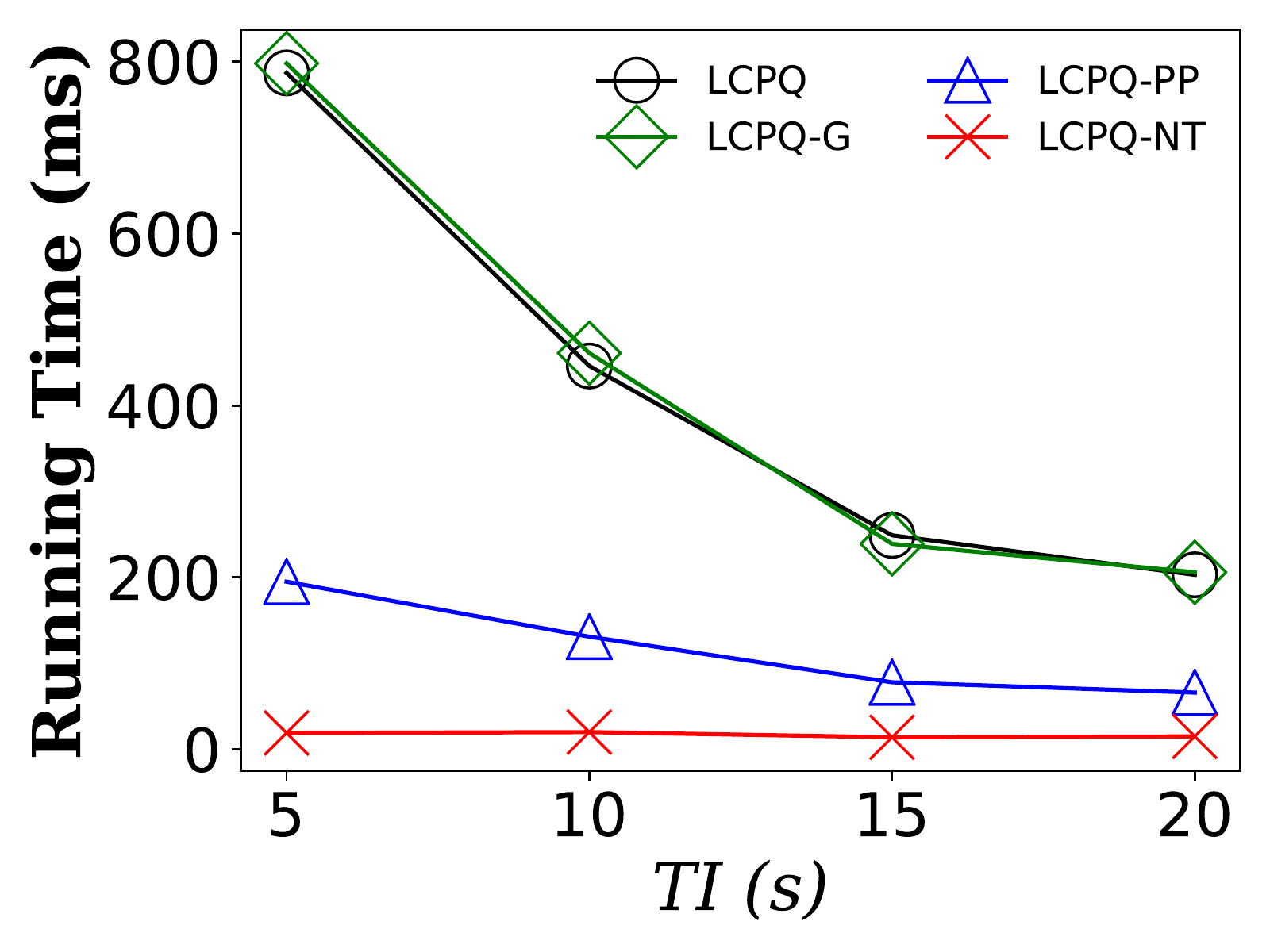}
\ExpCaption{\texttt{LCPQ} Time vs. $\mathit{TI}$}\label{fig:LCPQ_TI_time}
\end{minipage}
\begin{minipage}[t]{0.245\textwidth}
\centering
\includegraphics[width=\textwidth, height = 3cm]{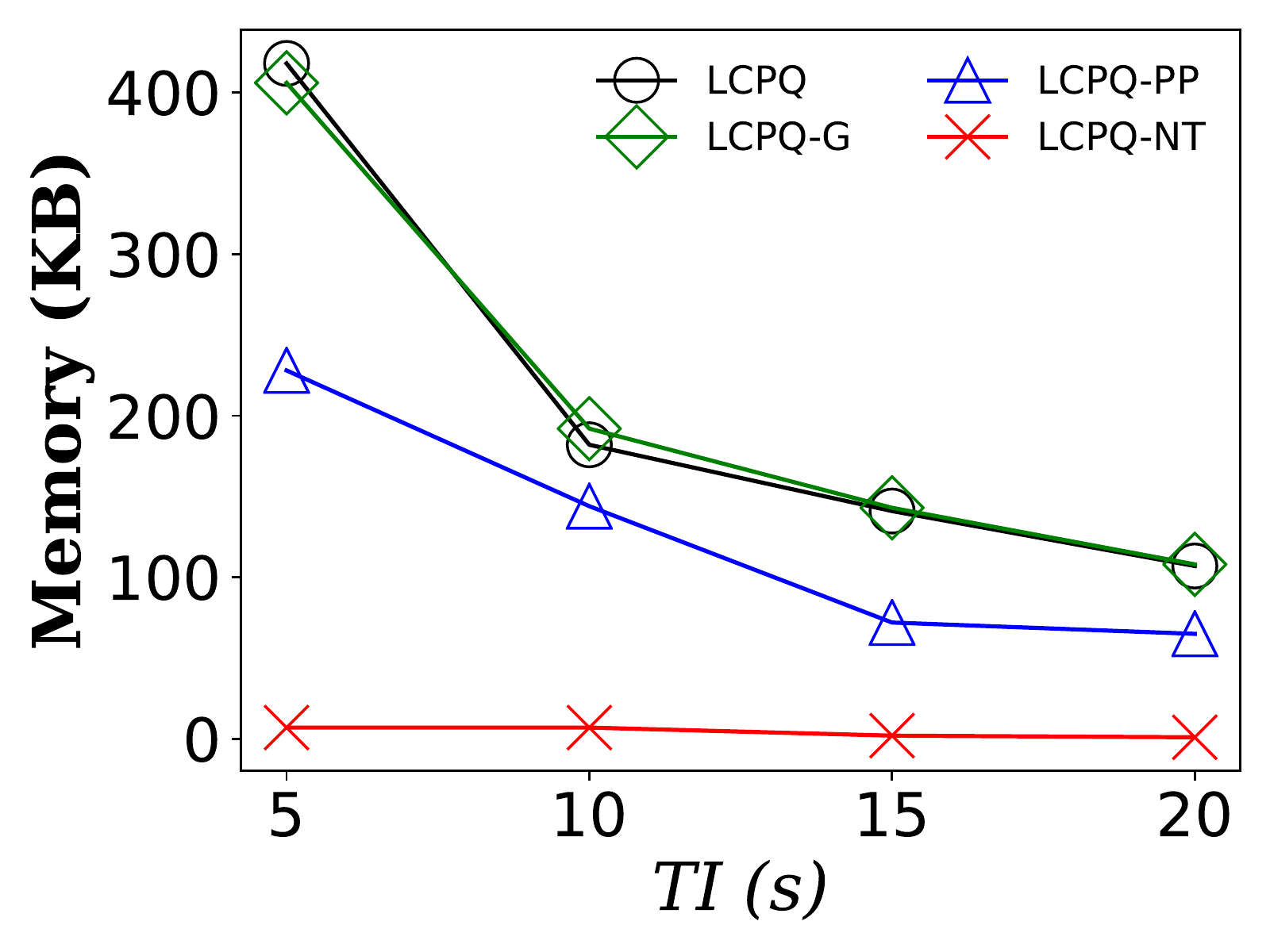}
\ExpCaption{\texttt{LCPQ} Memory vs. $\mathit{TI}$}\label{fig:LCPQ_TI_mem}
\end{minipage}
\end{figure*}

\begin{figure*}[!ht]
\centering
\begin{minipage}[t]{0.245\textwidth}
\centering
\includegraphics[width=\textwidth, height = 3cm]{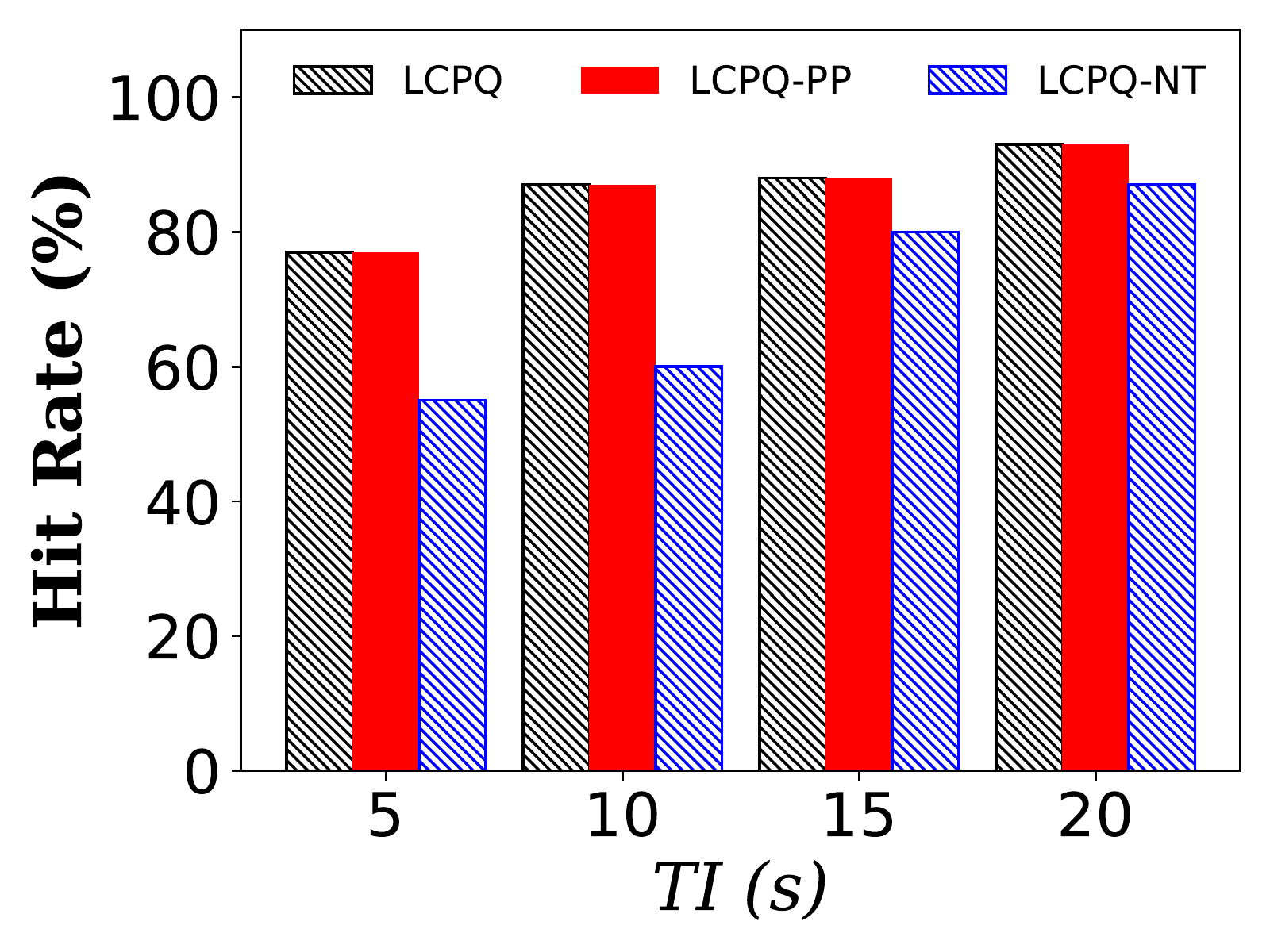}
\ExpCaption{\texttt{LCPQ} Hit Rate vs. $\mathit{TI}$}\label{fig:LCPQ_TI_hit}
\end{minipage}
\begin{minipage}[t]{0.245\textwidth}
\centering
\includegraphics[width=\textwidth, height = 3cm]{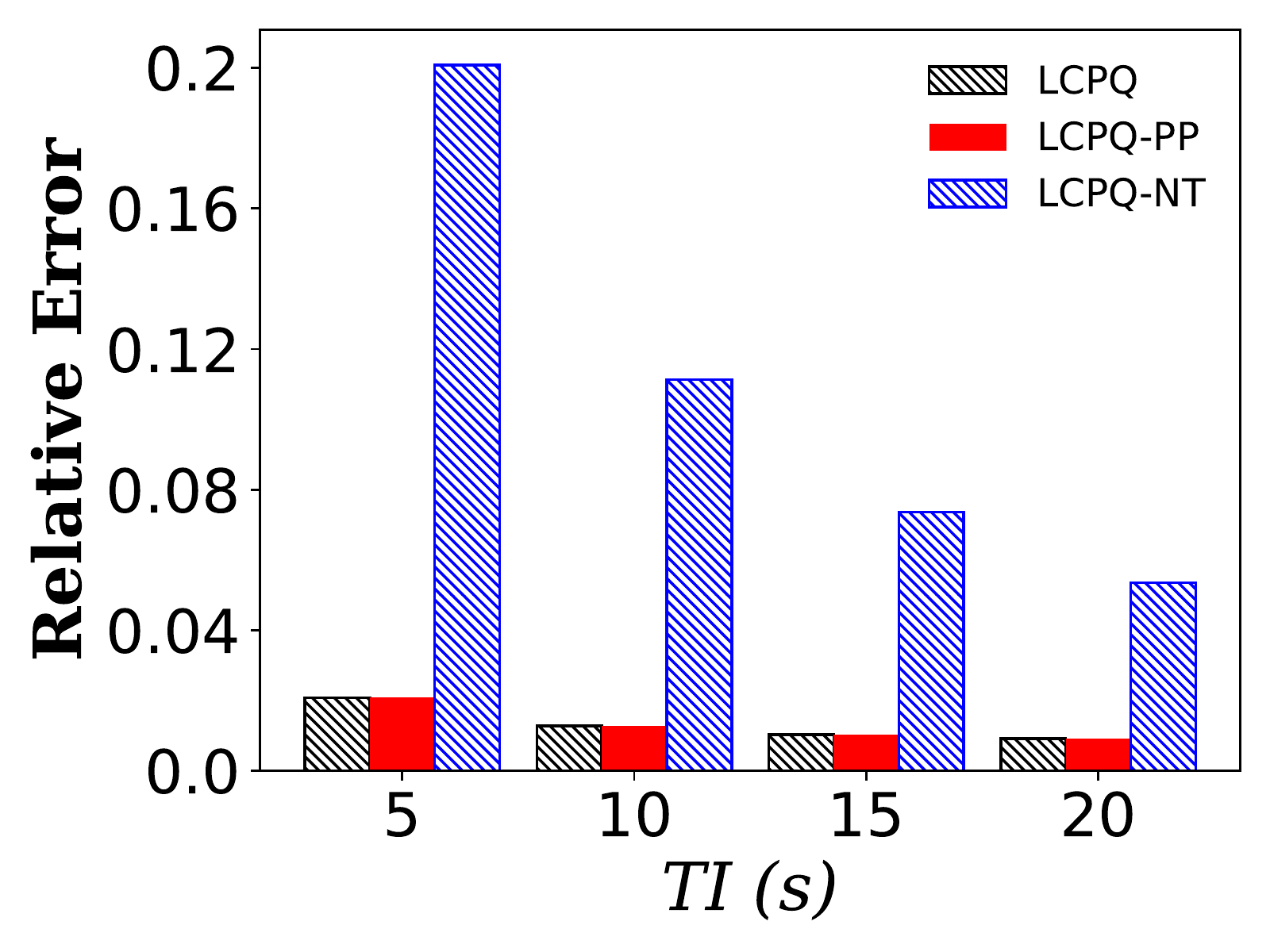}
\ExpCaption{\texttt{LCPQ}'s $\gamma$ vs. $\mathit{TI}$}\label{fig:LCPQ_TI_error}
\end{minipage}
\begin{minipage}[t]{0.245\textwidth}
\centering
\includegraphics[width=\textwidth, height = 3cm]{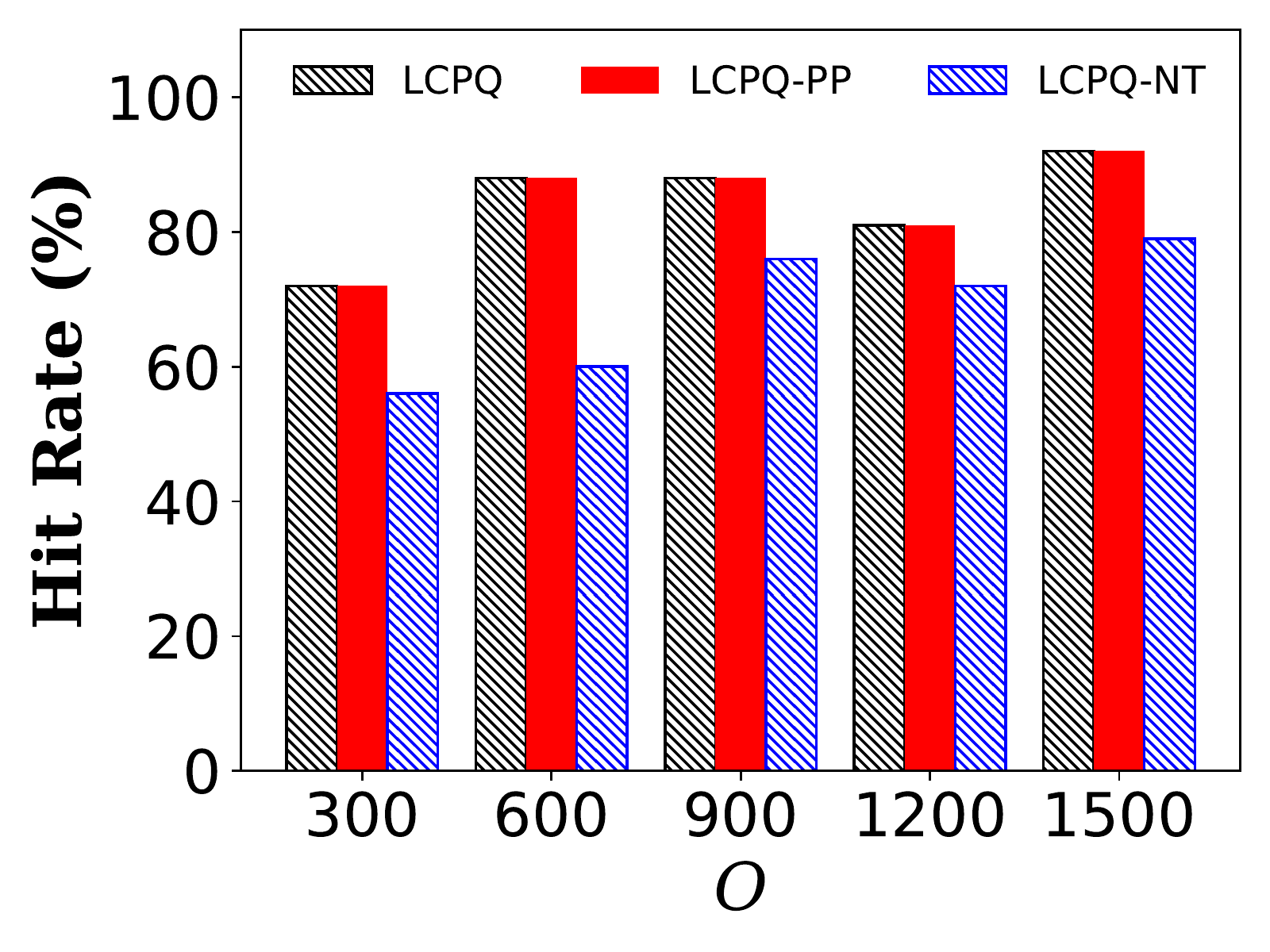}
\ExpCaption{\texttt{LCPQ} Hit Rate vs. $O$}\label{fig:LCPQ_O_hit}
\end{minipage}
\begin{minipage}[t]{0.245\textwidth}
\centering
\includegraphics[width=\textwidth, height = 3cm]{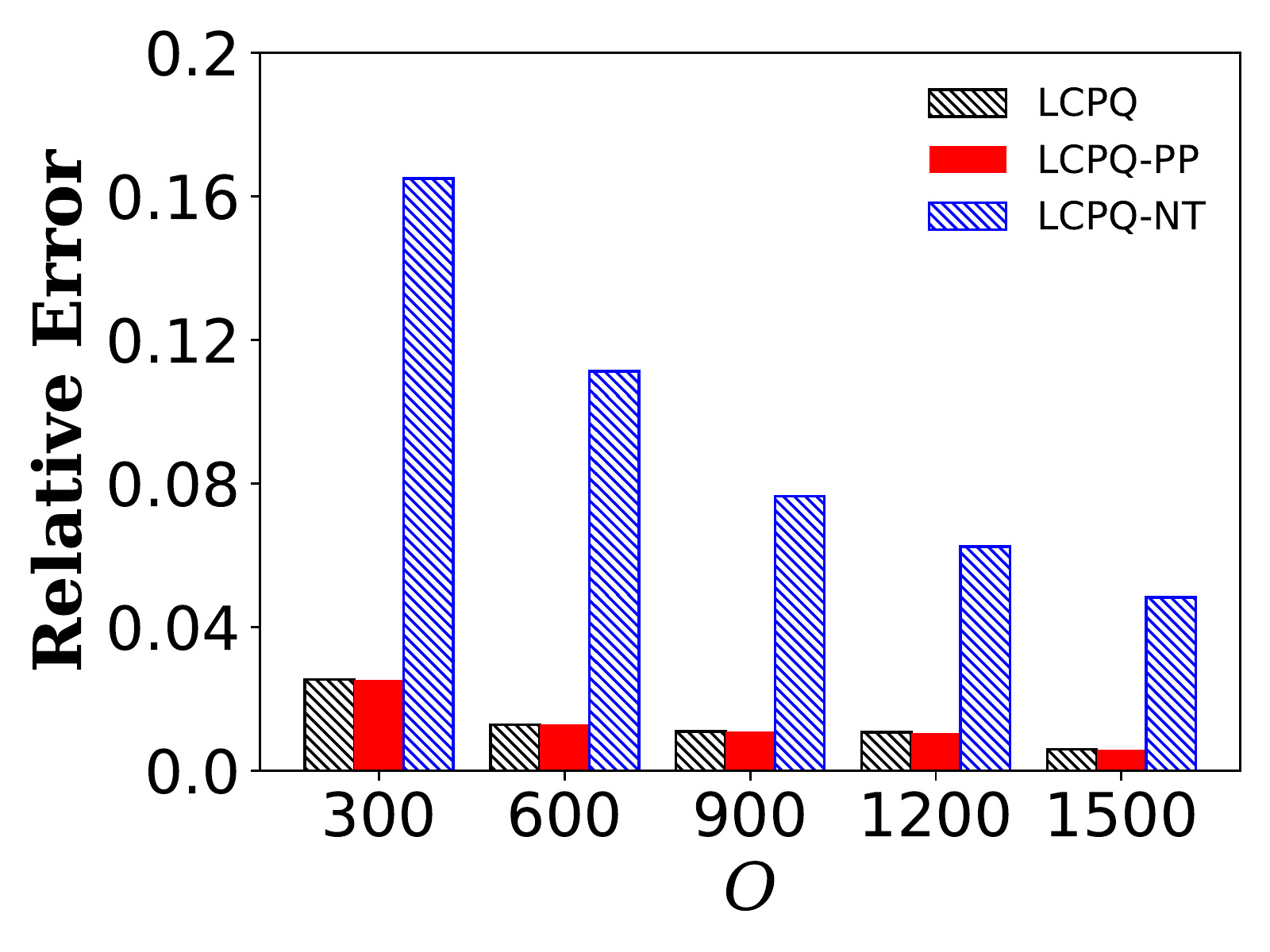}
\ExpCaption{\texttt{LCPQ}'s $\gamma$ vs. $O$}\label{fig:LCPQ_O_error}
\end{minipage}
\end{figure*}

\textbf{Effect of $\mathit{s2t}$.}
We vary the distance $\mathit{s2t}$ between $p_s$ and $p_t$ from 900m to 1700m and test the four \texttt{FPQ} algorithms.
Referring to Figure~\ref{fig:FPQ_s2t_time}, all algorithms' running time increases linearly with the source-target distance, since a larger $\mathit{s2t}$ involves a larger expansion range as well as more candidate path nodes.
Among all algorithms, \texttt{FPQ-NT} runs fastest because it skips the iterative population computation and directly estimates its population at the arrival time in each node.
Moreover, the time costs of approximate searches \texttt{FPQ-PP} and \texttt{FPQ-NT} increase very slowly with $\mathit{s2t}$ increases.
In contrast, the exact searches \texttt{FPQ} and \texttt{FPQ-G} are sensitive to $\mathit{s2t}$ because they need to compute more population.

Figure~\ref{fig:FPQ_s2t_mem} reports the memory consumptions.
The memory use of \texttt{FPQ} and \texttt{FPQ-G} grows faster than the others due to an extra cost of rigid population derivation.
Contrary to our intuition, \texttt{FPQ} incurs more memory cost than \texttt{FPQ-G}. In our test, the search framework needs to explore a large number of partitions. \texttt{FPQ-G}'s global population derivation shares intermediate results across all partitions. For a large $\mathit{s2t}$, \texttt{FPQ-G} may find more shared intermediate results, and thus consumes less memory than \texttt{FPQ}.

Figures~\ref{fig:FPQ_s2t_hit} and~\ref{fig:FPQ_s2t_error} reports the query hit rates and relative errors, respectively, for \texttt{FPQ}, \texttt{FPQ-PP} and \texttt{FPQ-NT}.\footnote {We exclude \texttt{FPQ-G} as its accuracy is the same with \texttt{FPQ}.}
For different $\mathit{s2t}$ values, \texttt{FPQ} achieves a higher hit rate and lower error.
As a sacrifice to improve search efficiency, \texttt{FPQ-NT} skips some intermediate update timestamps, so its accuracy of population derivation drops more significantly.
Moreover, as $\mathit{s2t}$ increases, \texttt{FPQ-NT} deteriorates rapidly while \texttt{FPQ} and \texttt{FPQ-PP} perform quite stably.
A larger $\mathit{s2t}$ leads to more update timestamps to derive populations. This makes \texttt{FPQ-NT}'s relatively aggressive strategy of skipping timestamps introduce more estimation errors. In contrast, \texttt{FPQ} and \texttt{FPQ-PP} derive populations timestamp by timestamp, and so the impact of $\mathit{s2t}$ is slight.

\textbf{Effect of $\mathit{TI}$.}
Referring to Figures~\ref{fig:FPQ_TI_time} and~\ref{fig:FPQ_TI_mem}, both running time and memory cost of all four algorithms decrease with a larger update time interval $\mathit{TI}$.
Still, \texttt{FPQ-NT} performs best in both measures as it approximates population derivation in both time and space aspects.
On the other hand, referring to Figures~\ref{fig:FPQ_TI_hit} and~\ref{fig:FPQ_TI_error}, \texttt{FPQ-NT}'s hit rate gets better while its relative error decreases as all doors' $\mathit{TI}$ enlarges.
This shows that one may consider skipping more timestamps when the flow update at doors is not that frequent.
Interestingly, while improving search efficiency, \texttt{FPQ-PP}'s hit rate is very close to \texttt{FPQ}, nearby 100\% when varying $\mathit{TI}$. This reflects that far-away partitions have little effect on the population of the target partition in queries.

\textbf{Effect of $\mathit{floor}$.}
We vary the number of floors to test the scalability of our algorithms.
Referring to Figure~\ref{fig:FPQ_floor_time}, all algorithms' search time increases steadily with more floors since more candidate path nodes are involved.
\texttt{FPQ-PP} and \texttt{FPQ-NT} run faster than \texttt{FPQ} and \texttt{FPQ-G}. Moreover, the running time of the two approximate searches grows more slowly.
Figure~\ref{fig:FPQ_floor_mem} reports the memory costs of the four algorithms. \texttt{FPQ-NT} costs less memory and is more scalable since it skips some timestamps.
We omit the results of the searches' hit rates and relative errors.
In the tests, both measures are insensitive to the floor number since the returned path stays unchanged for a given query instance.

We omit the results of varying $|o|$ on \texttt{FPQ} because different initial object numbers have little impact on the search performance.

\subsubsection{Search Performance of \texttt{LCPQ}}

\textbf{Comparison in default setting.}
The result trend of \texttt{LCPQ} is similar to that of \texttt{FPQ}. As reported in the right part of Table~\ref{tab:comparison_SYN}, \texttt{LCPQ-NT} is the best in terms of running time and memory due to its skipping strategy, while \texttt{LCPQ-GTG} costs the highest time and memory due to the large graph size.
\texttt{LCPQ-A} gets the best hit rate while the exact searches achieve a better result on the relative error.
Different from \texttt{FPQ} query, \texttt{LCPQ} query is highly sensitive to populations. A little error in population derivation can lead to a very different returned path.
Hence, the accuracy performance is slightly unstable for the tested algorithms.
Nevertheless, a good hit rate means the searches return satisfactory paths to users.
 
\textbf{Effect of $\mathit{s2t}$.}
Referring to Figure~\ref{fig:LCPQ_s2t_time}, the running time of each algorithm grows as $\mathit{s2t}$ increases.
In terms of memory, the results in Figure~\ref{fig:LCPQ_s2t_mem} show that \texttt{LCPQ} and \texttt{LCPQ-G} need more memory than the other two.
\texttt{LCPQ-NT} uses the least memory since it skips intermediate timestamps to reduce workload.

Figure~\ref{fig:LCPQ_s2t_hit} reports the hit rates of exact and approximate searches.
\texttt{LCPQ-PP} achieves a better hit rate than \texttt{LCPQ-NT}, and it is close to the exact search \texttt{LCPQ} since it corresponds to a more precise population derivation. This can be reflected in its relative error measures reported in Figure~\ref{fig:LCPQ_s2t_error}.
Compared to \texttt{LCPQ} and \texttt{LCPQ-PP}, \texttt{LCPQ-NT} incurs significantly higher errors and lower hit rates.
As we mentioned before, \texttt{LCPQ} query is highly sensitive to the population. A little error in population derivation can lead to a very different returned path.
Therefore, \texttt{LCPQ-NT} performs poorly when a larger $\mathit{s2t}$ is used.

\textbf{Effect of $\mathit{TI}$.}
Referring to Figures~\ref{fig:LCPQ_TI_time} and~\ref{fig:LCPQ_TI_mem}, all algorithms cost less time and memory as $\mathit{TI}$ increases since a larger $\mathit{TI}$ leads to fewer callings of population derivation.
The approximate approaches \texttt{LCPQ-PP} and \texttt{LCPQ-NT} always perform better in search efficiency.
Referring to Figures~\ref{fig:LCPQ_TI_hit} and~\ref{fig:LCPQ_TI_error}, all algorithms' search effectiveness deteriorates with an increasing $\mathit{TI}$.
As less flow information is observed when $\mathit{TI}$ becomes larger, the relative errors accumulate and the hit rates are thus decreased.
Likewise, \texttt{LCPQ}'s search effectiveness is worse than that of \texttt{FPQ}, due to its more stringent requirements on population derivation.

\begin{table*}[]
\scriptsize
\caption{Comparison of Algorithms for \texttt{FPQ} and \texttt{LCPQ} on Real Data (best result in bold)}\label{tab:comparison_HSM}
\centering
\resizebox{17.5cm}{0.7cm}{
\begin{tabular}{|l||l|l|l|l|l|l||l|l|l|l|l|l|}
\hline
               &\texttt{FPQ} &\texttt{FPQ-G} &\texttt{FPQ-PP} &\texttt{FPQ-NT} &\texttt{FPQ-GTG} &\texttt{FPQ-A}&\texttt{LCPQ}&\texttt{LCPQ-G}&\texttt{LCPQ-PP}&\texttt{LCPQ-NT}&\texttt{LCPQ-GTG}&\texttt{LCPQ-A} \\ \hline
Running Time (ms) & 1900      & 1997     & 67       & \textbf{11}& 25559    & 53    & 992         & 1047        & 28          & \textbf{10} & 13895       & 45      \\ \hline
Memory (KB)       & 367      & 393      & 61       & \textbf{1} & 669      & 2  & 307         & 341         & 30          & \textbf{1}  & 568         & 2         \\ \hline
Hit Rate (\%)     &\textbf{99}&\textbf{99}&\textbf{99}& 98       &\textbf{99}& 98  & 88 & 88 & 88 & 67          & 88 & \textbf{90}         \\ \hline
Relative Error    &\textbf{1.86E-15}&\textbf{1.86E-15}&\textbf{1.86E-15}& 4.38E-14 &\textbf{1.86E-15}& 0.1492& \textbf{0.0546} & \textbf{0.0546} & \textbf{0.0546} & 0.6606 & \textbf{0.0546} & 0.062 \\ \hline
\end{tabular}}
\end{table*}

\textbf{Effect of $|o|$.}
We test different initial object numbers on \texttt{LCPQ} query processing.
As an observation, the running time and memory cost are almost insensitive to $|o|$, since the algorithms do not process each individual object.
So we omit the results here.
Interestingly, increasing $|o|$ will affect the accuracy of the results.
Referring to Figures~\ref{fig:LCPQ_O_hit} and~\ref{fig:LCPQ_O_error}, as more objects are involved, all methods achieve a higher hit rate and a lower relative error.
We attribute it to that a larger population base is less affected by the flow estimation error and leads to a smaller relative error.

We omit the result about different floor numbers because it exhibits a trend similar to the counterpart of \texttt{FPQ} searches.

\begin{figure*}[!ht]
\centering
\begin{minipage}[t]{0.245\textwidth}
\centering
\includegraphics[width=\textwidth, height = 3cm]{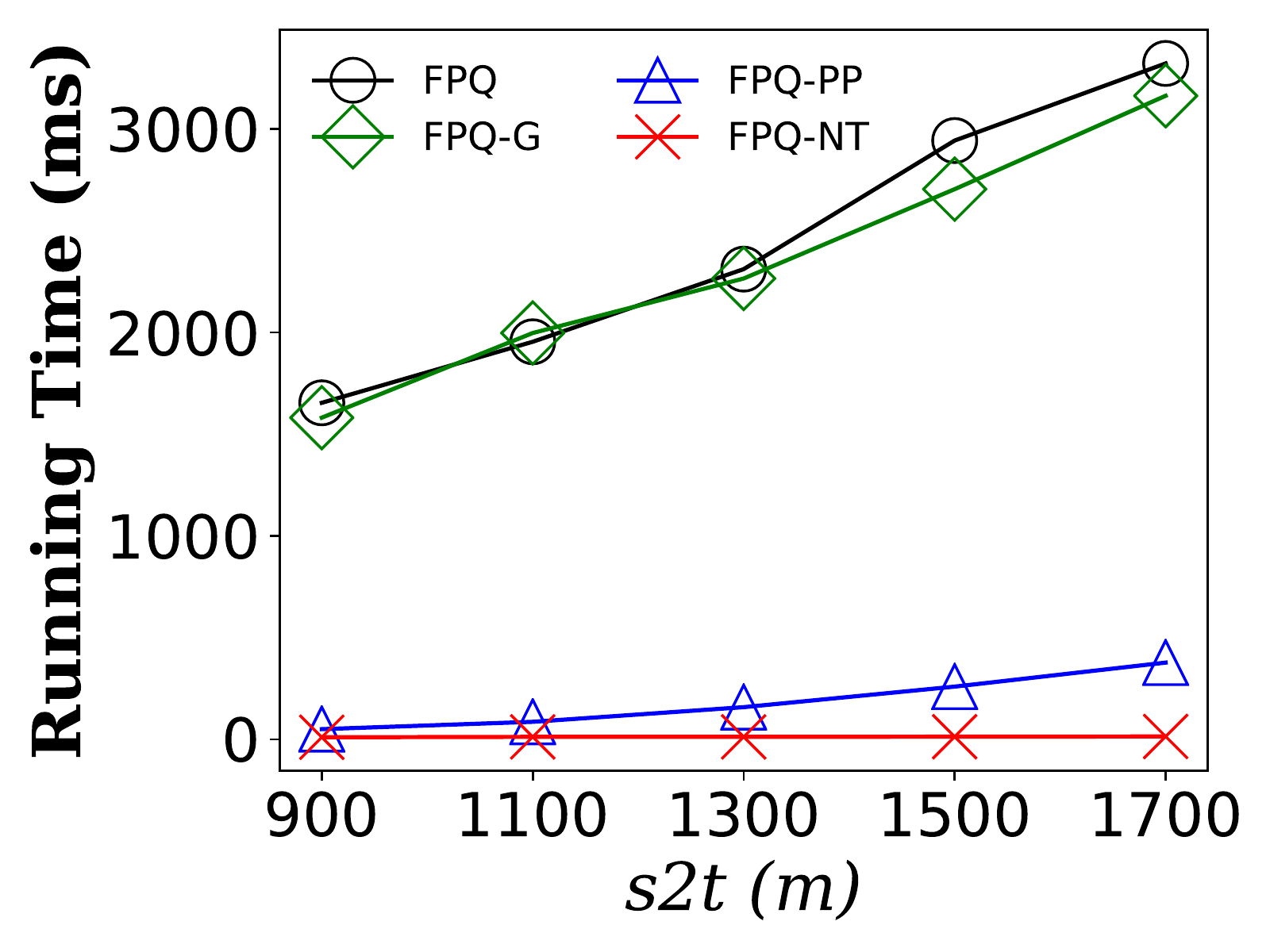}
\ExpCaption{\texttt{FPQ} Time vs. $\mathit{s2t}$ (Real)}\label{fig:FPQ_s2t_time_HSM}
\end{minipage}
\begin{minipage}[t]{0.245\textwidth}
\centering
\includegraphics[width=\textwidth, height = 3cm]{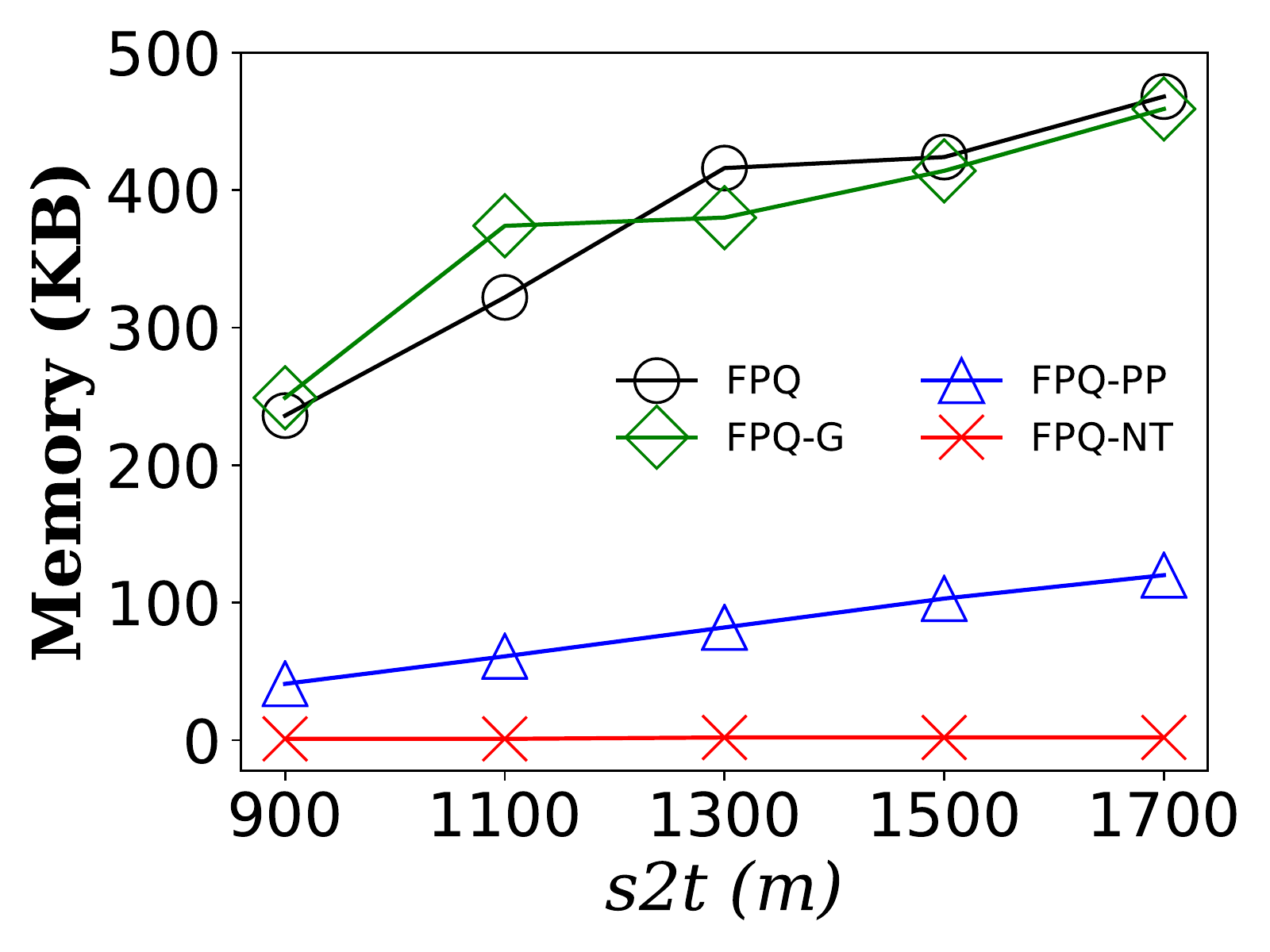}
\ExpCaption{\texttt{FPQ} Memory vs. $\mathit{s2t}$ (Real)}\label{fig:FPQ_s2t_mem_HSM}
\end{minipage}
\begin{minipage}[t]{0.245\textwidth}
\centering
\includegraphics[width=\textwidth, height = 3cm]{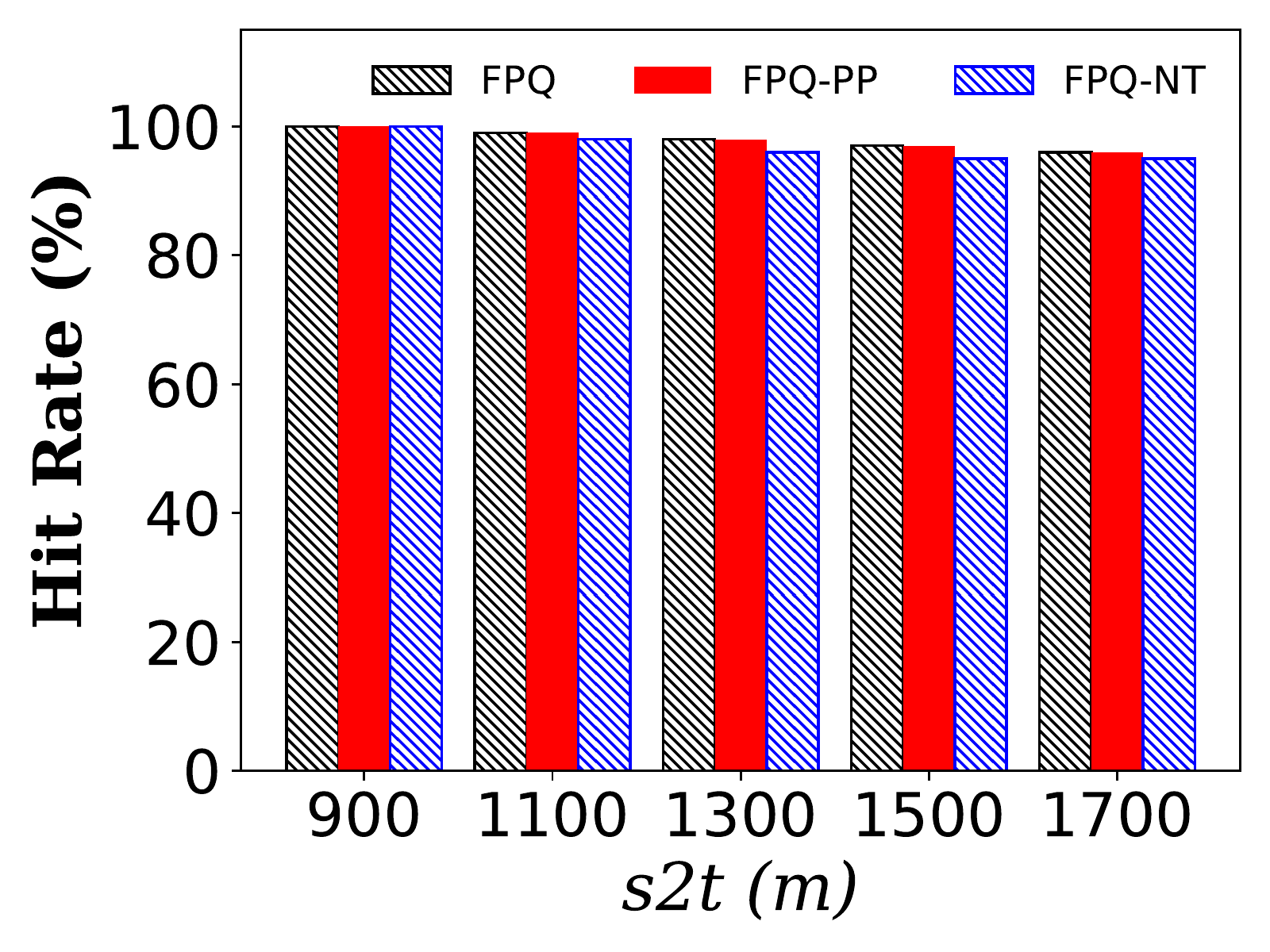}
\ExpCaption{\texttt{FPQ} Hit Rate vs. $\mathit{s2t}$ (Real)}\label{fig:FPQ_s2t_hit_HSM}
\end{minipage}
\begin{minipage}[t]{0.245\textwidth}
\centering
\includegraphics[width=\textwidth, height = 3cm]{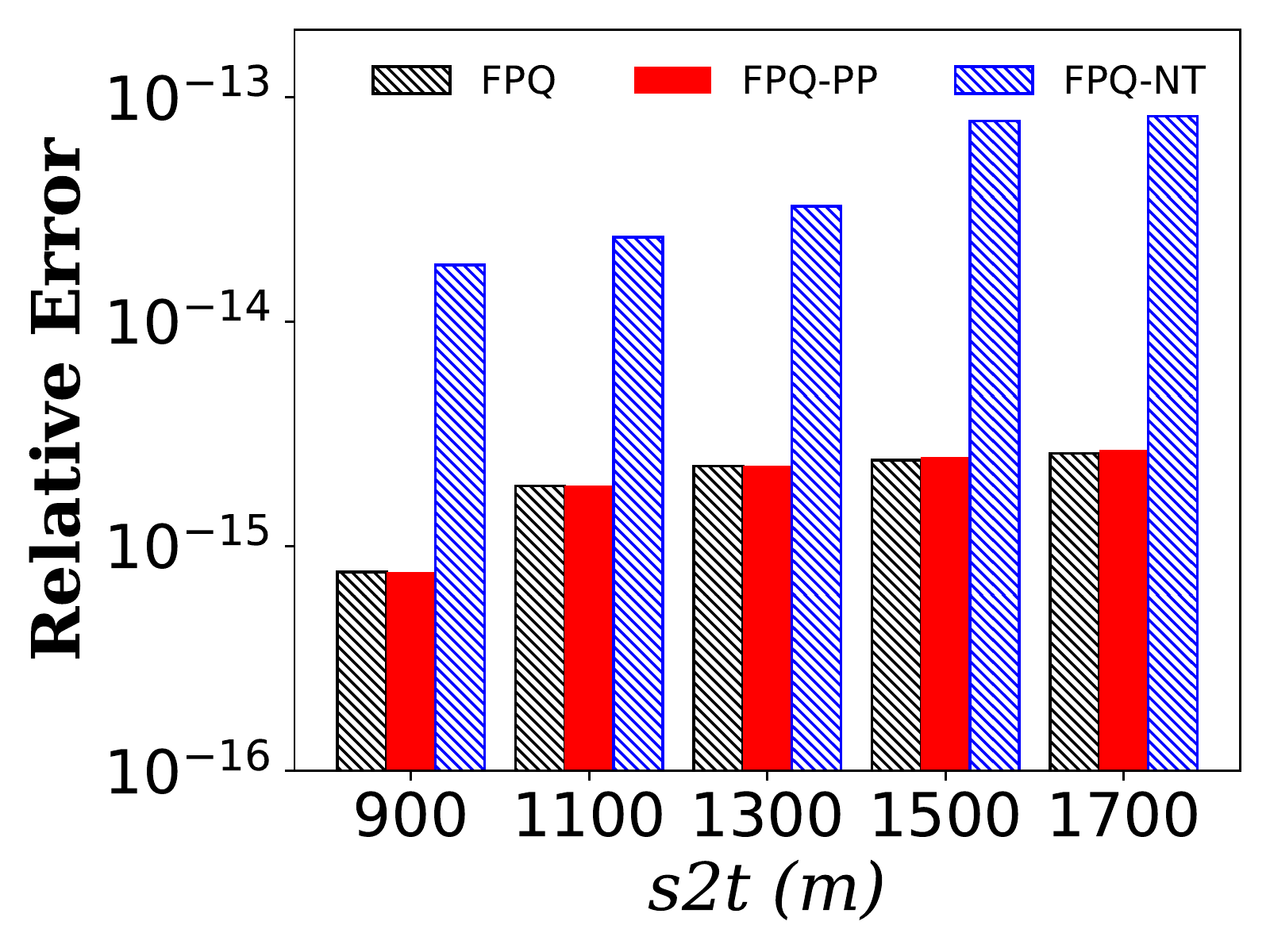}
\ExpCaption{\texttt{FPQ}'s $\gamma$ vs. $\mathit{s2t}$ (Real)}\label{fig:FPQ_s2t_error_HSM}
\end{minipage}
\end{figure*}

\begin{figure*}[!ht]
\centering
\begin{minipage}[t]{0.245\textwidth}
\centering
\includegraphics[width=\textwidth, height = 3cm]{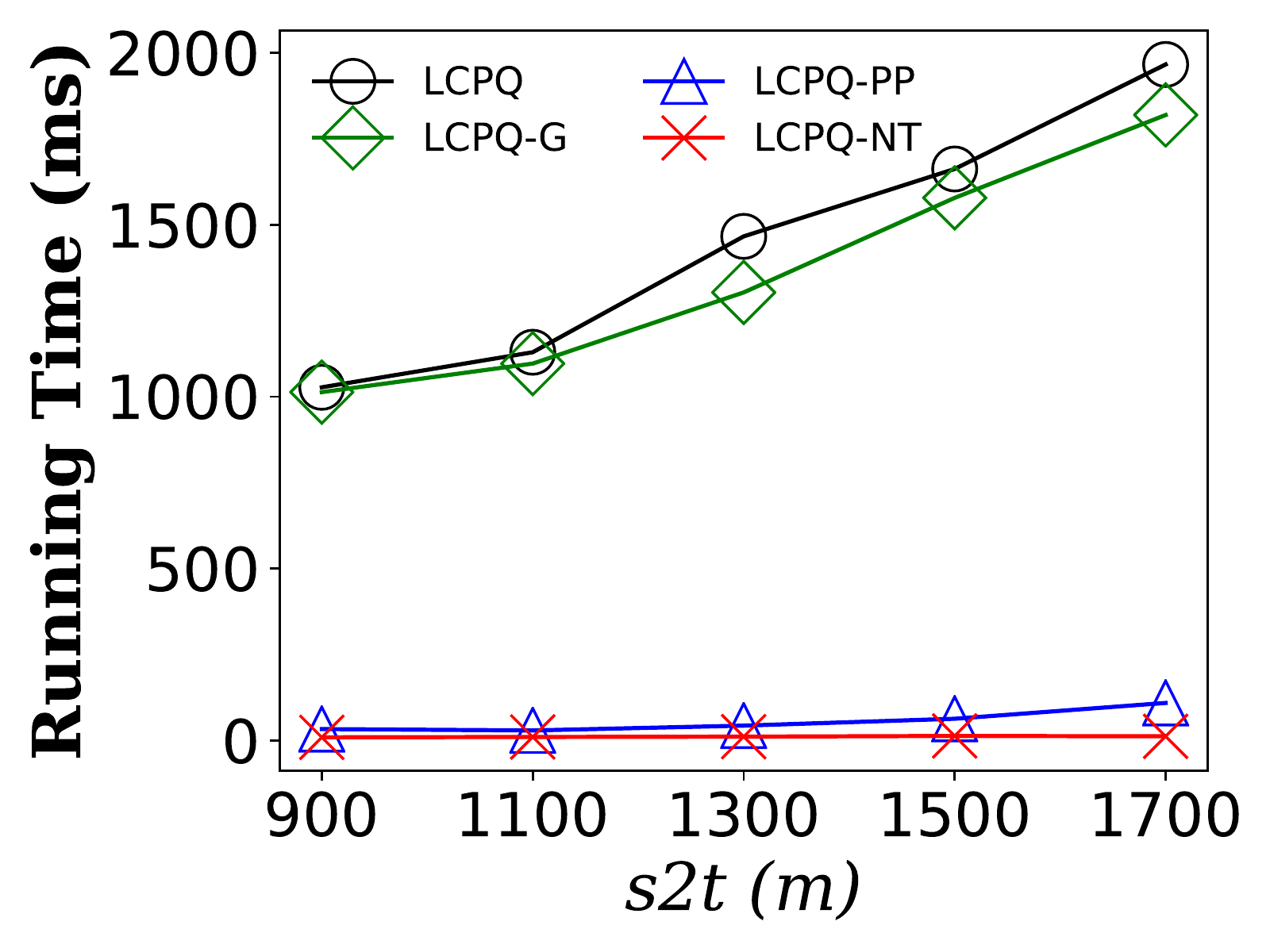}
\ExpCaption{\texttt{LCPQ} Time vs. $\mathit{s2t}$ (Real)}\label{fig:LCPQ_s2t_time_HSM}
\end{minipage}
\begin{minipage}[t]{0.245\textwidth}
\centering
\includegraphics[width=\textwidth, height = 3cm]{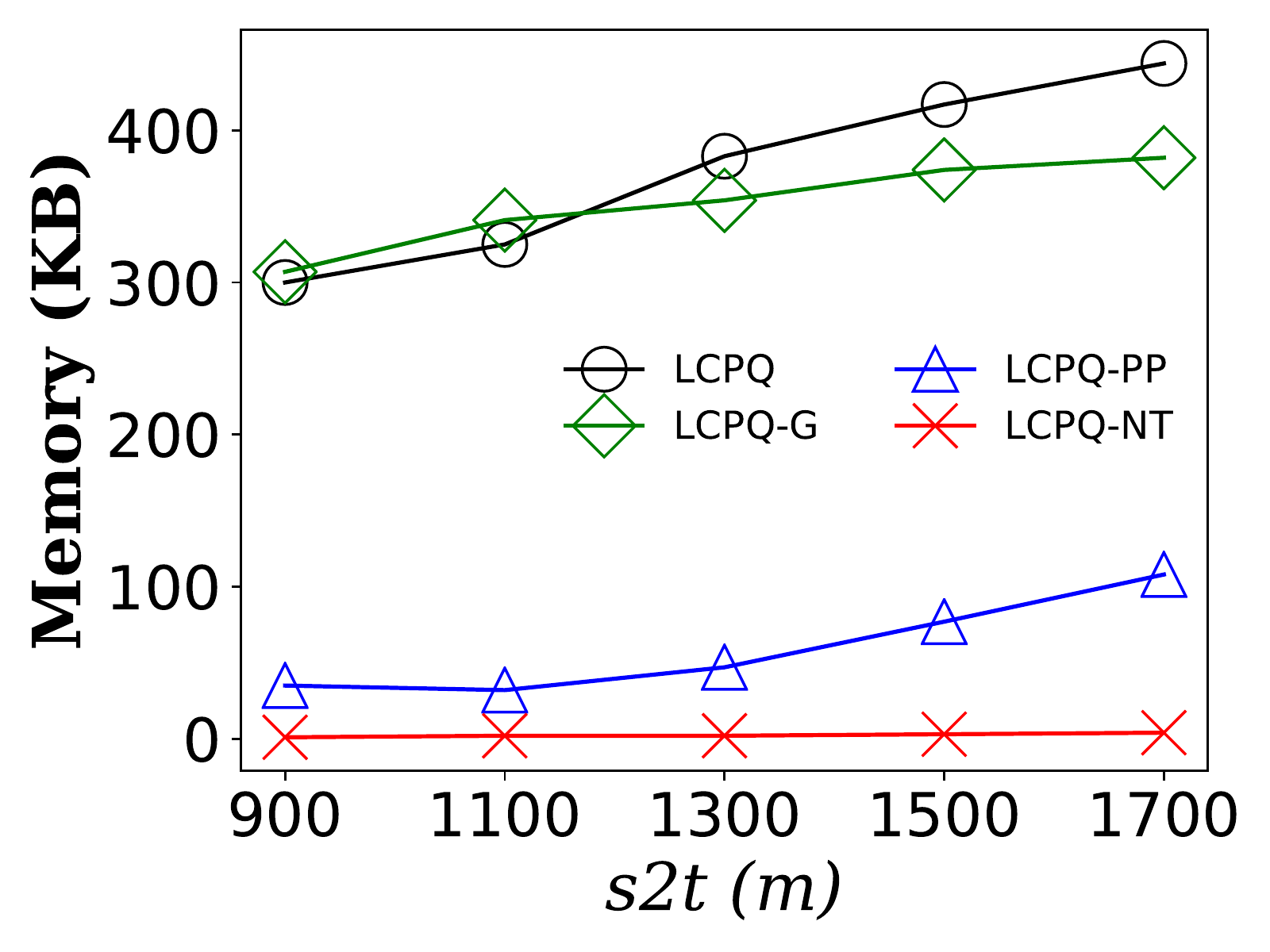}
\ExpCaption{\texttt{LCPQ} Memory vs. $\mathit{s2t}$ (Real)}\label{fig:LCPQ_s2t_mem_HSM}
\end{minipage}
\begin{minipage}[t]{0.245\textwidth}
\centering
\includegraphics[width=\textwidth, height = 3cm]{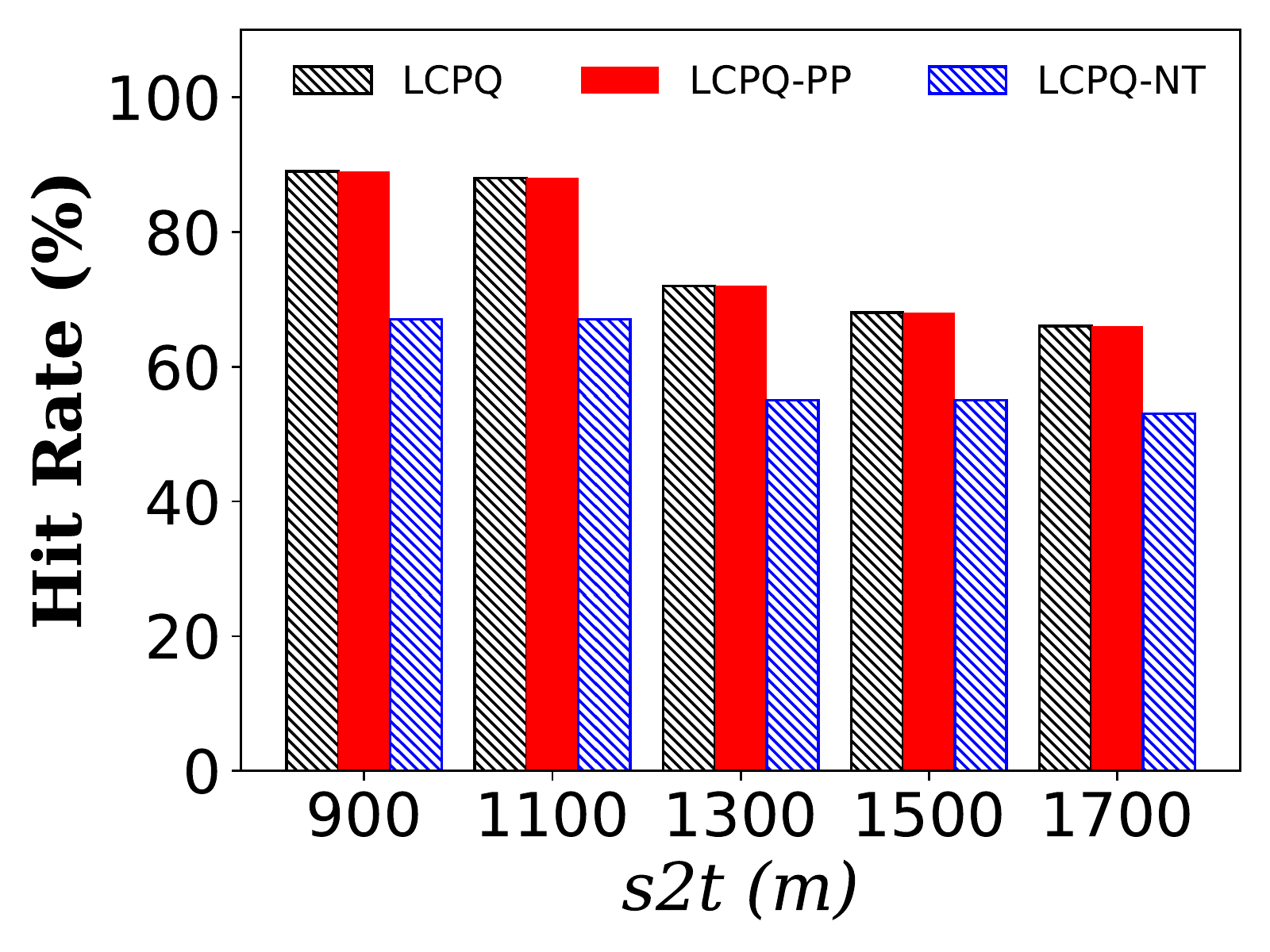}
\ExpCaption{\texttt{LCPQ} Hit Rate vs. $\mathit{s2t}$ (Real)}\label{fig:LCPQ_s2t_hit_HSM}
\end{minipage}
\begin{minipage}[t]{0.245\textwidth}
\centering
\includegraphics[width=\textwidth, height = 3cm]{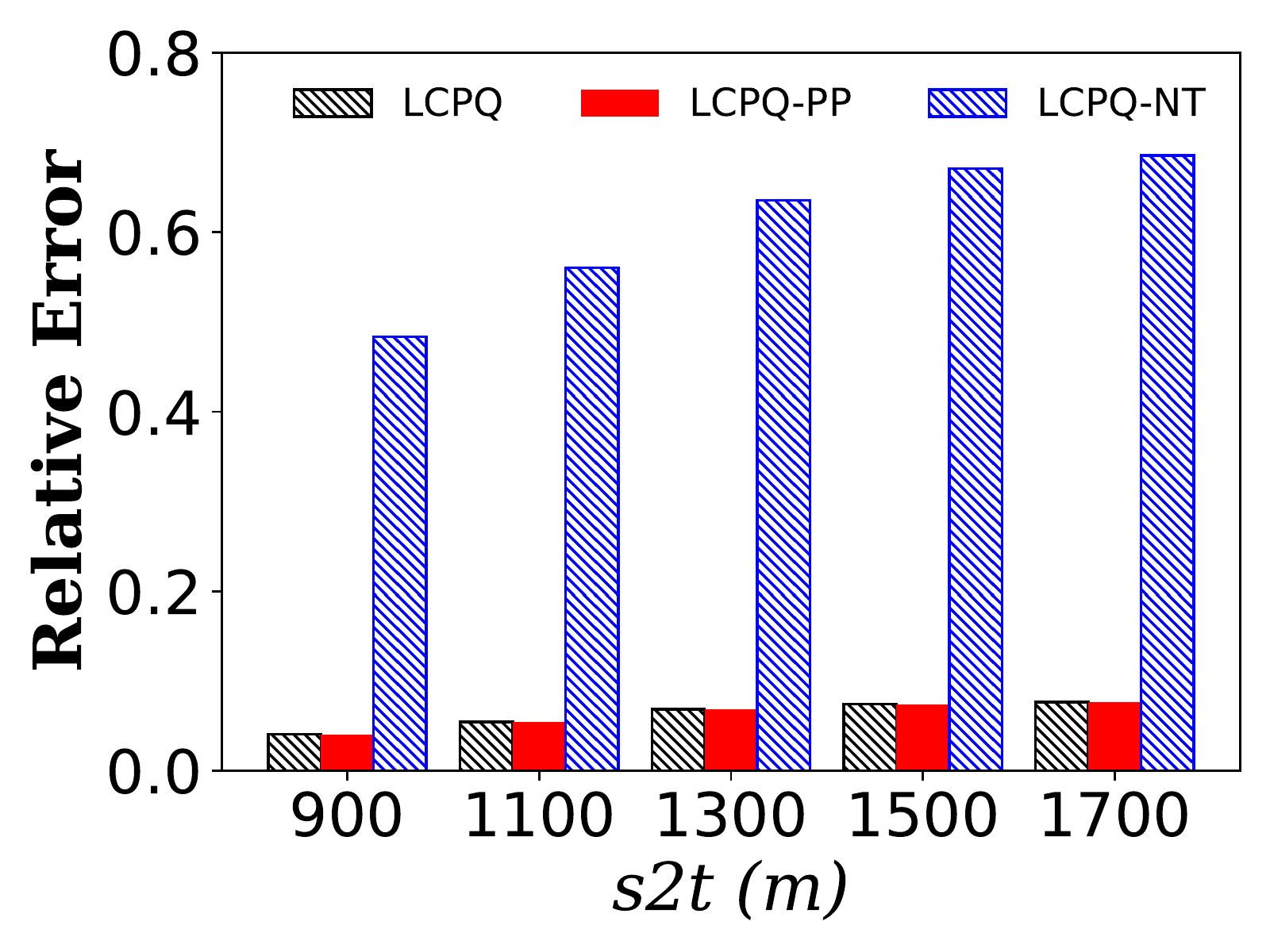}
\ExpCaption{\texttt{LCPQ}'s $\gamma$ vs. $\mathit{s2t}$ (Real)}\label{fig:LCPQ_s2t_error_HSM}
\vspace*{-6pt}
\end{minipage}
\end{figure*}

\subsection{Results on Real Data}

\begin{figure}[htbp]
    \centering
    \includegraphics[width=0.94\columnwidth]{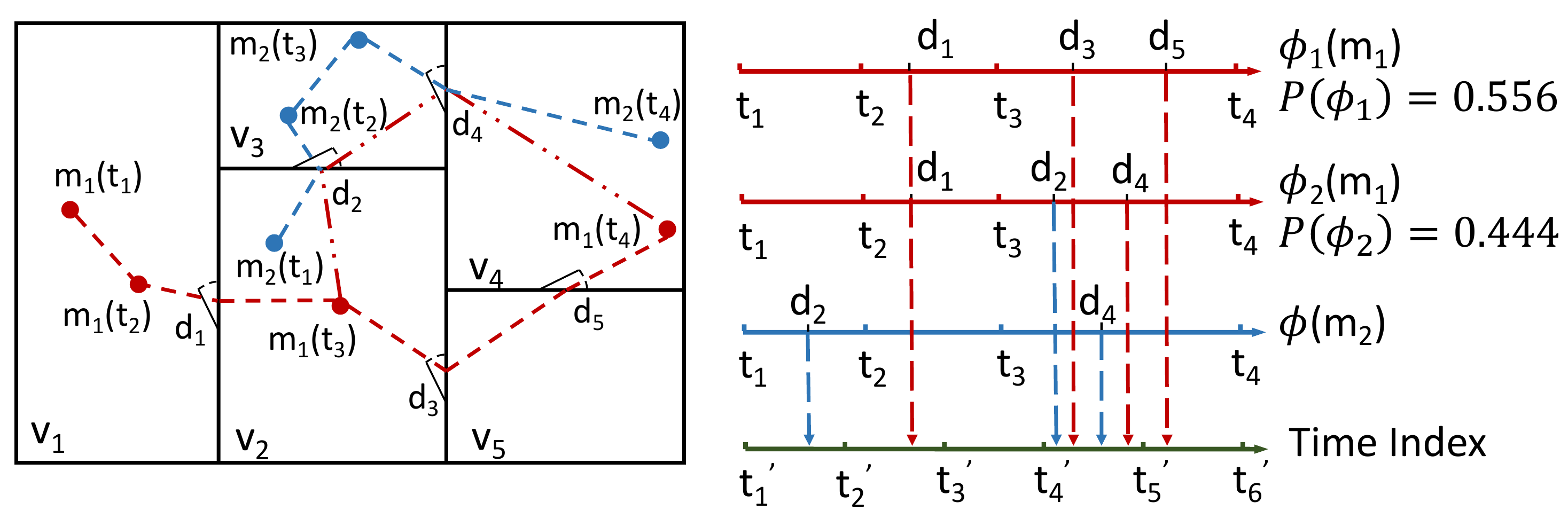}
    \caption{An Example of the Trajecory Data}
    \label{fig:trajectory}
\end{figure}

We collect a real dataset from a seven-floor, 2700m $\times$ 2000m shopping mall in Hangzhou, China. There are ten staircases each being roughly 20m long,
and 977 partitions connected by 1613 doors\footnote{We assume that there is no Q-partition in this shopping mall.
We varied the fraction of Q-partitions/R-partitions on synthetic data, but it shows little impact on all algorithms. The results are omitted.}.
The max capacity of a partition $v$ is $Area(v) \cdot 1 \text{~per~} m^2$.
We collected 1,598 object trajectories with totally more than 90,000 positioning records on 2017/01/05.
Nearly 12\% of two consecutive locations are not topologically-connected, i.e., not in the same partition or two adjacent partitions. The object movements in-between are uncertain.
To count flows against uncertainty, we applied a proven probabilistic method~\cite{li2018finding} as follows.
First, for every two consecutive locations not topologically-connected, a set $\Phi$ of valid sub-paths are found. Those sub-paths longer than twice of the shortest sub-path are excluded as the object unlikely took them.
Second, the probability that the object took sub-path $\phi_i \in \Phi$ is computed as $\mathsf{P}(\phi_i) = \frac{1/length(\phi_i)}{ \sum_{\phi_k \in \Phi} 1/length(\phi_k)}$. This way, a shorter sub-path has a higher probability to be taken.
Finally, the flow of a door $d$ is the sum of $\mathsf{P}(\phi_i)$s for all $\phi_i$s through $d$.
On top of the low-level flow computing, we sampled each door's flow every 10 seconds and used the samples to construct our indoor crowd model.
Figure~\ref{fig:trajectory} exemplifies a few trajectories, where $m_i(t_j)$ denotes the positioning location of a MAC address $m_i$ at time $t_j$.
For $m_1(t_3)$ and $m_1(t_4)$ that are not topologically-connected, two possible in-between paths are found, namely $\phi_1(m_1(t_3), d_3, d_5), m_1(t_4))$ of 20m long and $\phi_2(m_1(t_3), d_2, d_4), m_1(t_4))$ of 25m long. Their probabilities are $\mathsf{P}(\phi_1) = \frac{1/20}{1/20 + 1/25}$ $\approx$ $0.556$ and $\mathsf{P}(\phi_2) = \frac{1/25}{1/20 + 1/25}$ $\approx$ $0.444$. We sampled door flows as shown in the right part of Figure~\ref{fig:trajectory}. E.g., door $d_4$'s flow during $[t_4', t_5']$ is $1 + 0.444$ $=$ $1.444$ ($m_2$ with a probability of $1$ and $m_1$ with a probability of $0.444$).

\textbf{Comparison in default setting.}
We compare different methods for \texttt{FPQ} and \texttt{LCPQ} using real data and report the results in Table~\ref{tab:comparison_HSM}.
In terms of the running time and memory, \texttt{*PQ-NT} performs best while \texttt{*PQ-GTG} is the worst. This is because \texttt{*PQ-NT} skips the iterative population computations while \texttt{*PQ-GTG} uses an exact estimator but involves more node compared to our indoor crowd model.
In terms of hit rate and relative error, the results are similar to the counterparts in synthetic data.

\textbf{Effect of $\mathit{s2t}$.}
Figures~\ref{fig:FPQ_s2t_time_HSM} and~\ref{fig:FPQ_s2t_mem_HSM} report the results on running time and memory use of different \texttt{FPQ} searches.
All incur more time and memory as $\mathit{s2t}$ increases. Compared to \texttt{FPQ-PP} and \texttt{FPQ-NT}, \texttt{FPQ} and \texttt{FPQ-G} use more memory and more running time.
Compared to the counterparts on synthetic data, the search costs are higher since the shopping mall is a larger scale.
Referring to Figures~\ref{fig:FPQ_s2t_hit_HSM} and~\ref{fig:FPQ_s2t_error_HSM}, \texttt{FPQ-PP}/\texttt{FPQ-NT}'s search accuracy deteriorates. In terms of query hit rate, the search effectiveness of the two approximate algorithms is acceptable, considering that they have greatly improved the search efficiency.

The results in Figures~\ref{fig:LCPQ_s2t_time_HSM} and~\ref{fig:LCPQ_s2t_mem_HSM} exhibit similar trends as those in Figures~\ref{fig:FPQ_s2t_time_HSM} and~\ref{fig:FPQ_s2t_mem_HSM}.
\texttt{LCPQ-NT}'s relative error is much higher than that of \texttt{LCPQ} and \texttt{LCPQ-PP}, and it grows faster---more update timestamps involved due to a greater $\mathit{s2t}$ lead to more inaccurate cost estimates.
Compared to \texttt{FPQ} searches, \texttt{LCPQ} has lower hit rates and higher relative errors as reported in Figures~\ref{fig:LCPQ_s2t_hit_HSM} and~\ref{fig:LCPQ_s2t_error_HSM}.
In our definitions, partition-passing contact is more sensitive to the derived populations than the partition-passing time.

\subsection{Summary of Results}
We summarize our discoveries as follows.

First, in terms of running time and memory, the two approximate searches perform better than the two exact counterparts as workloads reduce.
Besides, Strategy \texttt{NT} costs less time and memory than Strategy \texttt{PP} since \texttt{NT} further utilizes historical information to skip timestamps.
For hit rate and relative error, \texttt{PP} outperforms \texttt{NT} in that \texttt{NT} skips many timestamps on the basis of \texttt{PP}, further decreasing the accuracy of intermediate results.

Second, the two approximate searches for \texttt{FPQ} perform better than those for \texttt{LCPQ} in terms of hit rate and error rate. The reason is that the partition-passing time is less sensitive to the populations compared to the partition-passing contact.

Third, a larger $\mathit{s2t}$ leads to more time and memory consumption but worse result accuracy, while a larger $\mathit{TI}$ exhibits an almost opposite trend.
In general, a larger $\mathit{s2t}$ or a smaller $\mathit{TI}$ means more timestamps to derive populations, which is critical to the cost estimation.
A larger floor number incurs more doors/partitions to explore and deteriorates the search efficiency. More initial object number $|o|$ leads to a higher hit rate and lower relative error as the population derivation is less variable.

Fourth, for the two baselines, \texttt{*PQ-GTG} performs poorly on efficiency because GTG contains more nodes to process. \texttt{*PQ-A} seems good both in efficiency and effectiveness. However, a user of \texttt{*PQ-A} cannot obtain the path before departure because \texttt{*PQ-A} needs to keep updating during expansion.

In general, the results show that the search algorithm with Strategy \texttt{PP} performs best. It costs relatively less time and memory and achieves good query result accuracy.
Strategy \texttt{NT} applies to the cases where door flows are updated frequently. In such a case, skipping some timestamps can improve efficiency without causing excessive errors in population estimates.

\section{Related Work}
\label{sec:related}

\textbf{Outdoor Time-Dependent Routing.}
In this setting, public transportation networks~\cite{hall1986fastest, brodal2004time, wang2015efficient} and road networks~\cite{ding2008finding, ardakani2012decremental, nannicini2012bidirectional, ardakani2015decremental, wei2020architecture} are modeled as discrete and continuous time-dependent graphs, respectively.
Hall~\cite{hall1986fastest} studies the least expected travel time path between two nodes in a public transportation network.
Brodal et al.~\cite{brodal2004time} propose efficient algorithms that find optimal itineraries for travelers using a train system with timetables.
Wang et al.~\cite{wang2015efficient} introduce an efficient indexing technique for route planning on timetable graphs.
Solutions for public transportation networks cannot solve our problem because they are mainly for a time-dependent graph with the static timetable for each station.

Ding et al.~\cite{ding2008finding} propose a Dijkstra-based algorithm to find the optimal path from A to B in a time-dependent road network, where the starting time is selected from a user-specified time interval.
Ardakani et al.~\cite{ardakani2012decremental} develop an adaptive approach to the continuous dynamic shortest path problem.
To speed up the shortest path query processing in continuous-time dynamic networks, Ardakani et al.~\cite{ardakani2015decremental} design an A$\ast$ algorithm using the decremental approach.
Wei et al.~\cite{wei2020architecture} design an efficient index for dynamic road networks.
Approaches for road networks do not work for indoor spaces because road network models do not support entities like doors, walls and rooms, altogether forming a complex topology.

The solutions for outdoor time-dependent routing are mainly Dijkstra-based algorithms~\cite{cooke1966shortest, brodal2004time, ding2008finding, yuan2019constrained}, A$\ast$ algorithms~\cite{nannicini2012bidirectional, ardakani2015decremental}, label-based methods~\cite{nachtigall1995time, wu2014path, wang2015efficient, wu2016reachability} (maily for time-dependent graph with timetable), adaptive approaches~\cite{gonzalez2007adaptive, ardakani2012decremental, ardakani2015decremental}.
Most of these works do not consider crowds that influence people's routing choices.
%

\textbf{Traffic-aware Routing.}
Shang et al.~\cite{shang2013finding} study traffic-aware fastest path (TAFP) query using a traffic-aware spatial network obtained by analysing uncertain trajectory data. It focuses on reconstructing trajectories with uncertain trajectory segments or positioning data rather than estimating traffic in the near future.
Demiryurek et al.~\cite{demiryurek2011online} propose an approach for the online computation of fastest paths in time-dependent spatial networks.
Nannicini et al.~\cite{nannicini2012bidirectional} study a bidirectional A$\ast$ search on time-dependent road networks.
Wang et al.~\cite{wang2019querying} propose a tree-structured index (TD-G-tree) to support fast route queries over time-dependent road networks.
The traffic information in these works~\cite{demiryurek2011online, nannicini2012bidirectional, wang2019querying} is prepared by mining historical trajectory data and known when routing.
Some adaptive approaches~\cite{gonzalez2007adaptive, ardakani2012decremental, ardakani2015decremental} can also solve traffic-aware routing problems, but they usually need continuous reevaluation.
Although these works take into consideration the traffic impact, none of them estimate the traffic in the near future while processing a query.


\textbf{Indoor Routing.}
Lu et al.~\cite{lu2012foundation} propose a distance-aware indoor space model and an indexing framework to facilitate indoor shortest path query.
To speed up distance-aware indoor pathfinding, Shao et al.~\cite{shao2016vip} design IP-tree and VIP-tree that enable more aggressive pruning. VIP-tree also supports indoor trip planning based on neighbour expansion~\cite{shao2018trip}.
Luo et al.~\cite{luo2016time} study the time-constrained sequenced route query (TCSRQ) in indoor space. The result of TCSRQ considers the stay-time period and types of indoor locations.
Zhou et al.~\cite{zhou2018indoor} propose an optimal indoor route planning method by introducing the navigation cost function and environment semantics into Dijkstra's algorithm.
Kim et al.~\cite{kim2018privacy} introduce an effective approach to compute top-$k$ routes between two points in an indoor environment by leveraging the indoor positioning datasets collected in a privacy-preserving manner.
Liu et al.~\cite{liu2019indoor} study indoor routing on a logical network that does not have notions for metrics but captures semantics and properties of indoor spaces.
Feng et al.~\cite{feng2020indoor} study the indoor top-$k$ keyword-aware routing query (IKRQ) to find $k$ routes that have optimal ranking scores integrating keyword relevance and spatial distance constraint.
Liu et al.~\cite{liu2020shortest} study the shortest path queries for indoor spaces where doors feature open and close times.
However, none of these works take moving objects into account, and thus their methods cannot be used for \texttt{LCPQ} and \texttt{FPQ} in this paper.


\textbf{Outdoor Flow, Crowd and Density.}
Tao et al.~\cite{tao2004spatio} use a novel technique to count spatio-temporal objects within a given spatial window during a given time interval.
Huang and Lu~\cite{huang2007snapshot} propose algorithms to find dense regions using location sensors distributed in a geographical area.
Tang et al.~\cite{tang2020general} design a general spatial-temporal graph attention based dynamic graph convolutional network model to predict traffic flow.
Castellano et al.~\cite{castellano2020crowd} introduce a crowd detection method for drone safe landing.
Wang et al.~\cite{wang2020sclnet} propose a spatial context learning network (SCLNet) for congested crowd counting.
Hao et al.~\cite{hao2008continuous} focus on continuously monitoring dense regions for moving objects.
These methods fall short in indoor spaces mainly for two reasons. First, indoor positioning techniques are usually RFID, Wi-Fi, and Bluetooth, which make coarser-grained location data. Second, the indoor topology is so different from the outdoor topology that indoor crowd modeling must consider carefully the connectivity among doors and partitions.

\textbf{Indoor Flow and Density.}
Ahmed et al.~\cite{ahmed2017finding} propose two graph-based models for indoor movement to map raw tracking records into mapping records with object entry and exit times in particular locations. To capture the historical object transitions and the duration of the transitions, they design two graphs, probabilistic flow graph (PFG) and aggregated probabilistic flow graph (APFG)~\cite{7517784}.
Li et al. propose to find the top-$k$ popular indoor semantic locations~\cite{li2018finding} with the highest flow values using probabilistic location samples, and the currently top-$k$ indoor dense regions~\cite{li2018search} by taking into account the uncertainty of online indoor positioning data.
These two works are different from our work.
On the one hand, the density analysis in this paper is based on coarse-grained flow values reported at door counters, not the point-based localization results count for individual moving objects.
On the other hand, our work focuses on path planning in presence of indoor crowds, while the previous works aim to find regions~\cite{li2018search} or locations~\cite{li2018finding} of interest based on mobility analytics.

\section{Conclusion and Future Work}
\label{sec:conclusion}

We study two types of crowd-aware indoor path planning queries. The $\texttt{FPQ}$ returns a path with the shortest travel time in the presence of crowds; the $\texttt{LCPQ}$ returns a path encountering the least objects en route.
To solve \texttt{FPQ} and \texttt{LCPQ}, we design a unified framework that consists of 1) an indoor crowd model that organizes indoor topology and captures indoor flows and densities; 2) a time-evolving population estimator that derives future time-dependent flows and populations for relevant partitions; 3) two exact and two approximate query processing algorithms that each can process both query types.
We conduct extensive experiments to evaluate our proposals. The results demonstrate the efficiency and scalability of the proposals and disclose the performance differences among all four algorithms.

There exist several directions for future research. First, it is interesting to consider other crowd models, e.g., learning crowd distributions and functions from historical data. 
Also, it is relevant to further speed up query processing by using an index, e.g., combining the object layer in the composite indoor index~\cite{xie2014distance, xie2013efficient} with a modified IP/VIP-Tree~\cite{shao2016vip} whose distance matrices are extended with time attributes.
Last but not least, it is possible to extend our proposals to support continuous monitoring of the fastest or least crowded paths.

\begin{acks}
 This work is an extension of the paper entitled ``Towards Crowd-aware Indoor Path Planning'' published at VLDB 2021.
 The work was supported by IRFD (No. 8022-00366B), ARC (No. FT180100140 and DP180103411), the Key R\&D Program (Zhejiang, China) (No. 2021C009) and NSFC (No. 62050099).
\end{acks}

\clearpage

\balance

\bibliographystyle{ACM-Reference-Format}
\bibliography{spatial.bib}

\clearpage
\begin{appendix}
	\section{Appendix}
\label{sec:appendix}

\begin{figure}[!ht]
    \centering
    \includegraphics[width=0.9\columnwidth]{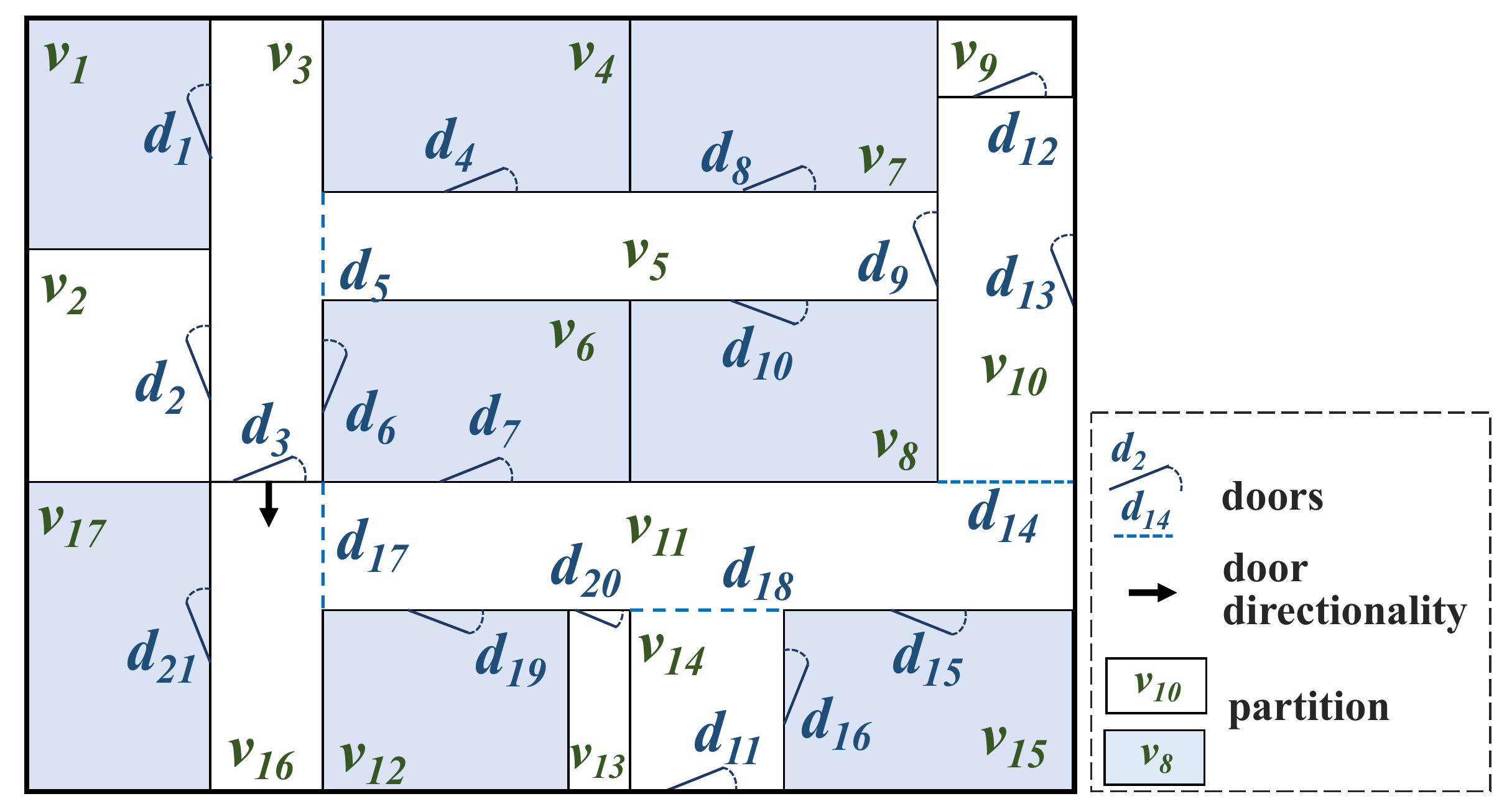}
    \caption{An Example of Indoor Floorplan}
    \label{fig:floorplanAppendix}
\end{figure}

\begin{figure}[!ht]
    \centering
    \includegraphics[width=0.7\columnwidth]{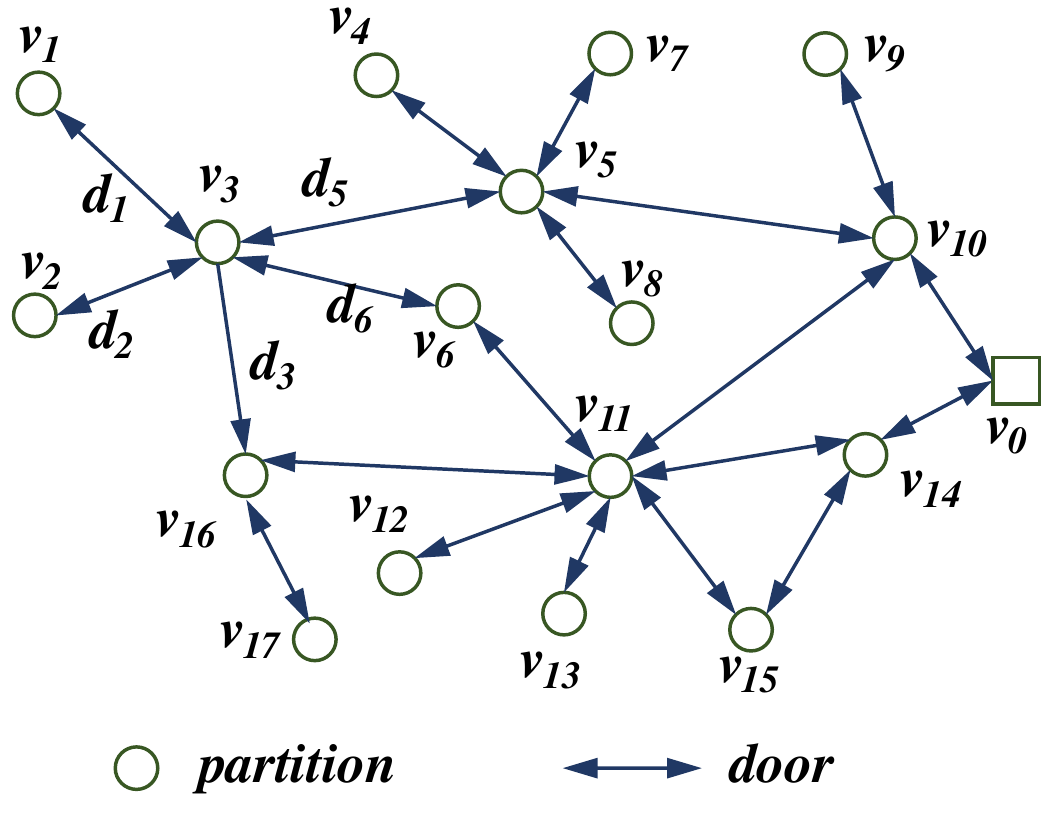}
    \caption{Indoor Model}
    \label{fig:modelAppendix}
\end{figure}

\begin{figure}[!ht]
    \centering
    \includegraphics[width=0.7\columnwidth]{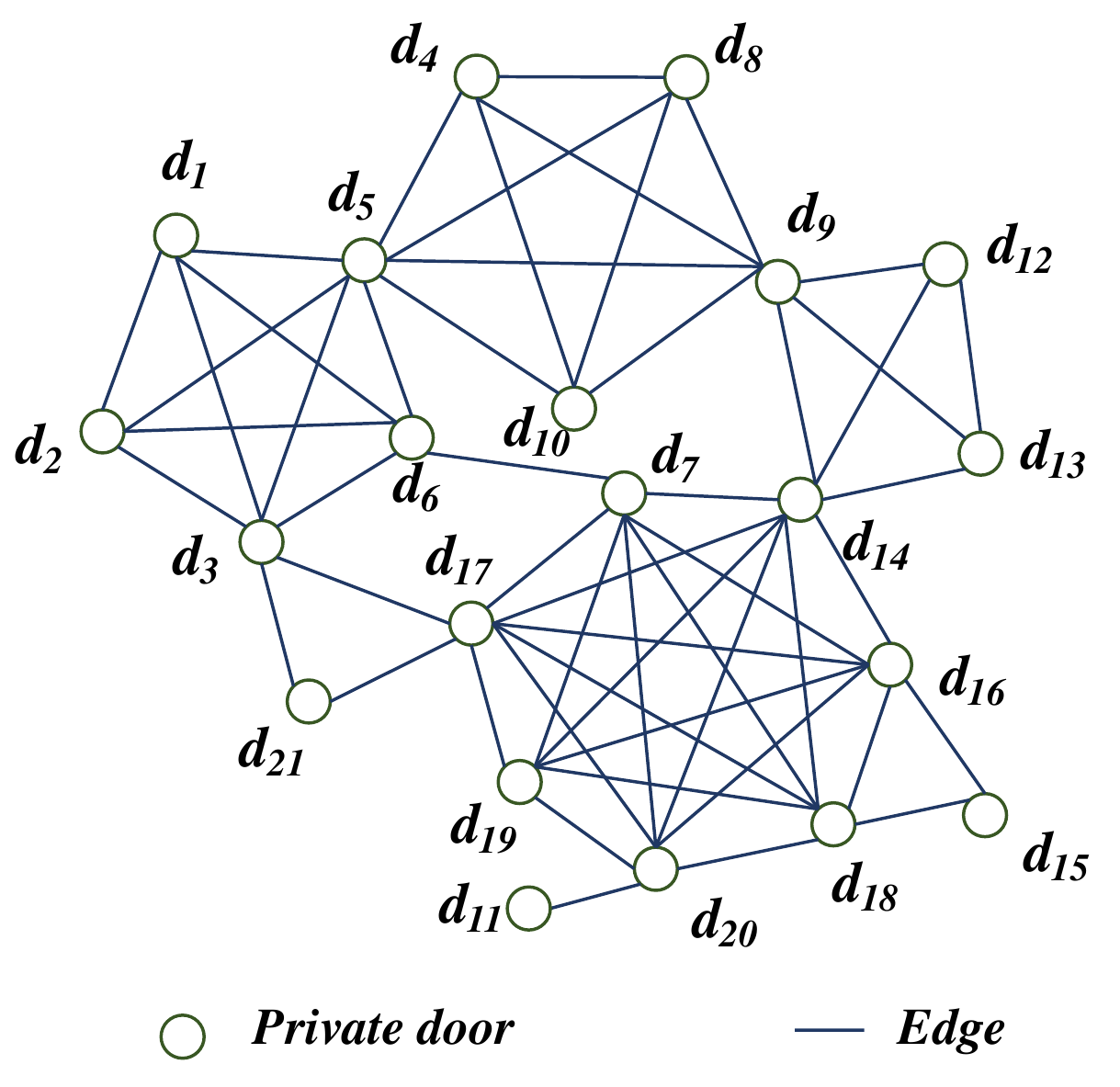}
    \caption{General Time-dependent Graph}
    \label{fig:modelAppendixGTG}
\end{figure}

We further compare the differences between the indoor model and the general time-dependent graph. Figure~\ref{fig:modelAppendix} and Figure~\ref{fig:modelAppendixGTG} show examples of the indoor model and the general time-dependent graph corresponding to Figure~\ref{fig:floorplanAppendix}. 

However, this way also falls short in our problem setting. First, it fails to represent the door directionality information directly as explained in Section~\ref{sec:model}.
Second, such a general time-dependent graph will result in many door-to-door edges for the same partition, which will render the graph based search inefficient. For example, in the model we use for indoor model (Figure~\ref{fig:modelAppendix}, there are 18 nodes and 21 edges. In contrast, the general graph model (Figure~\ref{fig:modelAppendixGTG}) contains 21 nodes and 54 edges.

\end{appendix}

\balance

\end{document}